\documentstyle[elsart12]{elsart}
\input psfig.tex
\def\met{\mbox{${\hbox{$E$\kern-0.6em\lower
               -.1ex\hbox{/}}}_{T}$}} 
\def\mpt{\mbox{${\hbox{$p$\kern-0.6em\lower
               -.05ex\hbox{/}}}_{T}$}} 
\def\mht{\mbox{${\hbox{$H$\kern
               -0.6em\lower-.1ex\hbox{/}}}_{T}$}} 
\def\vmet{\mbox{${\hbox{$\pol{E}$\kern
                -0.6em\lower-.1ex\hbox{/}}}_{T}$}} 
\def\vmpt{\mbox{${\hbox{$\pol{p}$\kern
                -0.4em\lower-.1ex\hbox{/}}}_{T}$}} 
\def\lsim{\mathrel{\rlap{\lower4pt\hbox{\hskip1pt$\sim$}}
    \raise1pt\hbox{$<$}}}         
\def\gsim{\mathrel{\rlap{\lower4pt\hbox{\hskip1pt$\sim$}}
    \raise1pt\hbox{$>$}}}         
\def\D0{D\O }

\def\PBARP{\mbox{$p\overline{p}$}}

\def\pt{\mbox{$p_{T}$}}

\begin{document}


\begin{frontmatter}

\title{Determination of the Absolute Jet Energy Scale 
in the \D0 Calorimeters}
%
%
%
\begin{flushleft}
\author{ \small                                    
B.~Abbott,$^{31}$                                                             
M.~Abolins,$^{27}$                                                            
B.S.~Acharya,$^{46}$                                                          
I.~Adam,$^{12}$                                                               
D.L.~Adams,$^{40}$                                                         
M.~Adams,$^{17}$}
\author{ \small                                                              
S.~Ahn,$^{14}$                                                                
H.~Aihara,$^{23}$                                                             
G.A.~Alves,$^{10}$                                                            
N.~Amos,$^{26}$
% ...
}                                               

\end{flushleft}
                                                                            
\vskip 0.70cm                                                                 
\centerline{(D\O\ Collaboration)}                                             
\vskip 0.70cm                                                                 

{\small

\begin{center}
                                                  
% ...
\address{$^{10}$LAFEX, Centro Brasileiro de Pesquisas F{\'\i}sicas,        
                  Rio de Janeiro, Brazil}                                     
% ...
\address{$^{12}$Columbia University, New York, New York 10027}             
% ...
\address{$^{14}$Fermi National Accelerator Laboratory, Batavia,            
                   Illinois 60510}                                            
% ...
\address{$^{17}$University of Illinois at Chicago, Chicago,                
                   Illinois 60607}                                            
% ...
\address{$^{23}$Lawrence Berkeley National Laboratory and University of    
                   California, Berkeley, California 94720}                    
% ...
\address{$^{26}$University of Michigan, Ann Arbor, Michigan 48109}         
\address{$^{27}$Michigan State University, East Lansing, Michigan 48824}   
% ...
\address{$^{31}$New York University, New York, New York 10003}             
% ...
\address{$^{40}$Rice University, Houston, Texas 77005}                     
% ...
\address{$^{46}$Tata Institute of Fundamental Research,                    
                   Colaba, Mumbai 400005, India}                              
% ...

\end{center}
            
}                                                                 

%
\begin{abstract}

The \D0 detector is used to study \PBARP\ collisions at the
1800~GeV and 630~GeV center-of-mass energies available at the Fermilab 
Tevatron.
To measure jets, the detector uses a sampling calorimeter composed of
uranium and liquid argon as the passive and active media respectively. 
Understanding the jet energy calibration is not only crucial
for precision tests of QCD, but also for the measurement of particle masses 
and the determination of physics backgrounds associated with new phenomena.
This paper describes the energy calibration of jets observed with
the \D0 detector at the two \PBARP\ center-of-mass energies 
in the transverse energy and pseudorapidity range 
$E_{T}>8\;{\mathrm GeV}$ and $|\eta|<3$.

\end{abstract}

\end{frontmatter}

\clearpage
\newpage

\tableofcontents

\clearpage
\newpage

\section{Introduction}\label{sec:intro}

Jet production is the dominant process in \PBARP\ collisions at
$\sqrt{s}$=1.8~TeV. Because almost every physics measurement at the Tevatron
involves events with jets, an accurate energy calibration is essential.
Currently, the jet energy scale is still the major source of systematic
uncertainty in both the \D0 inclusive jet cross section and top quark mass
measurements. 

This paper describes the determination and verification of the
jet energy calibration at \D0. The methods developed here 
are based on previous work~\cite{keh1,keh2}.
The calibration is performed for jets reconstructed with different
algorithms, from data taken in \PBARP\ collisions
at $\sqrt{s}=1800\;{\mathrm GeV}$ and 
630~GeV, but only representative plots are shown.
For detailed information see Refs.~\cite{note,630}.

The calorimeters are the primary tool for jet measurements at \D0. 
A brief summary of the characteristics and performance of these
subdetectors is given in Section~\ref{sec:calor}.
The definition of a jet and a description of the algorithm used in jet 
reconstruction
are included in Section~\ref{sec:jets}. Sections~\ref{sec:off}-\ref{sec:shower}
enumerate the different corrections involved in the jet energy 
calibration, explain the methods used in their derivation, and
display the numerical results. Section~\ref{sec:mc} is dedicated to  
verification tests based on a Monte Carlo derived correction. 
The results are summarized and conclusions are drawn in Section~\ref{sec:conc}.

\section{\D0 \hskip2mm Calorimeters}\label{sec:calor}

The \D0 Uranium-Liquid Argon sampling calorimeters~\cite{nim} are hermetic,
uniform, and provide nearly complete solid angle coverage.
The Central (CC) and End (EC) Calorimeters contain approximately 7 and
9 interaction lengths of material, respectively, ensuring
containment of nearly all particles except high-$p_{T}$ muons and 
neutrinos.
The intercryostat region (IC), between the CC and the EC
calorimeters, is covered by an intercryostat detector (ICD) and massless 
gaps (MG)~\cite{nim}. The ICD consists of an array of scintillator 
tiles located on the EC cryostat wall. The MG are separate single-cell 
structures installed in the CC and EC calorimeters between the module 
endplates and the cryostat wall. Given their fine segmentation
and excellent energy resolution,
the \D0 calorimeters are especially well suited to jet measurements and the
determination of missing energy in the plane transverse to the beams (\met). 

\begin{figure}[p]
\centerline{\psfig{figure=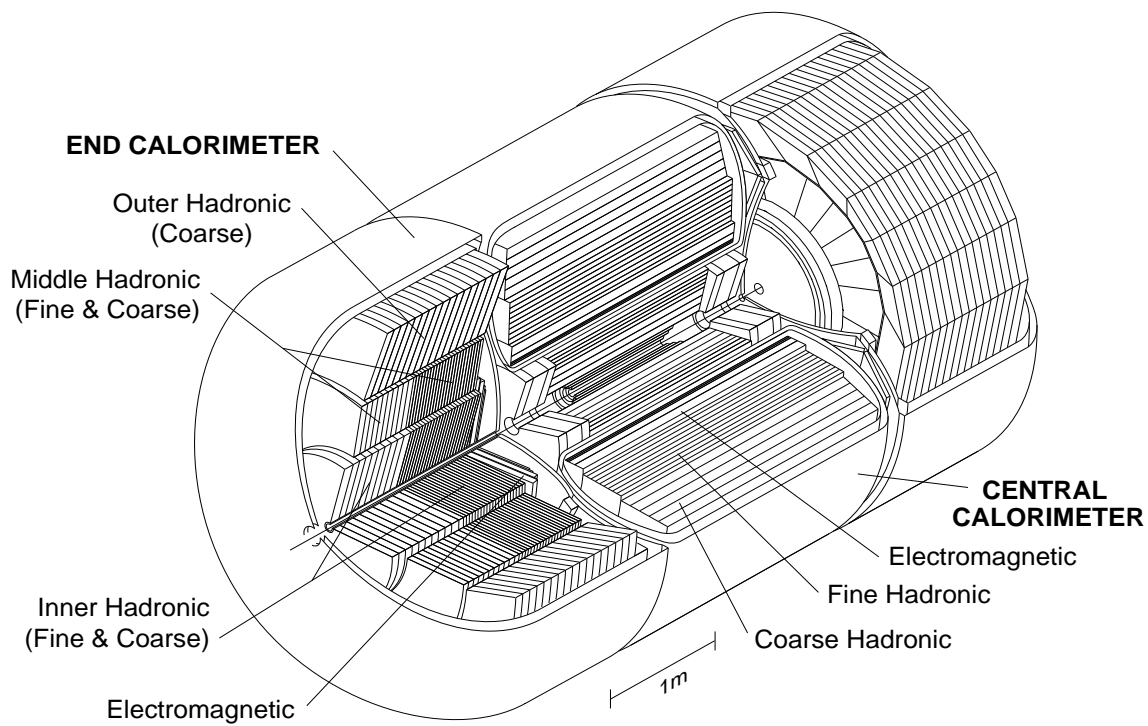,height=12cm,width=15cm}}
\caption{General view of \D0 \hskip1mm calorimeters.}
\label{fig:cal_general}
\end{figure}

\begin{figure}[p]
\centerline{\psfig{figure=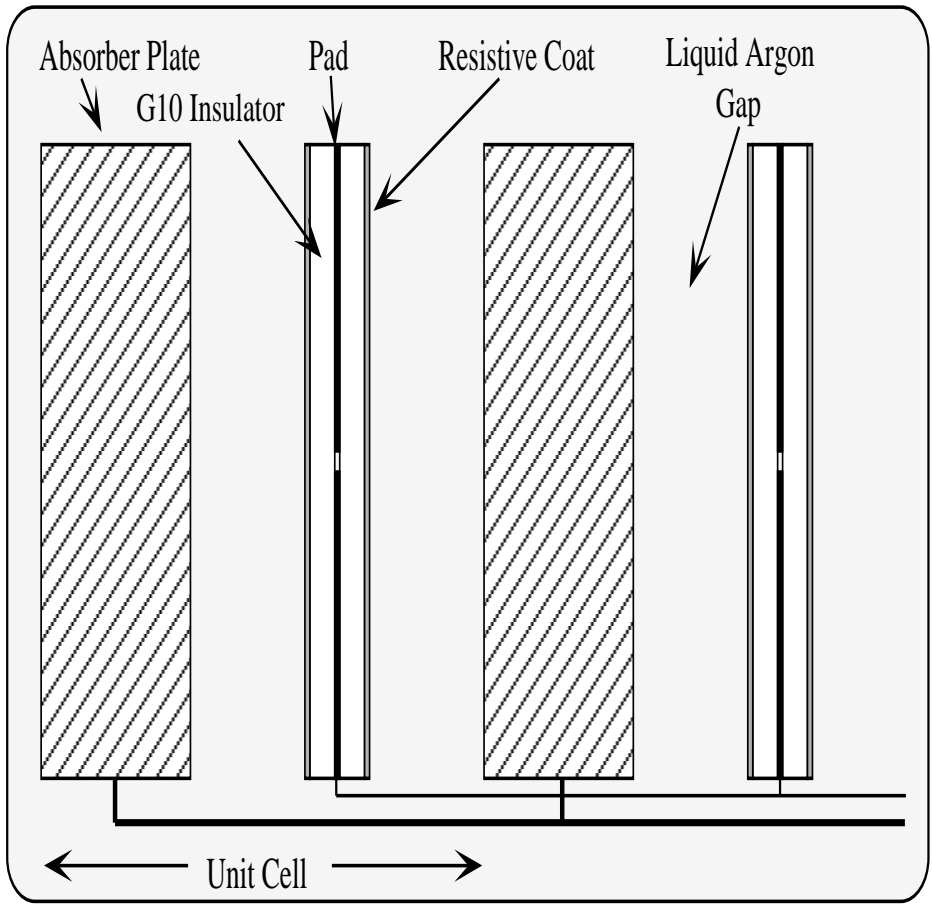,height=13cm,width=13cm}}
\caption{Diagram of a \D0 calorimeter unit cell.}
\label{fig:unit_cell}
\end{figure}

\begin{figure}[p]
\centerline{\psfig{figure=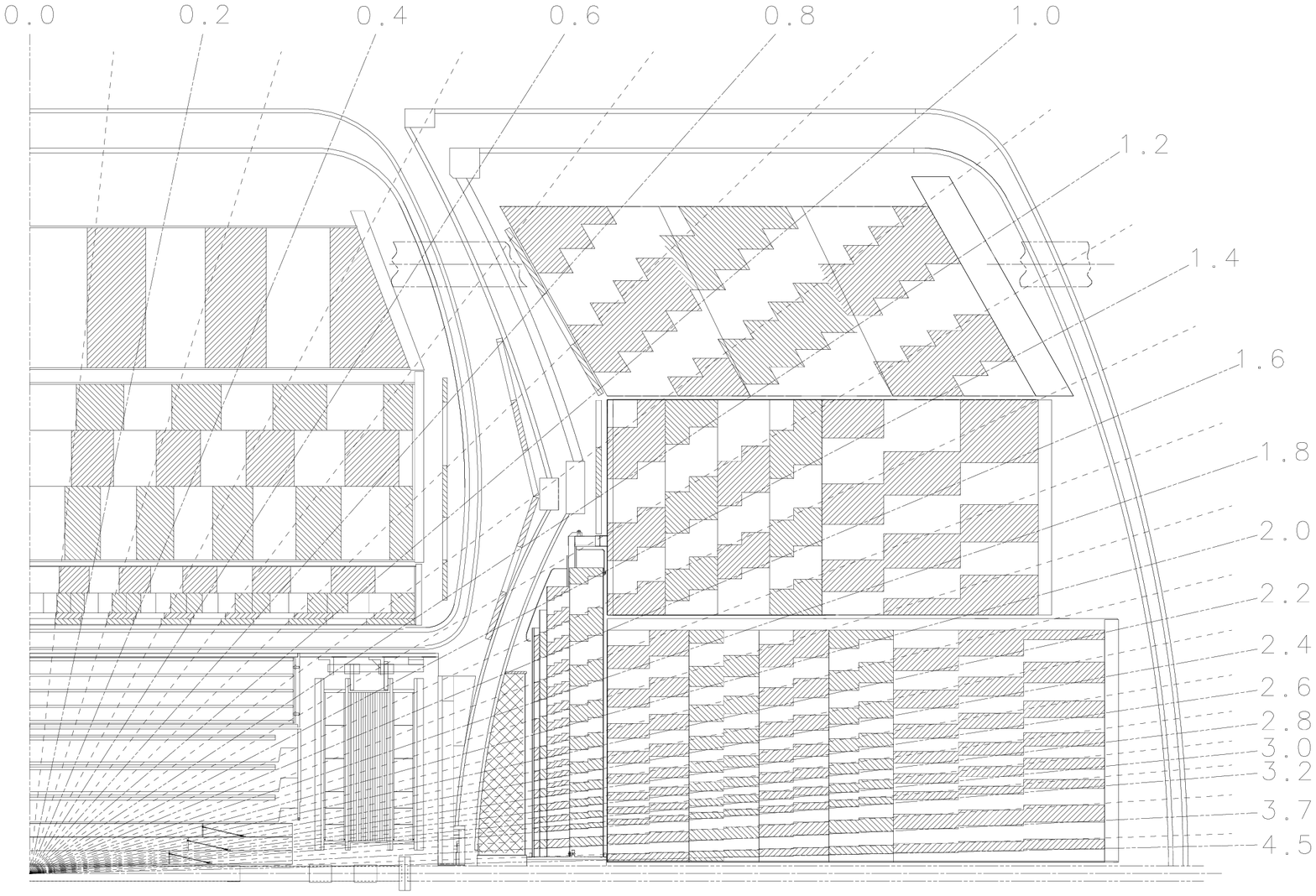,height=15cm,width=15cm}}
\caption{Side view of the \D0 calorimeters (one quadrant). 
The numbers indicate position in units of pseudorapidity.}
\label{fig:cal_side_eta}
\end{figure}

A general view of the \D0 calorimeters is displayed 
in Fig.~\ref{fig:cal_general}, showing the different
electromagnetic and hadronic sections.
The cells are constructed of uranium absorber plates interleaved with
Cu/G10 readout boards. The gaps are filled with LAr as the active medium.
A cross section of a unit cell is shown in
Fig.~\ref{fig:unit_cell}. The coarse hadronic sections are 
made of copper or steel instead of uranium. 
A side view of the \D0 calorimeters is shown in Fig.~\ref{fig:cal_side_eta}. 
\D0 defines a coordinate system with the origin at
the geometric center of the detector and the $z$ axis along the direction
of the proton beam. The transverse plane is defined by the 
cartesian axes $x$ and $y$.
In spherical coordinates, the three coordinates are described by the radius 
$r$,
the azimuthal angle $\varphi$, and the polar angle $\theta$. 
Pseudorapidity is defined
as $\eta=-\ln [{\mathrm tan} (\theta/2)]$. The segmentation in
$\Delta \eta \times \Delta \varphi$ space is 0.1$\times$0.1, or 
0.05$\times$0.05 at shower maximum (third electromagnetic layer).

The single particle energy resolutions for electrons ($e$) and charged
pions ($\pi$) were measured from test beam data, and parameterized with a
$(\sigma_{E}/E)^{2}= S^{2}/E+C^{2}$ functional form.
For electrons, the sampling term $S$ is 14.8 (15.7)\% in the CC (EC), 
and the constant term $C$ is 0.3\% in both the CC and EC. For pions, 
the sampling term is 47.0 (44.6)\%, and the constant term is 4.5 (3.9)\% 
in the CC (EC)~\cite{kot}.

The energy of a jet measured in the calorimeters is distorted by 
uninstrumented regions, phenomena which affect the response, energy 
deposits in the liquid argon due to uranium 
decays, spectator interactions, reconstruction and resolution effects. 
The calorimeter response to the different types of particles
is the most important aspect of the jet energy scale correction.
Electromagnetically interacting particles, like photons ($\gamma$) and 
electrons,
deposit most of their energy in the electromagnetic (EM) section of the 
calorimeters ($\sim 20.5$ radiation lengths thick).
Hadrons, by contrast, lose energy primarily through nuclear interactions.
The showers they cause are much longer than those from electromagnetic
particles, and extend through the three sections of the calorimeter:
EM, fine hadronic (FH), and coarse hadronic (CH). In general,
the calorimeter response to the EM ($e$) and non-EM ($h$) components of hadron
showers is not the same because there are different physics 
processes involved. {\em Non-compensating} 
calorimeters have a response ratio $e/h>1$, and suffer from non-Gaussian
event-to-event fluctuations in the fraction of the energy lost through
$\pi^{0}$ production ($f_{\mathrm{em}}$). In addition,
$\langle f_{\mathrm{em}} \rangle$
increases logarithmically with energy. As a result, such
calorimeters give a non-Gaussian signal distribution for
monoenergetic hadrons. Moreover, the signal is not linearly proportional to
the hadron energy ({\em non-linearity}), and $e/\pi$ 
(ratio of electron to pion response) is
energy dependent~\cite{wigmans}.

\begin{figure}[p]
\psfig{figure=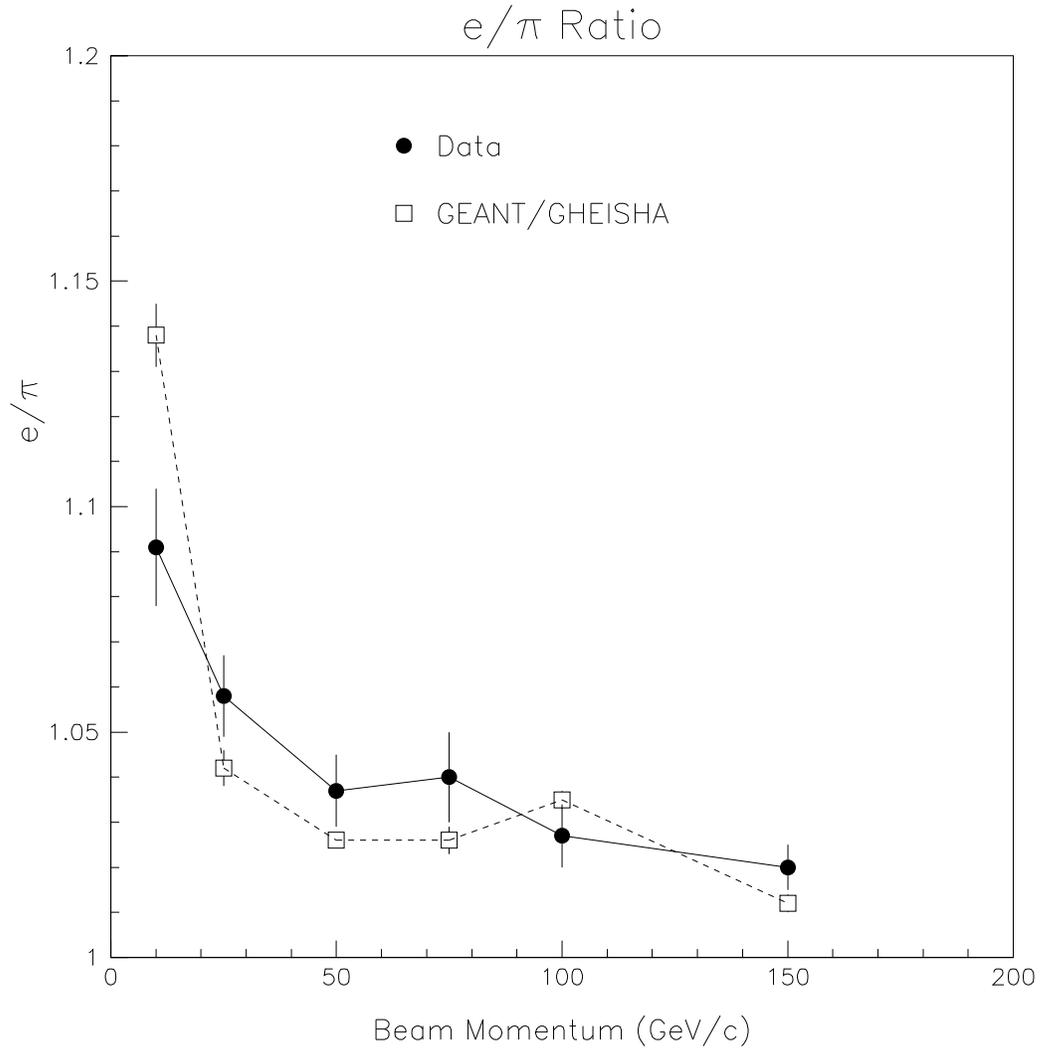,height=15cm,width=15cm}
\caption{ The ratio $e/\pi$ as measured from test beam data and determined
from a Monte Carlo simulation. Note that the Monte Carlo $e/\pi$ ratio 
(open squares) changes faster and flattens out earlier than the test beam 
ratio (full circles).}
\label{fig:etopi}
\end{figure}

\begin{figure}[p]
\psfig{figure=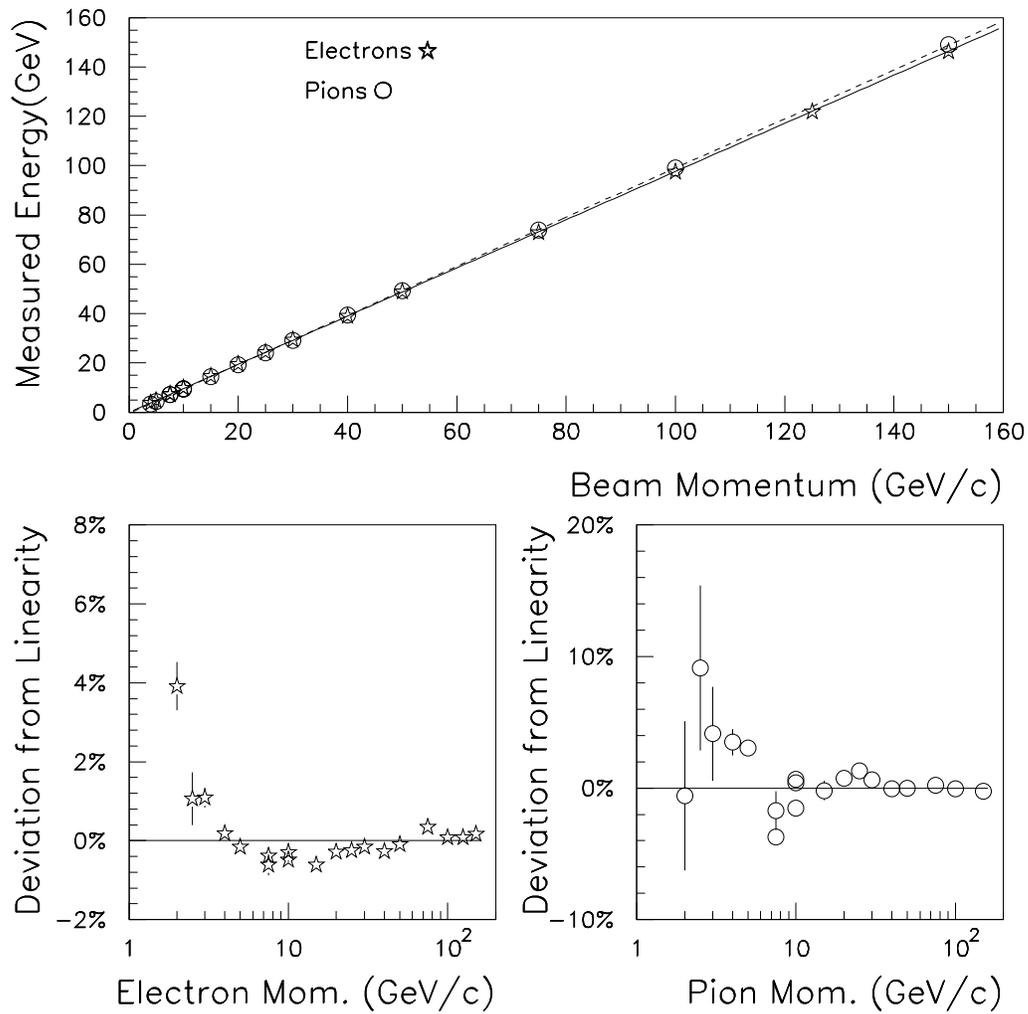,height=15cm,width=15cm}
\caption
{Measured test beam energy versus particle momentum for electrons and pions.
The solid and dashed lines are fits to the electron and pion data, 
respectively. Good linearity between reconstructed and test beam energy is 
achieved with the \D0 calorimeters.}
\label{fig:linearity}
\end{figure}

\begin{figure}[p]
\psfig{figure=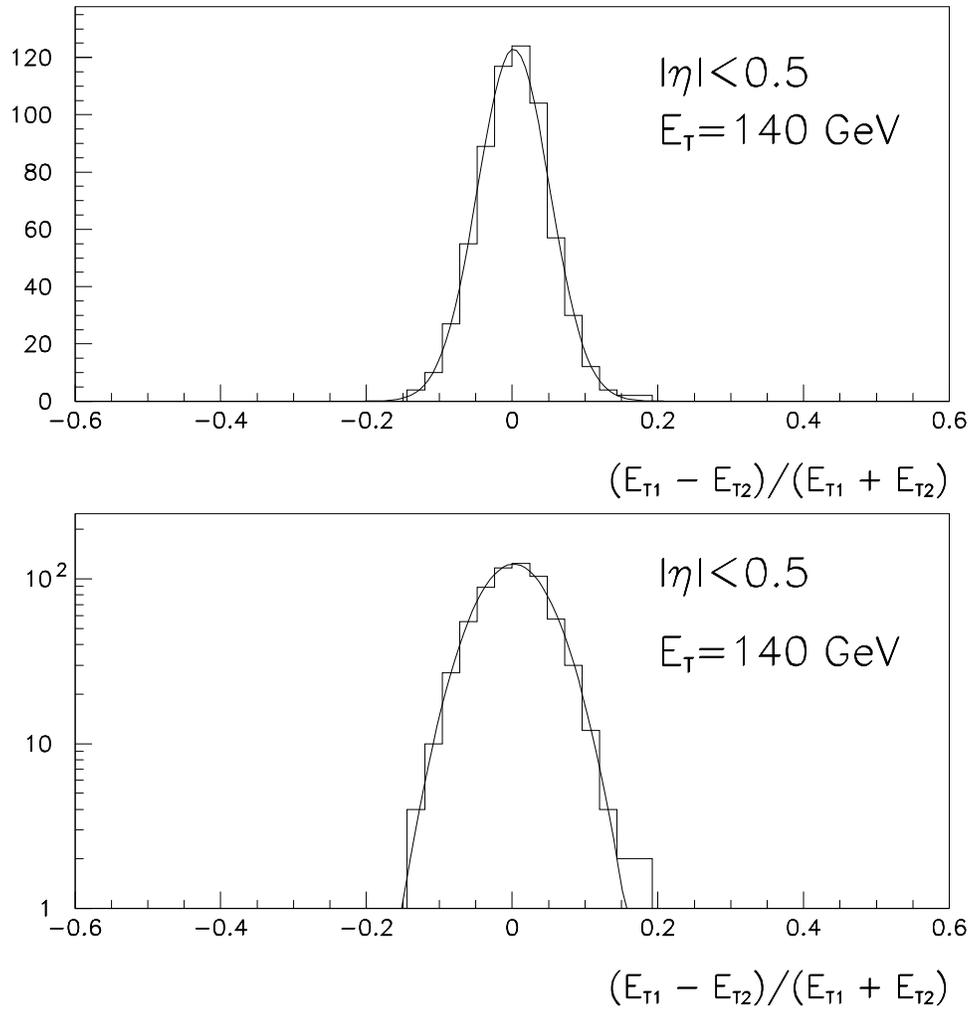,height=15cm,width=15cm}
\caption
{$(E_{T1}-E_{T2})/(E_{T1}+E_{T2})$ 
distribution from dijet events observed 
in the \D0 central calorimeter. The data are well described by a Gaussian,
showing the hermeticity and linearity of the calorimeters.}
\label{fig:delta_gauss}
\end{figure}

The \D0 calorimeters are nearly
compensating, with an $e/\pi$ ratio less than 1.05
above 30~GeV. Figure~\ref{fig:etopi} shows the $e/\pi$ ratio from a   
Monte Carlo~\cite{GEANT,geisha} simulation (open squares) and 
from test beam data (full circles). 
The calorimeters are also linear with energy, except at very low energy, 
as illustrated in Fig.~\ref{fig:linearity} from test beam data.
Figure~\ref{fig:delta_gauss} shows the Gaussian behavior of the variable
$(E_{T1}-E_{T2})/(E_{T1}+E_{T2})$
for dijet events with average jet transverse energy 
$E_{T}$ of 140~GeV  observed in the \D0 calorimeters,
where $E_{T1}$ and $E_{T2}$ are randomly ordered from the 
pair of leading transverse energy jets. This Gaussian signature is 
characteristic of the hermeticity and linearity of the \D0 calorimeters.

\section{Jet Measurement}\label{sec:jets}

Jets emerging from a hard interaction
deposit energy in numerous cells in the calorimeter.
To determine the jet energy, an algorithm is needed which assigns
calorimeter cells to jets. The final measured jet energy depends on this
algorithm or jet definition. In this paper, 
only a fixed cone algorithm is considered~\cite{defi}, with
jet centroid and cone size defined as 
($\eta_{\mathrm{jet}}$,$\varphi_{\mathrm{jet}}$) and
${\mathcal R}=\sqrt{(\varphi-\varphi_{\mathrm jet})^{2}+
(\eta-\eta_{\mathrm jet})^{2}}$, respectively. 

The process of reconstructing jets in the calorimeter is iterative.
First, towers ($\Delta \eta \times \Delta 
\varphi=0.1\times 0.1$) containing $E_{T}$ of
1~GeV or more are used as seeds for finding preclusters,
which are formed by adding neighboring towers within a radius of
${\mathcal R}=0.3$ to seed towers.
The $\eta_{i}$ of each tower is calculated with respect to 
the interaction vertex, which is determined from the 
tracking system~\cite{nim}.
Next, a fixed cone of radius ${\mathcal R}$
is drawn around each precluster centered at its centroid.
A new center is then calculated using the 
Snowmass Accord~\cite{snow} definitions for both the jet angles and
transverse energy:

\begin{eqnarray}
\hskip3.5cm E_{T\mathrm{jet}} & = &  \sum_{i} E_{T_{i}} \, , \\
\hskip3.5cm \varphi_{\mathrm{jet}} & = 
 & \sum_{i} E_{T_{i}} \varphi_{i}/E_{T\mathrm{jet}} \, , \\ 
\hskip3.5cm \eta_{\mathrm{jet}} & = 
 & \sum_{i} E_{T_{i}} \eta_{i}/E_{T\mathrm{jet}} \, .
\label{eq:snowmass}
\end{eqnarray}

The sums are over all towers contained in the cone.
This process is repeated iteratively, always using the Snowmass Accord
definition, replacing the previous seed or jet direction by the current
one until the center becomes stable.
In the final step, $\eta_{\mathrm{jet}}$ and $\varphi_{\mathrm{jet}}$
are recalculated using the following definition:

\begin{eqnarray}
\hskip2cm \varphi_{\mathrm{jet}} & = & \tan^{-1} \left( \sum_{i} 
E_{y_{i}} / \sum_{i} E_{x_{i}} \right) \\
\hskip2cm \theta_{\mathrm{jet}} & = & \tan^{-1} \left( \sqrt{\sum_{i} 
E_{x_{i}}^{2} + \sum_{i} 
E_{y_{i}}^{2}} / \sum_{i} E_{z_{i}} \right)
\label{eq:theta-phi}
\end{eqnarray}

where $E_{x_{i}}= 
E_{i}\sin \theta_{i} \cos \varphi_{i}$, 
$E_{y_{i}} =  
E_{i}\sin \theta_{i} \sin \varphi_{i}$, and
$E_{z_{i}} =  
E_{i}\cos \theta_{i}$.
The transverse energy of a jet is still the sum of the
transverse energies in each calorimeter tower inside the cone. 
Those jets
with $E_{T}<8\;{\mathrm GeV}$ are discarded.
Overlapping jets are merged into a single jet if more than 50$\%$ of the
$E_{T}$ of the lower energy jet is contained in the overlap
region. Otherwise, the energy of each cell in the overlap region is 
assigned to the nearest jet.

The {\em particle\/} level jet energy is defined as the energy 
of a jet
found from final state particles using a similar cone algorithm to that used 
at the {\em calorimeter\/} level. The particle jet is never observed in
data because it gets distorted when it interacts with the calorimeter material.
But it can be clearly defined in a Monte Carlo sample by clustering together 
the final state particles, 
instead of calorimeter cells, into a fixed cone size.
The particle jet includes only particles arising from partons participating 
in the hard scatter. The purpose of the jet energy scale calibration is to
correct observed jets to the particle level. 

\subsection{Jet Energy Scale}

The conversion from electronic ADC counts to energy and the layer-to-layer 
calibration are performed at the data reconstruction level and are based on 
test beam data. 
Detailed information on calorimeter 
performance and calibration can be found in Ref.~\cite{kot}.

This paper describes an {\em in situ\/} calibration based primarily on 
reconstructed collider data. 
This calibration relates the observed jet energy to that of the final 
state particle jet on average.
The linearity and hermeticity of the \D0 calorimeters provide Gaussian
response functions and, therefore, allow
separation of the jet energy resolution from the jet energy scale correction.
Thus, the following sections focus on the scale
correction and leave aside resolution effects. 
The \D0 jet energy calibration is based almost 
entirely on data measurements, exploiting conservation of transverse momentum 
through an accurate determination of the event \met.

The particle level or true jet energy $E_{\mathrm{jet}}^{\mathrm{ptcl}}$ is 
obtained from the measured jet energy 
$E_{\mathrm{jet}}^{\mathrm{meas}}$ using the following relation:

\begin{equation}
  E_{\mathrm{jet}}^{\mathrm{ptcl}} = 
  \frac{E_{\mathrm{jet}}^{\mathrm{meas}}- 
  E_{\mathrm{O}}({\mathcal R},\eta,{\mathcal L})}
  {R_{\mathrm{jet}}({\mathcal R},\eta,E)S({\mathcal R},\eta,E)} \, , 
  \label{eq:jes}
\end{equation}

where:

\begin{itemize}

\item {\bf $E_{\mathrm{O}}({\mathcal R},\eta,{\mathcal L})$}
is an offset, which includes the physics underlying event, noise from
the radioactive decay of the uranium absorber,
the effect of previous crossings (pile-up), and the 
contribution of additional \PBARP\ interactions.
The physics underlying
event is defined as the energy contributed by spectators to the hard parton
interaction which resulted in
the high-$p_{T}$ event. 
The offset increases as a function of the algorithm cone size ${\mathcal R}$.
The dependence on luminosity, ${\mathcal L}$,
arises from the contribution of additional interactions. $E_{\mathrm{O}}$ is
also parameterized as a function of $\eta$, measured with respect to the
vertex of the interaction (physics $\eta$). 

\item {\bf $R_{\mathrm{jet}}({\mathcal R},\eta,E)$} is the energy response
of the calorimeter to jets. It is parameterized
as a function of jet energy after the offset subtraction. 
$R_{\mathrm{jet}}$ is independent of the 
algorithm cone size; however, its parameterization versus
jet energy is cone size dependent because larger
cones encompass a larger fraction of the energy of the cluster.
Because the various detector components are not identical, 
$R_{\mathrm{jet}}$ is also dependent on detector pseudorapidity, that is
$\eta$ calculated from the geometric center of the detector and not the 
interaction vertex. 
$R_{\mathrm{jet}}$ is typically less than one, due to energy lost in
uninstrumented regions between modules, differences between electromagnetic
and nuclear interacting particles ($e/h>1$), and module-to-module
inhomogeneities. 

\item {\bf $S({\mathcal R},\eta,E)$} is the fraction of the particle jet
energy that is deposited inside the algorithm cone. The jet energy is
corrected back to the particle level, and therefore calorimeter showering 
effects must be removed. 
$S$ is less than one, meaning that the effect of
showering is a net flux of energy from the inside to outside the cone.
$S$ depends strongly on the cone size ${\mathcal R}$, energy, and $\eta$. 
It is parameterized as a function of jet energy after offset subtraction and
response correction, and binned in terms of physics $\eta$.

Most of the examples presented in this paper correspond to jets reconstructed
with a ${\mathcal R}=0.7$ fixed cone algorithm.  
Jets reconstructed with other cone sizes or at $\sqrt{s}=630 \; {\mathrm GeV}$ 
were also calibrated. This 
information is included in Refs.~\cite{note,630}.

\end{itemize}

\section{Offset Correction}\label{sec:off}

The \PBARP\ inelastic collisions are classified as non-diffractive,
single diffractive and double diffractive. 
An inelastic non-diffractive collision, also sometimes called hard core 
interaction,
is an event in which both the proton and the anti-proton break up. 
A particular case of a hard core interaction is high-$p_{T}$ or hard parton
scattering.

The offset correction is determined for non-diffractive events. It subtracts 
energy which is not associated with the high-$p_{T}$ interaction. 
This excess energy includes the contributions
of uranium noise, pile-up, and additional \PBARP\
interactions. Pile-up is the residual energy from previous \PBARP\ crossings,
which results from the long shaping times of electronic signals in calorimeter 
readout cells. Pile-up depends on the number of interactions in 
the previous
beam crossings, and is therefore luminosity dependent.
The underlying event includes soft interactions
between spectator partons that constituted the colliding proton and 
anti-proton. This energy can be subtracted or not, depending on the
characteristics of each particular physics measurement.

A crossing selected by a high-$p_{T}$ trigger can be modeled as the
sum of a hard parton scattering overlaid with a zero bias (ZB) event at the 
same luminosity. The ZB trigger accepts every \PBARP\ crossing, regardless of
whether or not it contains an actual collision.
The average number of hard interactions in a high-$p_{T}$ event, 
$\langle N \rangle$, may be written as:

\begin{equation}
  \langle N \rangle = 1 + \langle N_{\mathrm{ZB}} \rangle \, . \label{eq:off1}
\end{equation}

Here, $\langle N_{\mathrm{ZB}} \rangle$ is the average number of hard core 
interactions in a ZB event and  can be expressed as:

\begin{equation}
\langle N_{\mathrm{ZB}} \rangle = 
\sum_{n=0}^{\infty} n \, P(n) \, , \label{eq:off2}
\end{equation}

where the probability of having $n$ hard core collisions at a
given luminosity, $P(n)$, follows a Poisson distribution.

The total offset correction, $E_{\mathrm{O}}$, can be presented as the sum:

\begin{eqnarray}
\hskip3cm E_{\mathrm{O}} & = 
          & (1+\langle N_{\mathrm{ZB}} \rangle) E_{\mathrm{ue}}  + 
 E_{\mathrm{noise}} + E_{\mathrm{pile}} \nonumber \\
\hskip3cm       & = & E_{\mathrm{ue}} + \langle N_{\mathrm{ZB}} 
       \rangle E_{\mathrm{ue}} + 
             E_{\mathrm{noise}} + E_{\mathrm{pile}} \, , \label{eq:off3}
\end{eqnarray}

where $E_{\mathrm{ue}}$ is the energy associated with the
physics underlying event, $\langle N_{\mathrm{ZB}} \rangle E_{\mathrm{ue}}$
is the contribution 
to the underlying
event from additional \PBARP\ interactions, $E_{\mathrm{noise}}$ is the energy 
from uranium decay, and $E_{\mathrm{pile}}$ is the pile-up energy.
For the actual correction, the last three terms may be combined, yielding:

\begin{equation}
E_{\mathrm{O}} = E_{\mathrm{ue}} + E_{\Theta} \, . \label{eq:off4}
\end{equation}

\subsection{Physics Underlying Event}

The energy from spectator partons is measured using low luminosity 
minimum bias (MB) data 
(${\mathcal L}=0.1 \times 10^{30}\;{\mathrm cm}^{-2}{\mathrm sec}^{-1}$).
The MB trigger requires an inelastic 
collision in which both the proton and the anti-proton break up. 
MB events are, therefore, dominated by hard core interactions.  
The $E_{T}$ density per increment $\Delta \eta \times \Delta \varphi$ in MB 
events, $D_{\mathrm{MB}}$, is a measure of 
the transverse
$E_{\mathrm{ue}} + E_{\mathrm{noise}} + E_{\mathrm{pile}}$ density. The
contribution of uranium noise and pile-up, 
$D_{\mathrm{ZB}}^{{\mathrm{no \, HC} }}$,
is measured in a low luminosity ZB sample with the requirement that
there is no hard interaction. 
$D_{\mathrm{ue}}$ is then:

\begin{equation}
 D_{\mathrm{ue}} = 
 D_{\mathrm{MB}} - D_{\mathrm{ZB}}^{\mathrm{no \, HC}} \, . \label{eq:und1}
\end{equation}

\begin{figure}[p]
 \psfig{figure=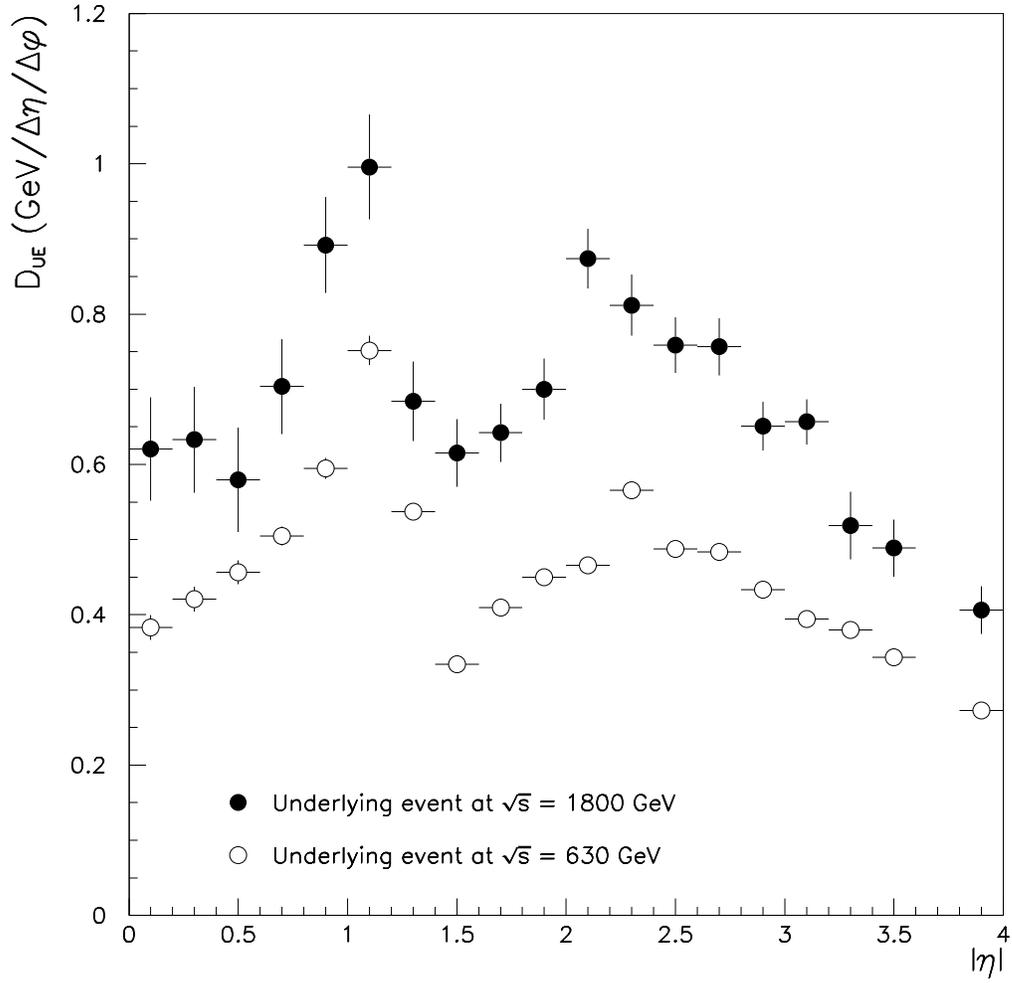,height=15cm,width=15cm}
 \caption{Physics underlying event $E_{T}$ density, 
$D_{\mathrm{ue}}$ ($\sqrt{s}=1800$ and $630\;{\mathrm GeV}$). The larger
samples available for the low center-of-mass energy measurement 
explain the smaller statistical errors.}
\label{fig:pue_630_1800}
\end{figure}

It depends both on the center-of-mass energy and pseudorapidity.
The energy to be subtracted from a jet, $E_{\mathrm{ue}}$, is
$D_{\mathrm{ue}} \times A_{\eta,\varphi}$, where $A_{\eta,\varphi}$ is the 
area of the jet in $\eta$-$\varphi$ space.
Figure~\ref{fig:pue_630_1800} shows the dependence 
of $D_{\mathrm{ue}}$ on physics $\eta$ for both 
$\sqrt{s}=1800\;{\mathrm GeV}$ and $\sqrt{s}=630\;{\mathrm GeV}$ collisions.
The uneven shape results from imperfect calibration of the intercryostat 
detector and loss of EM coverage in the intercryostat region.
Because the physics underlying event is associated
with the soft interactions in a single \PBARP\ collision,
it is independent of luminosity and of the number of \PBARP\ interactions
in the event.

\subsection{Uranium Noise, Pile-up, and Extra Interactions}

\begin{figure}[p]
\centerline{\psfig{figure=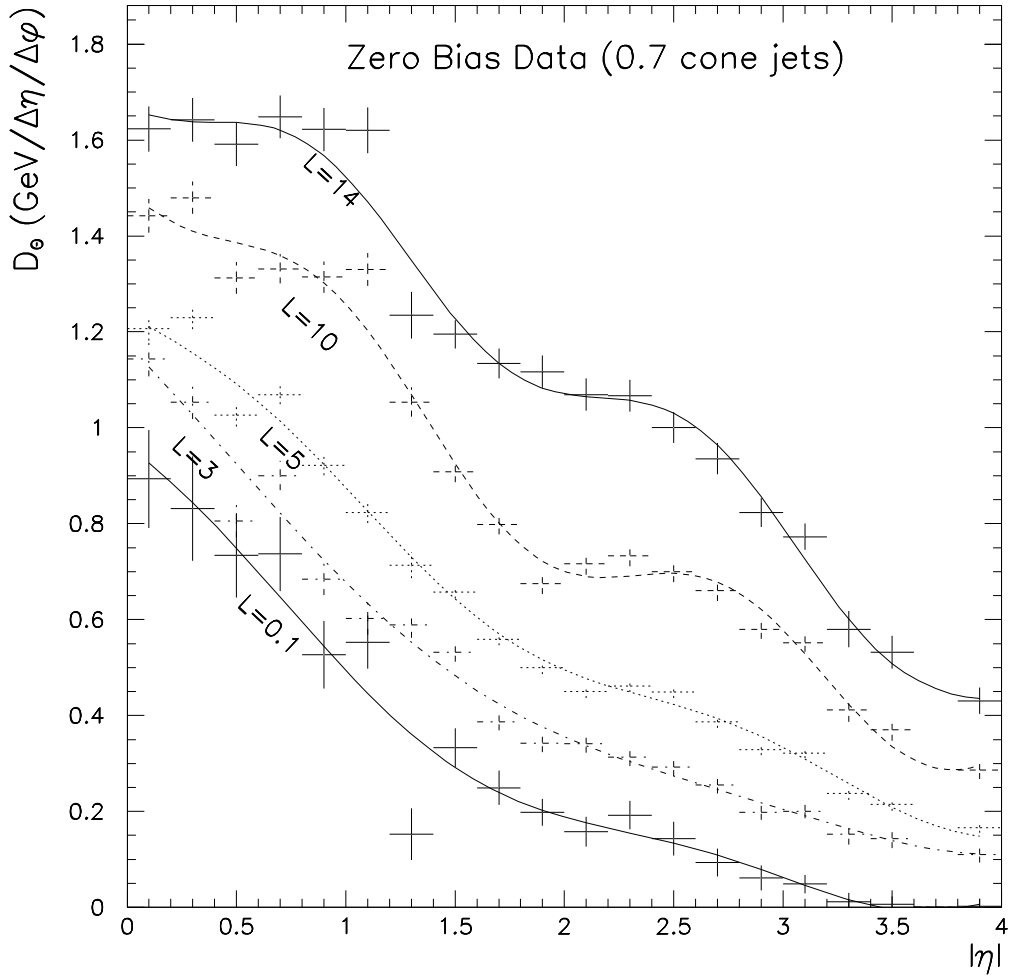,height=15cm,width=15cm}}
\caption{Transverse energy density, $D_{\Theta}$, 
contributed by uranium noise, pile-up, 
and additional hard interactions, for different luminosities in units of
$10^{30}\;{\mathrm cm}^{-2}{\mathrm sec}^{-1}$.}
\label{fig:totaloff}
\end{figure}

The contribution to the offset from uranium noise, 
pile-up, and additional \PBARP\ interactions, $D_{\Theta}$, is the
measured $E_{T}$ density in a ZB sample.
Figure~\ref{fig:totaloff} displays parameterizations of
$D_{\Theta}$ as a function of physics $\eta$ for different luminosities.

$D_{\Theta}$ also depends on the occupancy, defined as the number of 
calorimeter cells which were read out after zero-suppression divided by 
the total number of cells in the jet. Zero-suppression is the mechanism
that suppresses cells with energy depositions inside a 2$\sigma$ window
around the average pedestal value. Therefore, $D_{\Theta}$
from the ZB data must be extrapolated to a 
value consistent with average occupancies for jet data.
This difference in occupancy dominates the uncertainty in the offset
correction. It contributes an error of $\pm$ 0.25~GeV to $D_{\Theta}$.  
In addition, statistical and systematic errors in fitting the data are 
approximately 8\%.

\section{Response: The Missing $E_{T}$ Projection 
Fraction Method}\label{sec:mpf}

The jet reconstruction algorithm maps charge collected in the
liquid argon to energy. It is based
on single particle test beam data, assuming ideal instrumentation and a linear
response. The overall response after reconstruction is, however, less than
unity, due to a
non-linear response to low energy particles, dead material, and 
module-to-module response fluctuations. This section describes the missing
$E_{T}$ projection fraction method~\cite{mpfcdf} 
developed to measure the calorimeter response to jets in reconstructed data.

\subsection{Definitions}

\D0 makes a direct measurement of the jet energy
response, using  conservation of \pt\ in photon-jet ($\gamma$-jet) events. 
These events are composed of one photon balanced in \pt\ by one or more jets.
The missing transverse energy, \vmet, is defined as:

\begin{equation}
 \vmet = -\left(\sum_{i} E_{x_{i}},
 \sum_{i} E_{y_{i}}\right)
 \, , \label{eq:met}
\end{equation}

where $E_{x_{i}}$, $E_{y_{i}}$ are
defined as in Section~\ref{sec:jets}
and $i$ runs over all calorimeter cells with $|\eta|<4.5$. 
This definition
implies that the energy deposit in each cell may be treated as 
though it were deposited by a zero mass particle,
so that \vmet$=$\vmpt. In an ideal
calorimeter, any non-zero \met\ indicates the presence of
particles which failed to deposit their energy in the detector, such as
neutrinos or high-$p_{T}$ muons.
For a photon-jet event in a real detector, a non-zero \met\ 
measures the overall imbalance of transverse energy in the calorimeter due
to differences in response to photons and jets.
This can be used to measure the calorimeter response to jets, 
$R_{\mathrm{jet}}$, relative to the precisely known photon response.

In $\gamma$-jet events, the particle level or true photon and recoil
transverse energies $E_{T\gamma}$ and $E_{T\mathrm{recoil}}$, satisfy:

\begin{equation}
  \pol{E}_{T\gamma} + \pol{E}_{T\mathrm{recoil}} 
  = 0  \, . \label{eq:mpf1}
\end{equation}

In a real calorimeter, however, the photon and jet responses 
($R_{\mathrm em}$ and $R_{\mathrm recoil}$) are both less
than unity, and the equation is modified to:

\begin{equation}
  \pol{E}_{T\gamma}^{\mathrm meas} + 
  \pol{E}_{T{\mathrm recoil}}^{\mathrm meas} =
   - \pol{\met}^{\mathrm meas}  \, , \label{eq:mpf2}
\end{equation}

where $\pol{E}_{T\gamma}^{\mathrm meas}=R_{\mathrm em}\pol{E}_{T\gamma}$ and 
$\pol{E}_{T{\mathrm recoil}}^{\mathrm meas}=
R_{\mathrm recoil}\pol{E}_{T{\mathrm recoil}}$.
The energy scale for electromagnetically interacting particles is
determined~\cite{wmass} from the $Z \rightarrow e^{+}e^{-}$, $J/\psi$, and 
$\pi^{0}$ data samples, using the known masses of these 
resonances. If
$\pol{E}_{T\gamma}^{\mathrm meas}$ is corrected for energy
scale in the $\gamma$-jet data sample, Eq.~\ref{eq:mpf2} transforms into:

\begin{eqnarray}
\hskip2.5cm  \pol{E}_{T\gamma} + R_{\mathrm{recoil}} 
  \pol{E}_{T\mathrm{recoil}} & = &
   - \pol{\met} \, \nonumber \\
\hskip2.5cm  \pol{E}_{T\gamma} + R_{\mathrm{recoil}} \hat{n}_{T\gamma} 
  \cdot \pol{E}_{T\mathrm{recoil}} & = &
   - \hat{n}_{T\gamma} \cdot \vmet \, \nonumber \\
\hskip2.5cm  1 + R_{\mathrm{recoil}} \frac{\hat{n}_{T\gamma} \cdot 
   \pol{E}_{T\mathrm{recoil}}}{E_{T\gamma}} & = &
   - \frac{\hat{n}_{T\gamma} \cdot \vmet}{E_{T\gamma}} \, , 
\end{eqnarray}

where $\hat{n}_{T\gamma}=
\pol{E}_{T\gamma}/|\pol{E}_{T\gamma}|$, and \vmet\ is the missing transverse
energy recalculated after the photon correction. 
Eq.~\ref{eq:mpf1} can be written as
$\pol{E}_{T\gamma}=-\hat{n}_{T\gamma} \cdot \pol{E}_{T\mathrm{recoil}}$; then
$R_{\mathrm{recoil}}$ is:

\begin{equation}
 R_{\mathrm{recoil}}
 = 1 + \frac{\vmet \cdot \hat{n}_{T\gamma}}{E_{T\gamma}} 
 \, . \label{eq:mpf3}
\end{equation}

In the special case of a $\gamma$-jet two body process, and in the
absence of offset and showering effects,
$R_{\mathrm{recoil}}$ would be the ratio 
$E_{T\mathrm{jet}}^{\mathrm{meas}}/E_{T\mathrm{jet}}^{\mathrm{ptcl}}$ of
measured to particle jet transverse energies.
In the presence of offset and showering losses, 
$R_{\mathrm{recoil}}$ is the energy response of the calorimeter 
to jets, $R_{\mathrm jet}$, where ``jet'' is the leading jet of the event.
This is a good approximation if the difference in azimuth between the 
$\gamma$ and the leading jet is close to $\pi$.

\subsection{The Energy Estimator, $E^{\prime}$ }

$R_{\mathrm{jet}}$ is measured at \D0 as $R_{\mathrm recoil}$ from
a $\gamma$-jet sample, using
conservation of transverse momentum (the terms transverse momentum and
transverse energy are used interchangeably).
The response is, however, dependent on jet energy rather than its transverse
component. This is because $e/\pi$ 
and the particle
composition of jets are energy dependent. 

\begin{figure}[p]
\psfig{figure=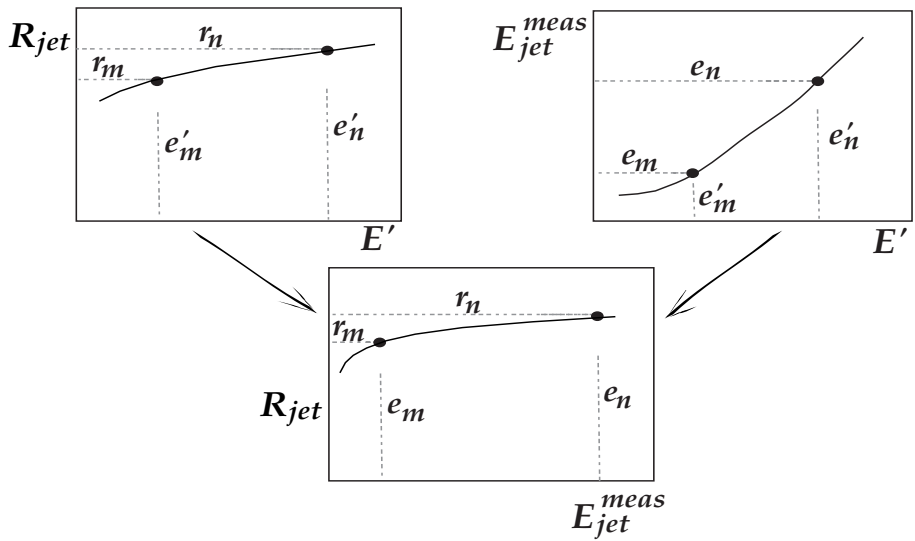,height=8cm,width=13cm}
\caption{Derivation of $R_{\mathrm{jet}}$ versus 
$E_{\mathrm{jet}}^{\mathrm{meas}}$ using the energy
estimator $E^{\prime}$.}
\label{fig:eprime_sketch}
\end{figure}

Measuring $R_{\mathrm{jet}}$ directly as a function of 
$E_{\mathrm{jet}}^{\mathrm{meas}}$
is problematic. 
The finite jet and photon energy resolutions,
steeply falling photon cross section, trigger and
reconstruction thresholds, and event topology contribute
biases which must be removed. 
Most of these biases and smearing effects are reduced to 
negligible levels by binning the response 
not in $E_{\mathrm{jet}}^{\mathrm{meas}}$ but instead in some better
measured quantity 
which is strongly correlated with $E_{T\mathrm{jet}}^{\mathrm{ptcl}}$.
For this quantity the jet energy estimator $E^{\prime}$ is chosen 
and defined as:

\begin{equation}
 E^{\prime} = 
 E_{T\gamma} \cdot \cosh (\eta)
 \, , \label{eq:mpf4}
\end{equation}

where $E_{T\gamma}$ includes the electromagnetic scale correction.
Both $E_{T\gamma}$ and the jet pseudorapidity
are measured with high resolution compared to
$E_{\mathrm{jet}}^{\mathrm{meas}}$ . After
binning the response in terms of $E^{\prime}$, the dependence of 
$R_{\mathrm{jet}}$ on
$E_{\mathrm{jet}}^{\mathrm{meas}}$ is obtained by
measuring the average $E_{\mathrm{jet}}^{\mathrm{meas}}$ in each
$E^{\prime}$ bin.
Figure~\ref{fig:eprime_sketch} illustrates the
$E^{\prime} \rightarrow E_{\mathrm{jet}}^{\mathrm{meas}}$ mapping procedure.

\subsection{Tests of the Method}

The method is tested using a parametric
simulation. The simulation generates $\gamma$-jet events according
to a cross section with a given $E_{T}$ dependence.
The energies of the photon and the jet are smeared and scaled with
energy resolutions and responses measured from data.

\begin{figure}[p]
\psfig{figure=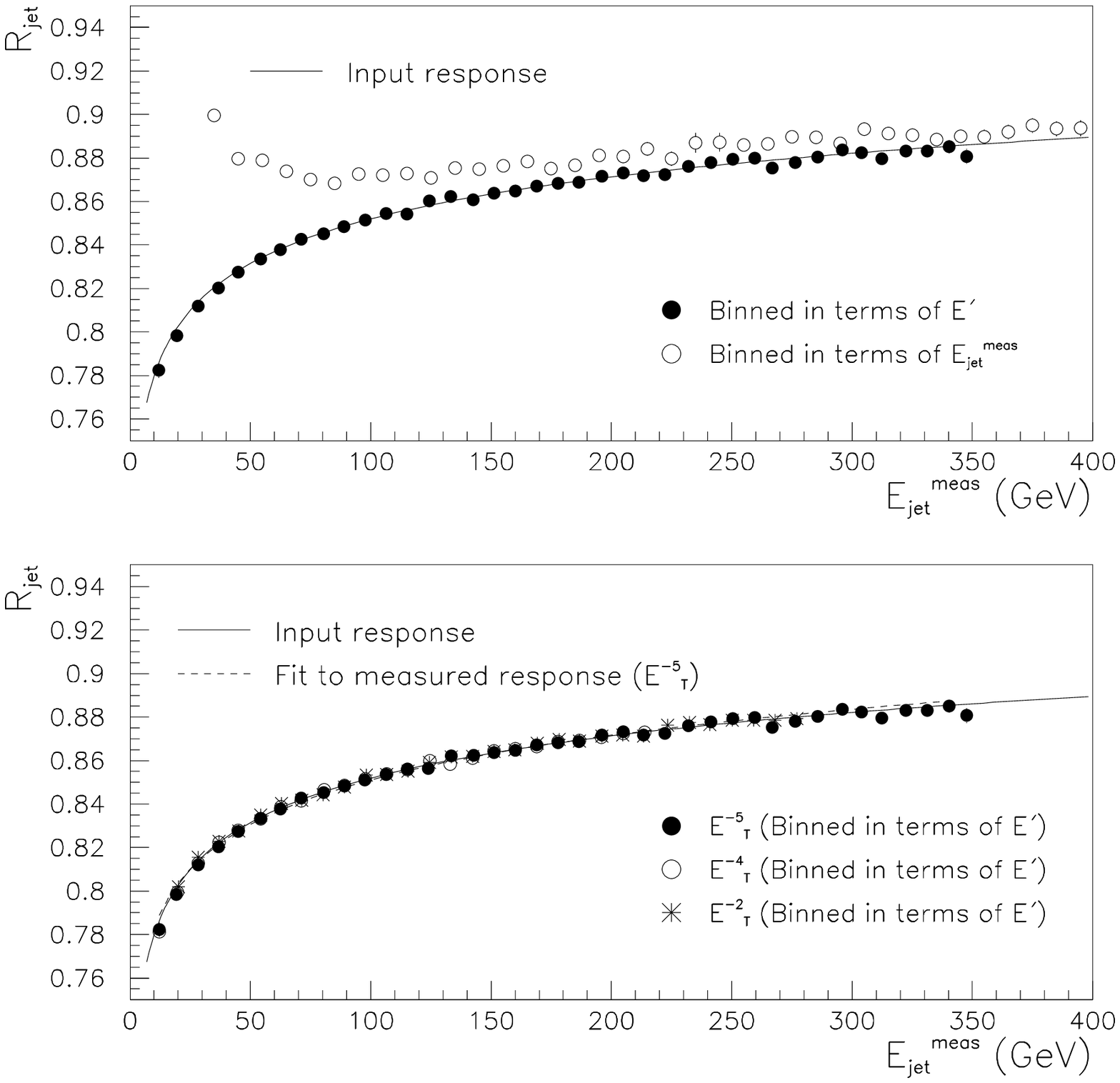,height=15cm,width=15cm}
\caption
{Parametric simulation of the $R_{\mathrm{jet}}$ measurement. A fit 
to $R_{\mathrm{jet}}$ ``measured'' from the simulated data, when binned 
in terms of $E^{\prime}$ to
remove resolution effects, agrees very well with the input response (top).
The agreement between the fit and the input function is also excellent given 
different reasonable assumptions for the $E_{T}$
dependence of the $\gamma$ cross section, such as 
$E_{T}^{-5}$, $E_{T}^{-4}$, and $E_{T}^{-2}$ (bottom).}
\label{fig:simul_pho_jet} 
\end{figure}

Figure~\ref{fig:simul_pho_jet} (top) shows a comparison between 
$R_{\mathrm{jet}}$
determined from the simulation directly as a function of 
$E_{\mathrm{jet}}^{\mathrm{meas}}$ (open circles), and the same quantity 
binned in terms of $E^{\prime}$
and then mapped onto $E_{\mathrm{jet}}^{\mathrm{meas}}$ (closed circles).
While the closed circles are in very good
agreement with the input response (solid line), the open circles exhibit
the smearing effect arising from a finite jet energy resolution.
Figure~\ref{fig:simul_pho_jet} (bottom) shows excellent agreement
between the fits to $R_{\mathrm{jet}}$ versus $E^{\prime}$ and 
the input function given different reasonable assumptions
for the $E_{T}$ dependence of the photon cross section, such as
$E_{T}^{-5}$,$E_{T}^{-4}$, and $E_{T}^{-2}$.

\section{Response: Photon Sample Selection}\label{sec:selec}

A large event sample is necessary
to measure the jet response of each calorimeter with a reasonably small
uncertainty over a large range of energies.
Thus, $R_{\mathrm{jet}}$ is measured from a collider data sample
obtained by applying a set of cuts which retain
a large sample while limiting the systematic biases. 
For the purpose of calibration, the ``$\gamma$-jet'' sample is not
restricted to direct photon events. It also includes events with one
jet well contained in the EM calorimeter and little nearby hadronic energy.

\subsection{General Cuts}

From the events with photon candidates accepted during data collection,
only those with one or more reconstructed EM clusters are used.
Events with noisy cells which fake or distort jets are rejected. 
The Main Ring, the injector for 
the Tevatron, passes through the hadronic calorimeter and 
is active during data taking for $\overline{p}$ 
production. Events containing particles related to Main Ring activity are, 
therefore, removed from the sample. The muon system is used as a loose veto
to avoid events with bremsstrahlung photons from cosmic ray muons.
If the leading photon $E_{T}$ is
less than 30~GeV, there must be no reconstructed muons in the
event; otherwise the event 
is accepted if muons have $p_{T\mu}<100\;{\mathrm GeV/c}$.
Because events with primary vertices far from the center of the detector
can distort $E_{T}$ measurements,
the $z$-coordinate of the vertex must satisfy $|z|<70\;{\mathrm cm}$. 

Fiducial cuts on detector $\eta$ are applied to ensure good containment of 
the EM cluster inside the calorimeters ($|\eta|<1$ or $1.6<|\eta|<2.5$).
Candidates near detector boundaries are also rejected.
There must be at least one reconstructed jet in the event, and
the leading jet must be contained in the CC ($|\eta|<0.7$) or the 
EC ($1.8<|\eta|<2.5$) for the CC and EC calibrations, respectively. 

There are two types of background to the calibration sample, removed
as described below.

\subsubsection{ Instrumental Background}

The instrumental background is contributed by events in which
the photon, or pair of photons from a highly EM jet (mostly $\pi^{0}$ jets),
is not well isolated from a significant amount of hadronic energy.
The following variables are used in the selection procedure to remove
background:

\begin{itemize}

\item The fraction of the cluster $E_{T}$ deposited in the 
EM layers of the calorimeters, EMF.

\item The cluster isolation in the plane transverse to the beam, 
$f_{\mathrm iso}$, defined as:

 \[ f_{\mathrm iso} \equiv 
\frac{E_{\mathrm tot}({\cal R} = 0.4)-E_{\mathrm em}({\cal R} = 0.2)}
           {E_{\mathrm em}({\cal R} = 0.2)} \, , \]

\hskip-4mm where $E_{\mathrm tot}$ and $E_{\mathrm em}$ are the total 
energy and the EM energy
within a cone radius ${\cal R}$.

\item The total charge in the transition
      radiation detector~\cite{nim}, $\varepsilon_{t}$, measured in units of
      the charge of the electron.

\item The presence of a track in a road projected from the EM calorimeter 
      cluster through the tracking chambers to the event vertex. Track match
      significance, $\sigma_{\mathrm trk}$, is defined as:

 \[ \sigma_{\mathrm trk} = 
\sqrt{ \left( \frac{\Delta \varphi}{\sigma_{\Delta \varphi}} \right)^{2} +
\left( \frac{\Delta z}{\sigma_{\Delta z}} \right)^{2} } \, , \]

\hskip-4mm 
where $\Delta \varphi$ and $\Delta z$ are the distances between the track
and the cluster along the azimuthal and beam directions. 

\item The ionization in the central tracking detectors, dE/dx, in units of
minimum ionizing particles.

\end{itemize}

Table~\ref{tab:phocut} shows the cuts on photon candidates to reject 
background. The three cuts based on tracking information, 
$\varepsilon_{t}$, $\sigma_{\mathrm trk}$, and
dE/dx must be satisfied for events with 
$E_{T\gamma}>25 \; {\mathrm GeV}$ only if
$\met/E_{T\gamma}>0.65$. Instead, only one of the three is required if 
$E_{T\gamma}<25 \; {\mathrm GeV}$ or $\met/E_{T\gamma}<0.65$.

\begin{figure}[p]
\centerline{\psfig{figure=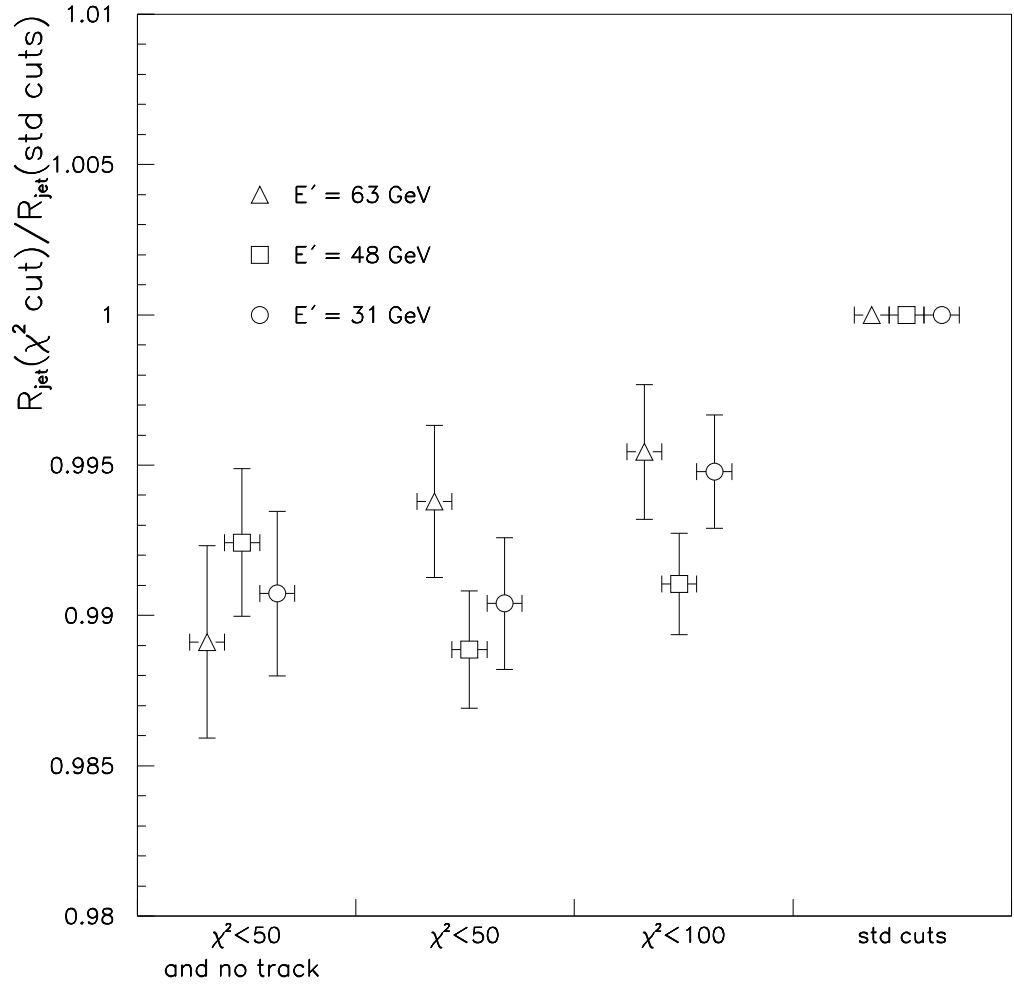,%
height=15cm,width=15cm}}
\caption{Change in the measured $R_{\mathrm{jet}}$ as the
$\chi^{2}$ cut is tightened, for three different $E^{\prime}$ bins. 
$R_{\mathrm{jet}}$ (std cuts) is the measured
response from the $\gamma$-jet sample selected with the cuts 
listed in Table~\ref{tab:phocut}. 
An additional cut on $\chi^{2}$ (matrix test)
is applied for the other bins to study the effect of highly EM jets
on $R_{\mathrm jet}$.}
\label{fig:response_vs_cuts}
\end{figure}

There is a bias in $R_{\mathrm{jet}}$ due to the presence of highly 
electromagnetic jets in the sample, as compared to photons.
This bias is studied using a 
$\chi^{2}$ test~\cite{topprd}, which measures the
consistency of the shower shape of the calibration sample 
with that expected from an electromagnetic shower. 
A matrix is constructed with transverse and longitudinal shower 
shape information from the photon candidate,
and then compared with the same type of matrix built from electromagnetic 
clusters in a Monte Carlo sample. A cut on $\chi^{2}$ removes a 
significant
number of events but is a good discriminator between real photons 
and highly electromagnetic jets.
Figure~\ref{fig:response_vs_cuts} shows the magnitude of the 
instrumental background bias on the jet response. As the
cut on $\chi^{2}$ tightens, the fraction of real photons in
the sample increases, but the response changes only slightly.
So that a large sample can be retained, a $\chi^{2}$ cut is not applied 
in the event selection procedure, and a bias on the order
of 0.7\% towards a larger response is
present. A large fraction of this bias cancels against the topology bias,
as will be discussed in Section~\ref{sec:enedep}.

\subsubsection{Physics Backgrounds} 

The physics backgrounds include Drell-Yan, $Z \rightarrow e^{+}e^{-}$, and
$W \rightarrow e\nu$ events.
In events with two electromagnetic clusters, such as
$Z$ $\rightarrow$ $e^{+}e^{-}$
production, $R_{\mathrm{jet}}$ is approximately unity.
The effect of the $Z$ boson background can
be determined by applying a cut on the dielectron invariant mass,
$m_{ee}$. This background is very small and, therefore, the cut 
$75\;{\mathrm GeV/c}^{2}<m_{ee}<105\;{\mathrm GeV/c}^{2}$ 
produces a negligible 
change in the measured response.

$W \rightarrow e \nu$ events have a large \vmet\ in the direction of
the neutrino, so they
are not useful for calibration purposes.
They are removed using the tracking cuts listed in Table~\ref{tab:phocut},
combined with a \met\ cut, as described before.
After the cuts, the remaining bias on
$R_{\mathrm{jet}}$ due to physics background, determined from Monte Carlo
and data samples, is $\lsim$ 0.5\%.

\subsection{Topology Cuts}

The relation $R_{\mathrm{jet}}=R_{\mathrm{recoil}}$, where the subscript
jet refers to the leading jet in the
$\gamma$-jet event, is exact only in the case of a two body process. 
Photon events
may contain more than one reconstructed jet along with a number of energy 
clusters which are not reconstructed as jets. 
Variations in topology, therefore, contribute a systematic error to 
the response measurement. 

\begin{figure}[p]
\centerline{\psfig{figure=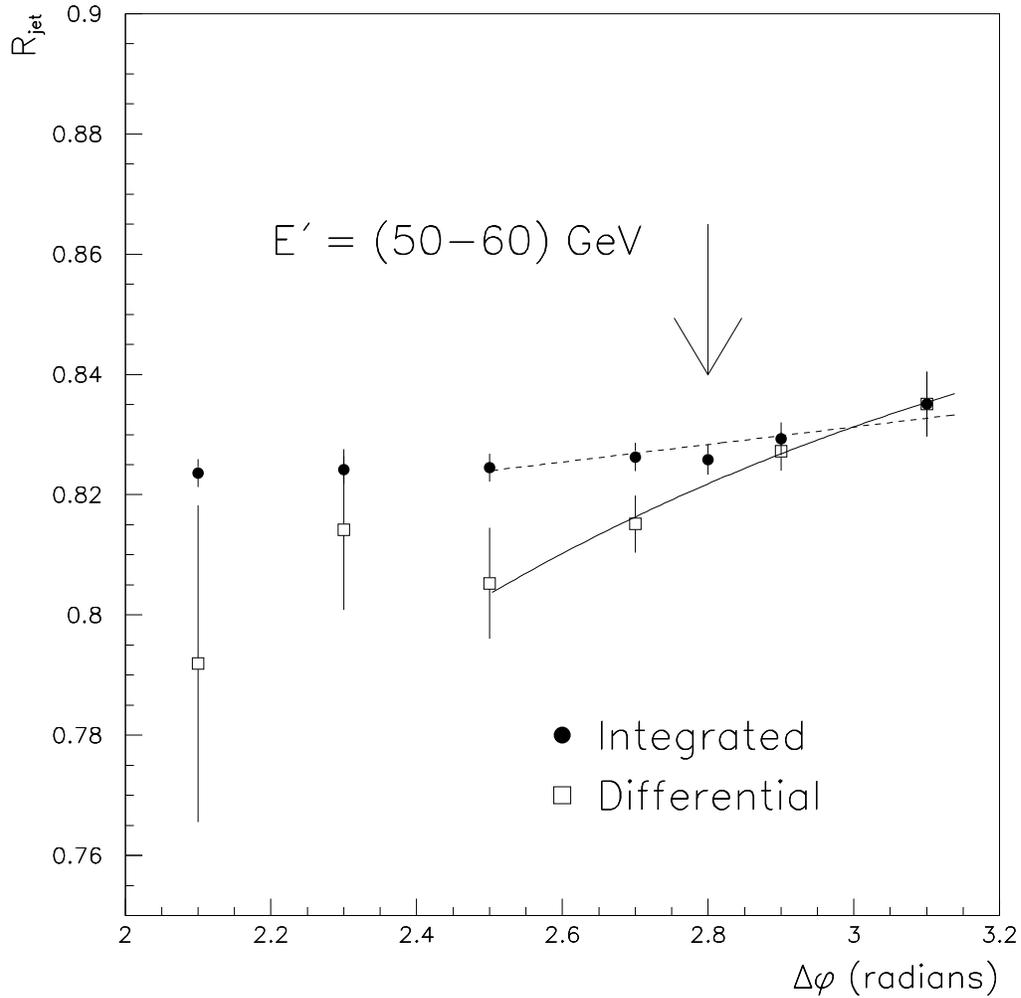,height=15cm,width=15cm}}
\caption
{$R_{\mathrm{jet}}$ as a function of the $\Delta \varphi$ cut threshold for
$50<E^{\prime}<60\;{\mathrm GeV}$. The circles correspond to an integrated
distribution (lower $\Delta \varphi$ bins include the events of the
higher $\Delta \varphi$ bins), while the open squares represent
the differential distribution.
The solid and dashed lines are parameterizations of
the differential and integrated distributions. The cut is at 
$\Delta \varphi=2.8\;{\mathrm radians}$ (arrow).}
\label{fig:dphi1}
\end{figure}

\begin{figure}[p]
\centerline{\psfig{figure=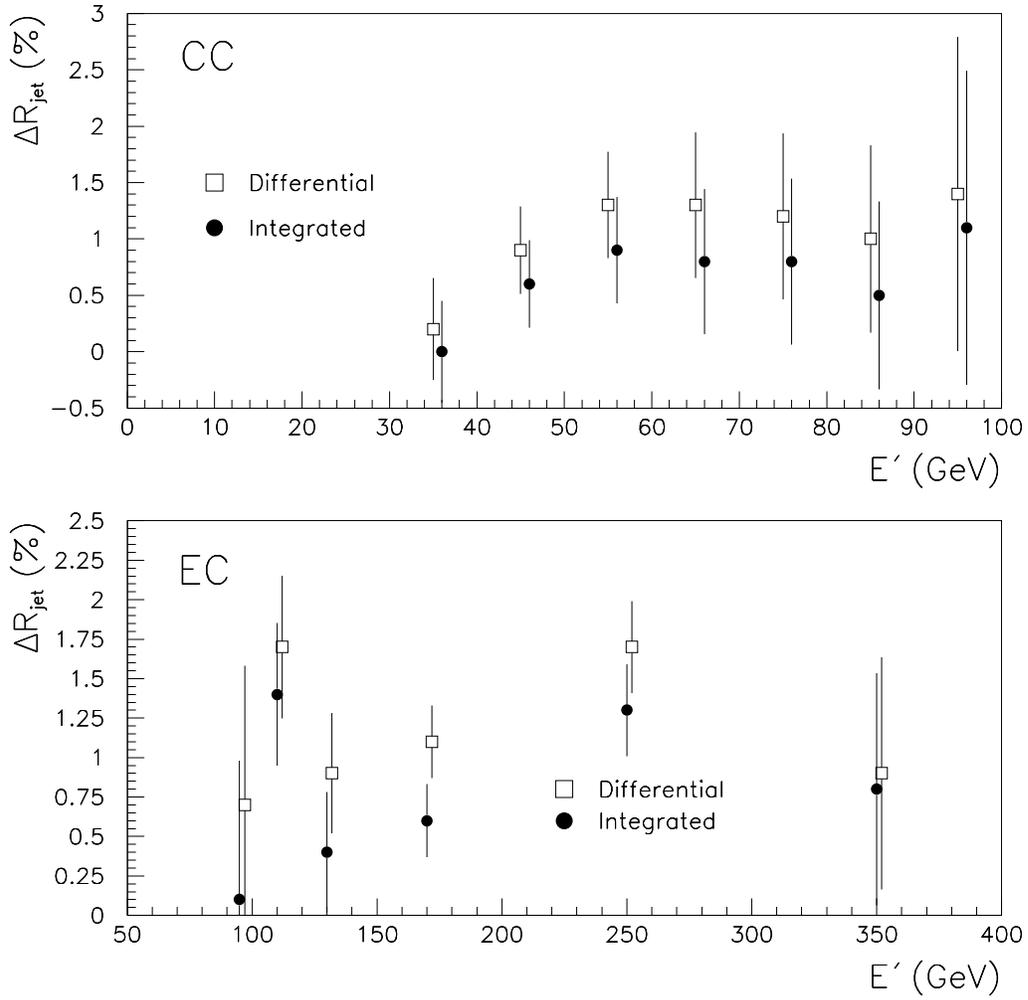,height=15cm,width=15cm}}
\caption
{Difference between
$R_{\mathrm{jet}}$ at $\Delta \varphi=\pi$ and $\Delta \varphi>2.8$ radians
versus $E^{\prime}$.
The circles are the result from the integrated distribution
and the squares from the differential distribution.}
\label{fig:dphi2}
\end{figure}

To remove this topology bias, the photon and the leading jet must have
a difference greater than 2.8 radians in azimuth.
The residual error is investigated by studying the change in response 
for different energy bins as  
the $\Delta \varphi$ cut is varied from $2.1$ to $3.1$ radians.
Figure~\ref{fig:dphi1} shows the change in $R_{\mathrm{jet}}$ as
the $\Delta \varphi$ cut is tightened for the 
$50<E^{\prime}<60\;{\mathrm GeV}$ 
bin. The circles
correspond to an integrated distribution (lower $\Delta \varphi$ bins include 
the events of the higher $\Delta \varphi$ bins), 
while the open squares represent
the differential distribution. Requiring
$\Delta \varphi>2.8$ radians is a good compromise between a large event sample
and bias minimization.
The residual error after the cut is estimated with the parameterizations to the
integrated (linear fit) and differential distributions (exponential fit).
The fractional difference between
$R_{\mathrm{jet}}$ at $\Delta \varphi=\pi$ and
$\Delta \varphi>2.8$ radians is an estimate of the remaining
topology bias. This difference as a function of $E^{\prime}$ is plotted in
Fig.~\ref{fig:dphi2} for both the CC and the EC. 
The plots show that, after requiring $\Delta \varphi>2.8$ radians,
the remaining bias is approximately 0.5-1\% towards a lower response and 
consistent with a constant dependence on $E^{\prime}$. A large 
fraction of this bias cancels against the effect of instrumental
background, as will be discussed in Section~\ref{sec:enedep}.

\subsection{Multiple Interaction Cuts}

Events which consist of several overlaid interactions are called
multiple interaction events. Usually, only the ``primary'' interaction will
produce high-$E_{T}$ objects. The additional interactions can, 
however, degrade the accuracy of the vertex determination for the primary 
interaction.
On average, jets reconstructed with an incorrect
vertex are assigned a higher pseudorapidity, yielding a larger
event $E^{\prime}$ and a lower jet $E_{T}$.
The overall effect is an increase of the \met\ in the direction of the jet, 
thus lowering the measured response.
This effect grows as luminosity increases, as a result of the increasing
proportion of multiple interaction crossings.

\begin{figure}[p]
\centerline{\psfig{figure=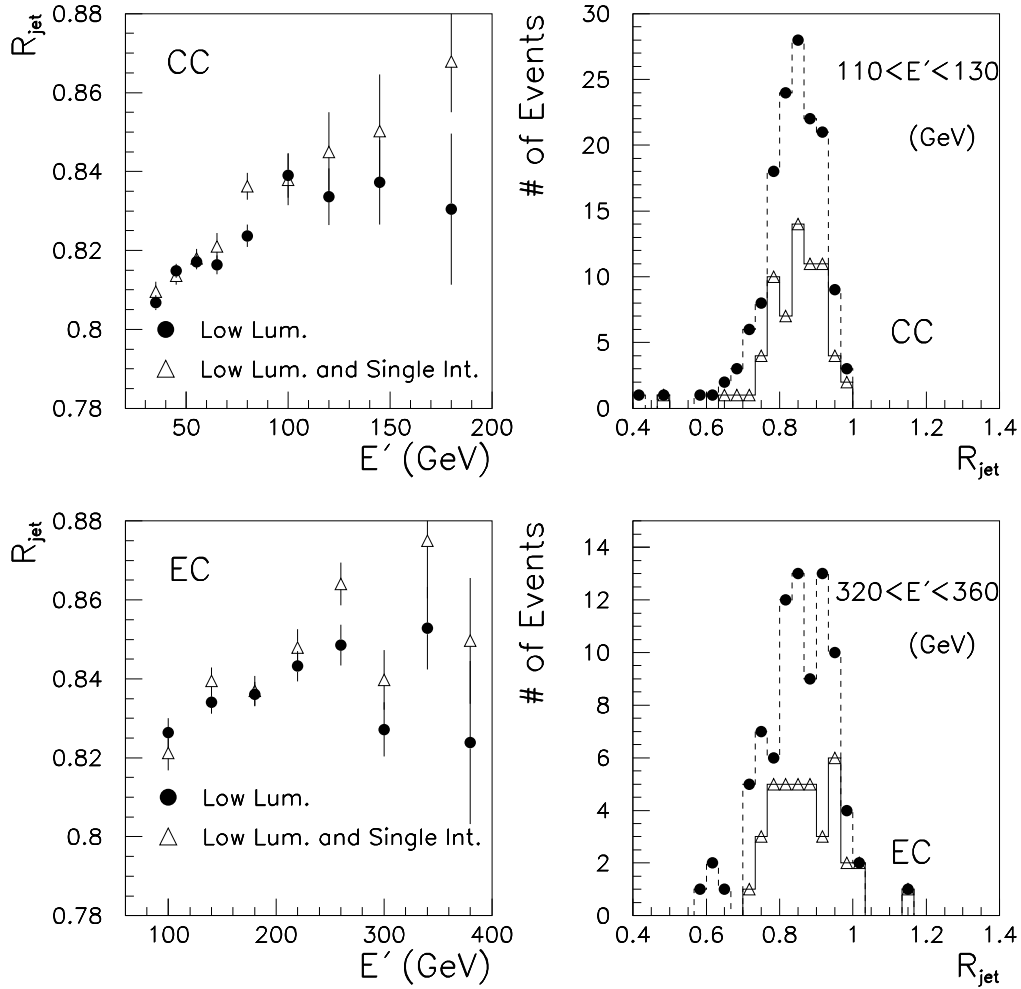,height=15cm,width=15cm}}
\caption
{$R_{\mathrm{jet}}$ versus $E^{\prime}$ in the CC and EC from
a low luminosity sample 
(${\mathcal L}<5\times 10^{30}\;{\mathrm cm}^{-2}{\mathrm sec}^{-1}$). 
If the single interaction requirement is not enforced,
the response distributions are biased towards lower values, due to the effect
of misvertexing (full circles). The bias vanishes if only single 
interaction events are accepted (open triangles).}
\label{fig:lum_micut}
\end{figure}

To reduce this bias, a low luminosity sample is used. This sample, however, 
still contains a number of multiple interaction events. 
These are removed by requiring the trigger,
tracking detectors, and calorimeter information to be consistent with a single
interaction. 
Figure~\ref{fig:lum_micut}
shows the change in the measured response from a low luminosity sample
when the single interaction requirement is applied. 
The low tails in the $R_{\mathrm{jet}}$
distributions are removed by this cut.

\begin{figure}[p]
\centerline{\psfig{figure=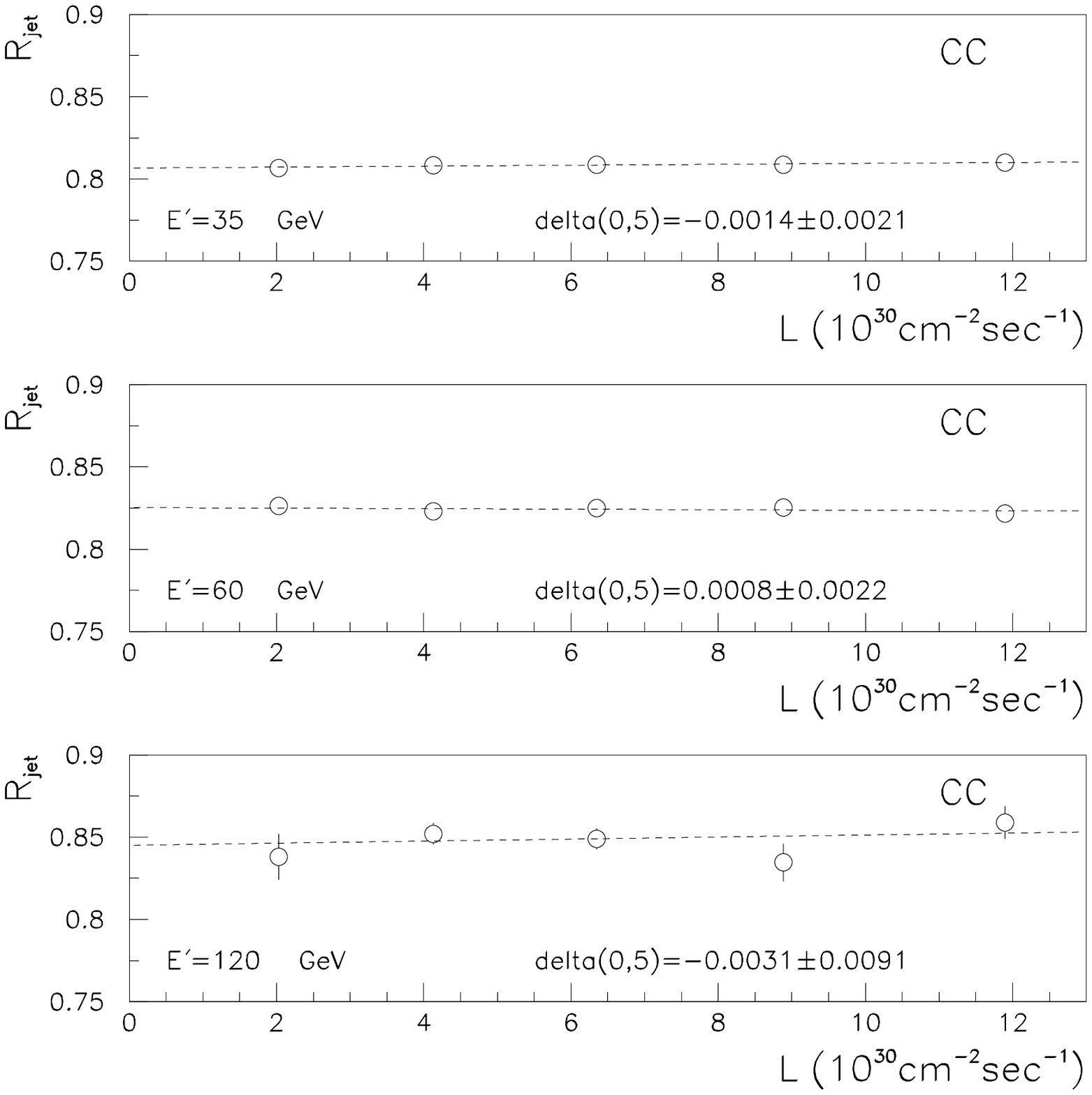,height=15cm%
,width=15cm}}
\caption
{Response as a function of luminosity for CC jets with
$E^{\prime}=$ 35, 60, and 120~GeV. The variable delta(0,5) is the difference
in $R_{\mathrm{jet}}$ measured from samples taken at luminosities of zero
and $5\times 10^{30}\;{\mathrm cm}^{-2}{\mathrm sec}^{-1}$.
The single interaction requirement is always enforced.}
\label{fig:lum1}
\end{figure}

\begin{figure}[p]
\centerline{\psfig{figure=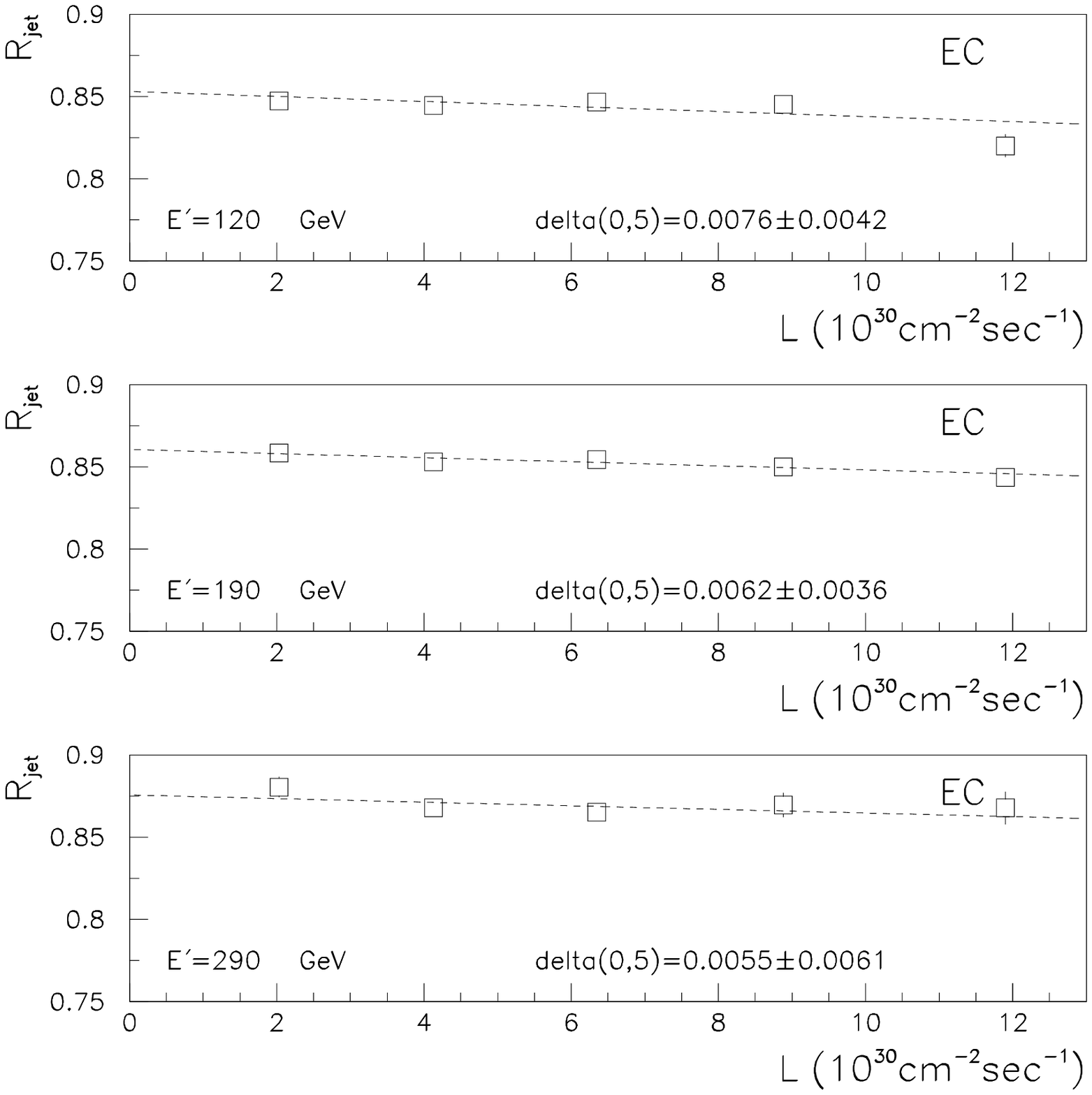,height=15cm%
,width=15cm}}
\caption
{Response as  a  function of  luminosity for EC jets with
$E^{\prime}=$ 120,  190, and 290~GeV. The variable delta(0,5) is the difference
in $R_{\mathrm{jet}}$ measured from samples taken at luminosities of 
zero and $5\times 10^{30}{\mathrm cm}^{-2}{\mathrm sec}^{-1}$.
The single interaction requirement is always enforced.}
\label{fig:lum2}
\end{figure}

To measure any residual luminosity dependence of the response after
the single interaction cut, the response is
measured as a function of luminosity for various $E^{\prime}$
bins for CC and EC jets, as shown in
Figs.~\ref{fig:lum1}~and~\ref{fig:lum2}.
A linear fit of $R_{\mathrm{jet}}$ versus ${\mathcal L}$ allows an 
extrapolation to zero luminosity.
In Fig.~\ref{fig:lum1}, the CC data
show very little sensitivity to luminosity.  The EC data, however, show a 
slight trend to lower response with increasing luminosity.  
The estimated difference
in response between a luminosity of 
$5\times 10^{30}\;{\mathrm cm}^{-2}{\mathrm sec}^{-1}$ 
and zero is approximately 0.5\%
for the EC jets. Because the sample used in this analysis is integrated over
this range of luminosities, any residual luminosity effects are
less than 0.25\%.  

\section{Response: Rapidity Dependence}\label{sec:etadep}

After test beam calibration, the \D0 calorimeter system still shows 
non-uniform regions in pseudorapidity. 
Given that most physics measurements need a high level of 
accuracy at all rapidities, $\eta$-dependent corrections 
become necessary. These are the cryostat factor and the
IC corrections applied in consecutive order  
to both the jet energy and the event \vmet. 

The $\eta$-dependent corrections are applied as multiplicative factors to
the jet energy.   
Given that the energy corrections are not derived for calorimeter cells but
for physics objects, it is not possible to recalculate \vmet\ using the 
definition in Section~\ref{sec:mpf}. Instead, the corrected
missing transverse energy, $\vmet^{\mathrm corr}$, is determined 
by adding to the measured or reconstructed missing transverse energy, 
$\vmet^{\mathrm meas}$, 
the difference between the uncorrected and
the corrected vector sums 
of the $E_{T}$ of the individual physics objects (jets and photons
in the calibration sample). In other words:

\begin{equation}
 \vmet^{\mathrm{corr}}=
 \vmet^{\mathrm{meas}} + \sum_{\gamma} 
         \left( \pol{E}_{T\gamma}^{\mathrm{meas}}
              - \pol{E}_{T\gamma}^{\mathrm{corr}} \right)
         + \sum_{\mathrm{jet}} 
         \left( \pol{E}_{T\mathrm{jet}}^{\mathrm{meas}}
              - \pol{E}_{T\mathrm{jet}}^{\mathrm{corr}} \right) 
              \, . \label{eq:etadep3}
\end{equation}

This formula depends on the jet algorithm because it is
based on reconstructed physics objects instead of calorimeter cells.
It is, however, a good approximation to the \met\ corrected for the
photon scale and the rapidity
dependence of jet response, if the jet cone size is large (${\mathcal R}$=0.7).

\subsection{Cryostat Factor Correction}

\begin{figure}[p]
\centerline{\psfig{figure=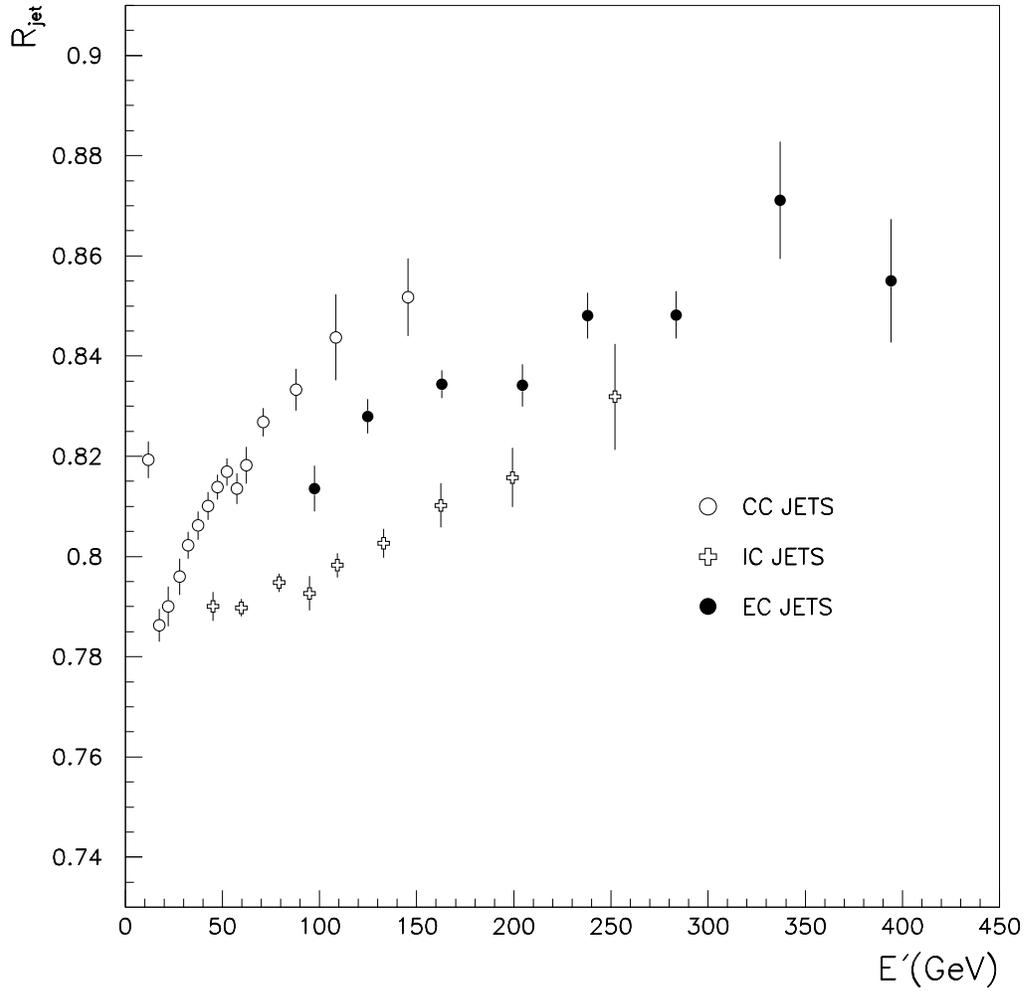,height=15cm,%
width=15cm}}
\caption
{$R_{\mathrm{jet}}$ versus $E^{\prime}$ for a 
sample with no $\eta$-dependent corrections. 
CC jets correspond to $|\eta|<0.7$,
IC jets to $0.7<|\eta|<1.8$, and EC jets to 
$1.8<|\eta|<2.5$. The lowest $E^{\prime}$ point in the CC is affected by 
low-$E_{T}$ resolution bias, as explained in Section~\ref{sec:enedep}.}
\label{fig:R_v_EP_nocor}
\end{figure}

The cryostat factor, $F_{\mathrm{cry}}$, is defined as the ratio 
$R_{\mathrm{jet}}^{\mathrm{EC}}/R_{\mathrm{jet}}^{\mathrm{CC}}$.
$F_{\mathrm{cry}}$ is expected to be independent of $E^{\prime}$ 
because the construction of the CC and the EC is similar.
The measured response versus $E^{\prime}$ 
before corrections is shown in
Fig.~\ref{fig:R_v_EP_nocor}. $R_{\mathrm{jet}}$ is plotted
in three detector $\eta$ regions:
CC ($|\eta|<0.7$), IC ($0.7<|\eta|<1.8$), and EC ($1.8<|\eta|<2.5$).

\begin{figure}[p]
\psfig{figure=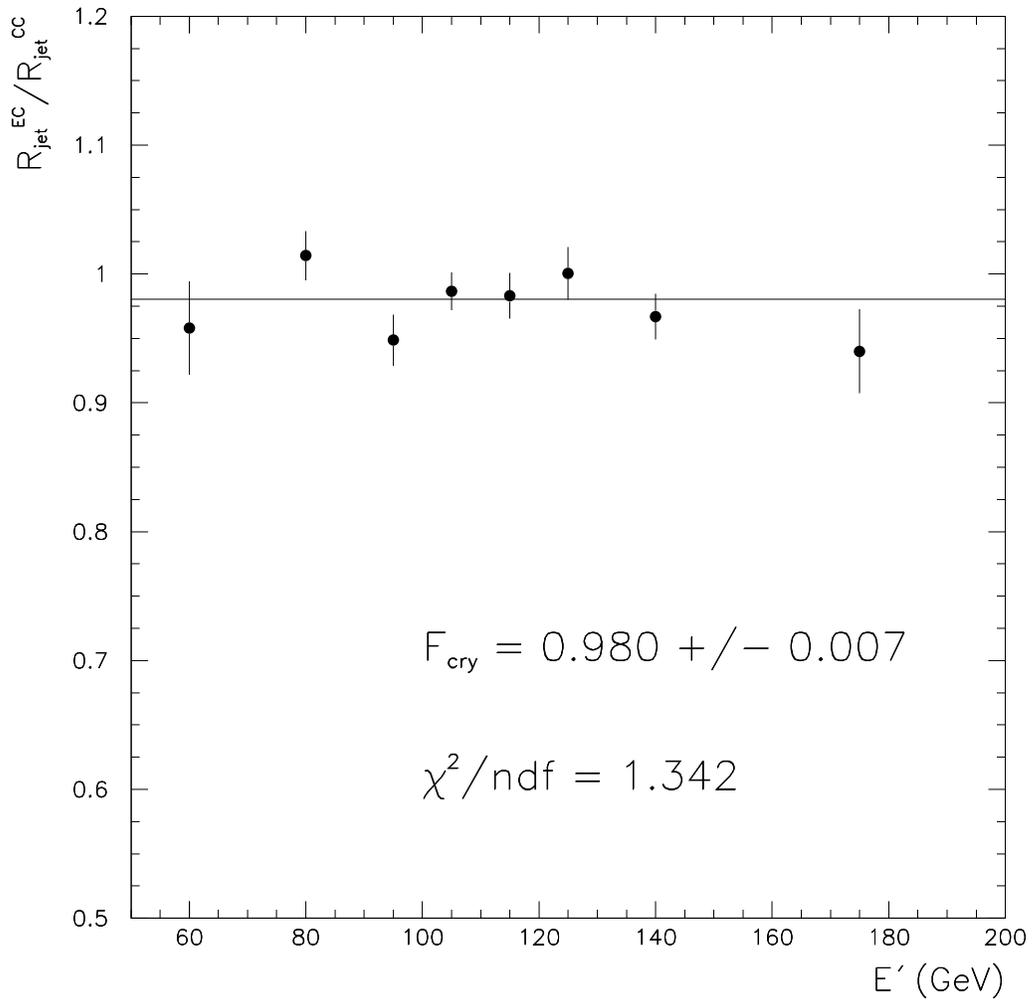,height=15cm,width=15cm}
\caption
{Cryostat factor as a function of $E^{\prime}$.
The solid line is a fit to a constant function.}
\label{fig:cc_vs_ec_ia}
\end{figure}

$F_{\mathrm{cry}}$ is measured where the CC and the EC data overlap, in the 
range $60\;{\mathrm GeV}<E^{\prime}<180\;{\mathrm GeV}$. In the EC data, 
an additional cut $E_{T\gamma}>25\;{\mathrm GeV}$ is necessary to
remove the reconstruction threshold bias due to jet finite energy resolution.
$F_{\mathrm{cry}}$ is measured separately in the EC north ($-2.5<\eta<-1.8$) 
and the EC south ($1.8<\eta<2.5$) calorimeters. The
ratio $F_{\mathrm{cry}}^{\mathrm{N}}/F_{\mathrm{cry}}^{\mathrm{S}}=0.997 
\pm 0.003$ shows that $F_{\mathrm{cry}}$ is the
same in both EC's within errors.
Figure~\ref{fig:cc_vs_ec_ia} shows the measured value for the cryostat factor,
$F_{\mathrm{cry}}=0.980 \pm 0.007$ (stat), and illustrates its derivation
from a fit of a constant function to the measured 
$R^{\mathrm{EC}}_{\mathrm{jet}}/R^{\mathrm{CC}}_{\mathrm{jet}}$ versus 
$E^{\prime}$.

The most important consequence of the independence of $F_{\mathrm{cry}}$
on $E^{\prime}$ is the possibility of using EC data to extend the range of 
the CC response measurement.
This is possible because forward jets have higher
energies than central jets with the same $E_{T}$.
Thus, the energy range of the response
measurement can be extended from $\sim 150\;{\mathrm GeV}$ to nearly 300~GeV.

\subsection{IC Correction}

The intercryostat region, covering the pseudorapidity range
$0.8<|\eta|<1.6$, is the least well instrumented region of
the calorimeter system. The IC is a non-uniform region
covered by different types of detectors.
A substantial amount of energy is lost in the
cryostat walls, module endplates, and support structures. Additionally,
in the range $1.2<|\eta|<1.4$, the system lacks
electromagnetic calorimetry and the total depth falls slightly
below 6 interaction lengths. 
As a consequence,
the response has a residual $\eta$ dependence
in this region, even after the cryostat factor has been
applied to EC jets.

\subsubsection{Event Selection and Method}

The IC correction to the response is measured using both $\gamma$-jet and
jet-jet events. Because the dependence of the response
on pseudorapidity is related to detector inhomogeneities,
$R_{\mathrm{jet}}$ is measured as a function of detector pseudorapidity.
For the $\gamma$-jet sample, the photon is kept central
($|\eta|<0.5$), while the jet $\eta$ is unconstrained.
For the jet-jet sample, one of the
two leading $E_{T}$ jets is required to be central 
($|\eta|<0.5$), and there is no
restriction on the $\eta$ of the other jet. In this case, 
$R_{\mathrm{jet}}$ is defined as:

\begin{equation}
 R_{\mathrm{jet}} \equiv
 1+\frac{\vmet \cdot \hat{n}_{\mathrm{central \, jet}}}
 {E_{T \, \mathrm{central \, jet}}} \, , \label{eq:etadep4}
\end{equation}

and is measured as a function of the
jet detector $\eta$.
The $\gamma$-jet events are used only
at low jet $E_{T}$, as the sample is statistically limited
at high $E_{T}$. The correction for high $E_{T}$ jets is
determined from the jet-jet data. 
For two identical $E_{T}$ jets,
one central and one far forward,
it is more likely for the central
$\eta$ jet to
fluctuate to a higher measured $E_{T}$ than the
forward jet. This is because, for
the same $E_{T}$, forward jets have higher energies and, therefore
better resolutions than central jets. 
In addition, forward jets are kinematically suppressed.
This causes a bias in the jet-jet sample above $|\eta|=2$
which is removed with a cut on the product of the transverse energies of
the two leading jets.

The IC correction is performed before the energy
dependent response correction. Because the energy dependence of
$R_{\mathrm{jet}}$ is folded into $R_{\mathrm{jet}}$ versus $\eta$, this 
function
is not a constant even for a uniform detector (ideal $\eta$ dependence);
it must be, however, smooth.
After the cryostat factor correction is applied, 
the IC correction is determined by means of a
smooth interpolation through the IC of a fit to the measured 
$R_{\mathrm{jet}}$ versus $\eta$ in the CC and the EC.

\subsubsection{Results}

To determine the ideal $\eta$ dependence of the response in the IC, the
response is fit to the function
$R_{\mathrm{jet}}=a+b \cdot \ln [\cosh (\eta)]$ in the CC and EC regions.
This functional form is
derived from the energy dependence of the
response, which is well described by:

\begin{equation}
 R_{\mathrm{jet}} = \alpha + b \cdot \ln E \, . \label{eq:etadep5}
\end{equation}

Thus:

\begin{eqnarray}
\hskip3.5cm R_{\mathrm{jet}} & = &
  \alpha + b \cdot \ln (E_{T} \cdot \cosh \eta) 
  \nonumber \\
\hskip3.5cm         & = & \alpha + b \cdot \ln E_{T} +
   b \cdot \ln (\cosh \eta) \, . \label{eq:etadep6}
\end{eqnarray}

Reordering terms, and for a fixed $E_{T}$ bin:

\begin{equation}
 R_{\mathrm{jet}}=a+b \cdot \ln ( \cosh \eta) \, .  \label{eq:etadep7}
\end{equation}

\begin{figure}
\centerline{\psfig{figure=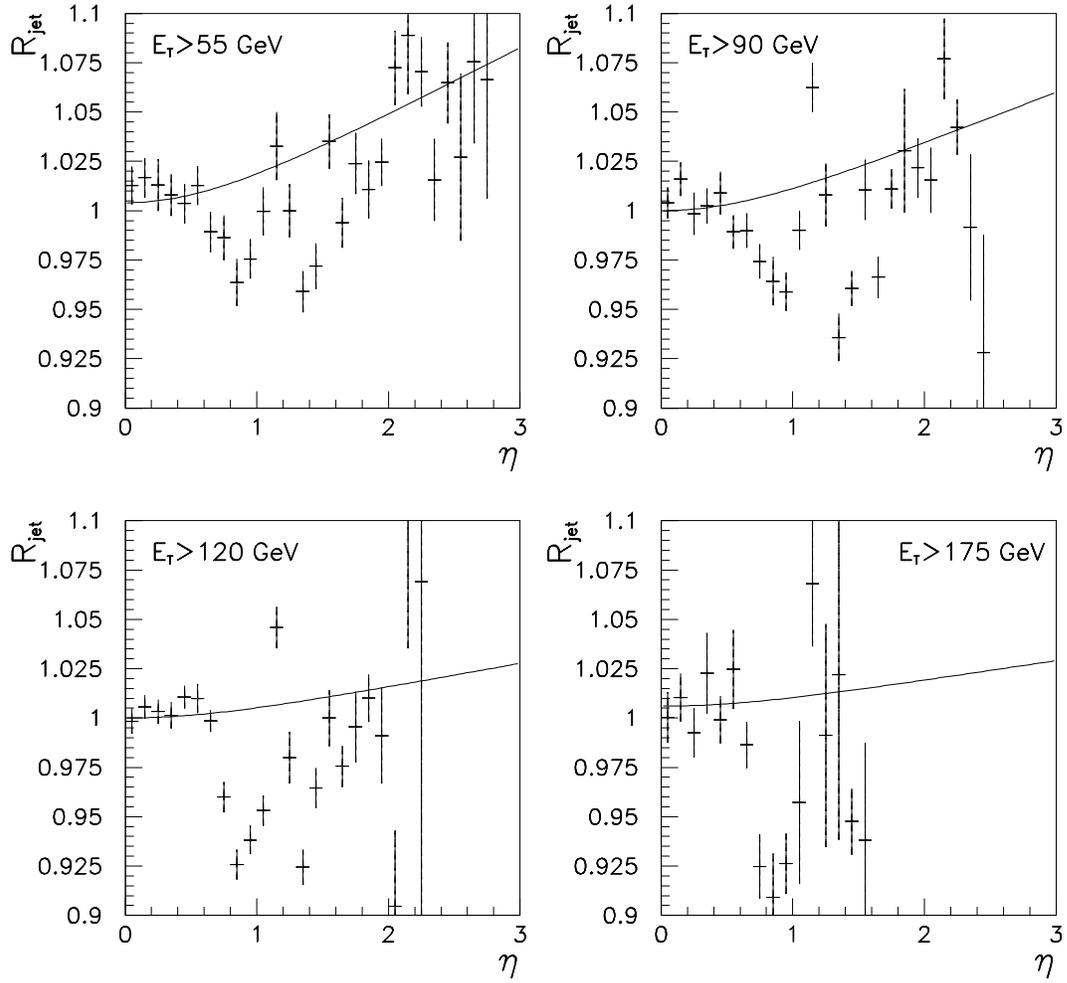,height=15cm,width=15cm}}
\caption
{Response versus detector $\eta$ for jet-jet data ($\eta>0$).
The line is the fit to the ideal $\eta$ dependence.}
\label{FIG:1}
\end{figure}

\begin{figure}
\centerline{\psfig{figure=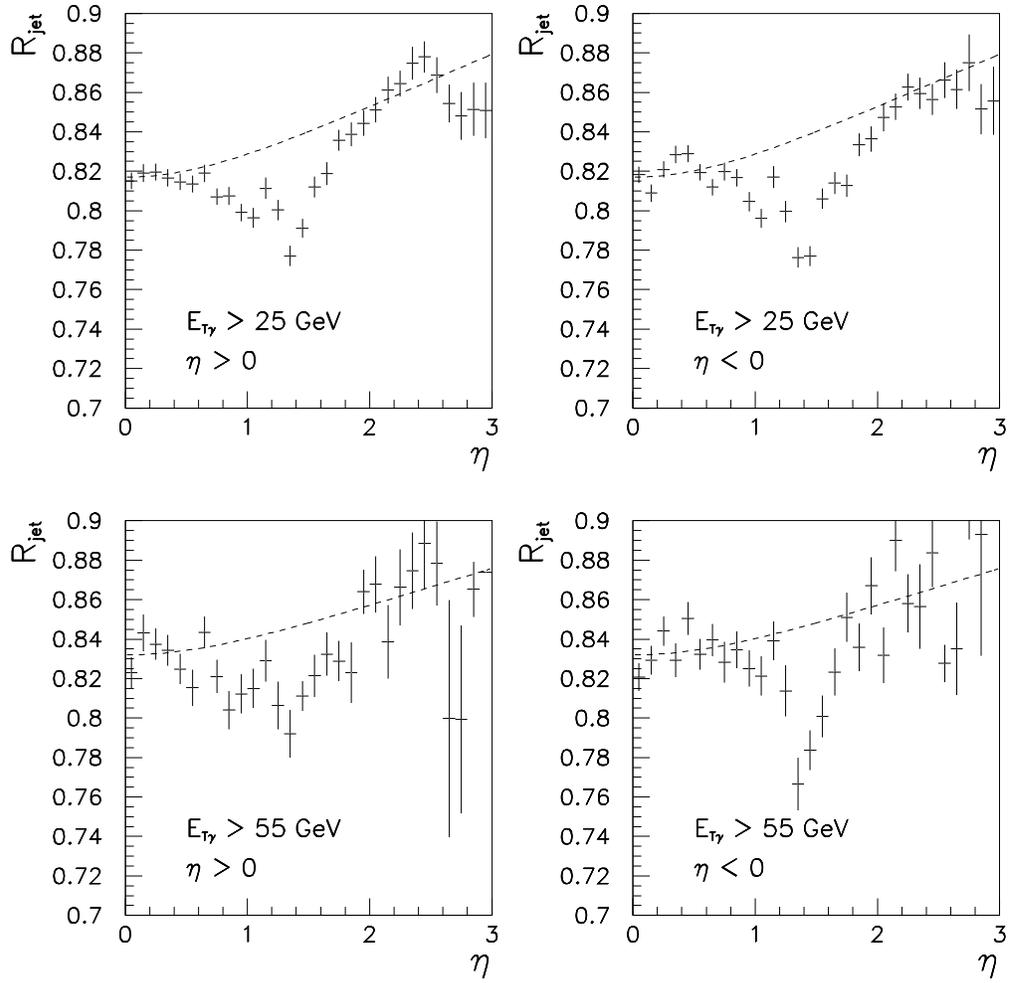,height=15cm,width=15cm}}
\caption
{Response versus detector $\eta$ for $\gamma$-jet data.
The dashed line is the fit to the ideal $\eta$ dependence.}
\label{FIG:3}
\end{figure}

The data are fit in the 
range $|\eta|<0.5$ and $2<|\eta|<2.5$.
A  cut on $E_{T1} \cdot E_{T2}$, which requires this product to be greater
than the $E_{T}$ threshold for which a particular jet trigger is fully
efficient, is applied to avoid the resolution
bias in jet-jet data above $\eta=2$.
The two halves of the detector, $\eta>0$ and $\eta<0$, are
treated separately.  Figures~\ref{FIG:1}~and~\ref{FIG:3}
show the measured $\eta$ dependence
for jet-jet and $\gamma$-jet data respectively, 
along with the fit to an ideally uniform detector. 

\begin{figure}
\centerline{\psfig{figure=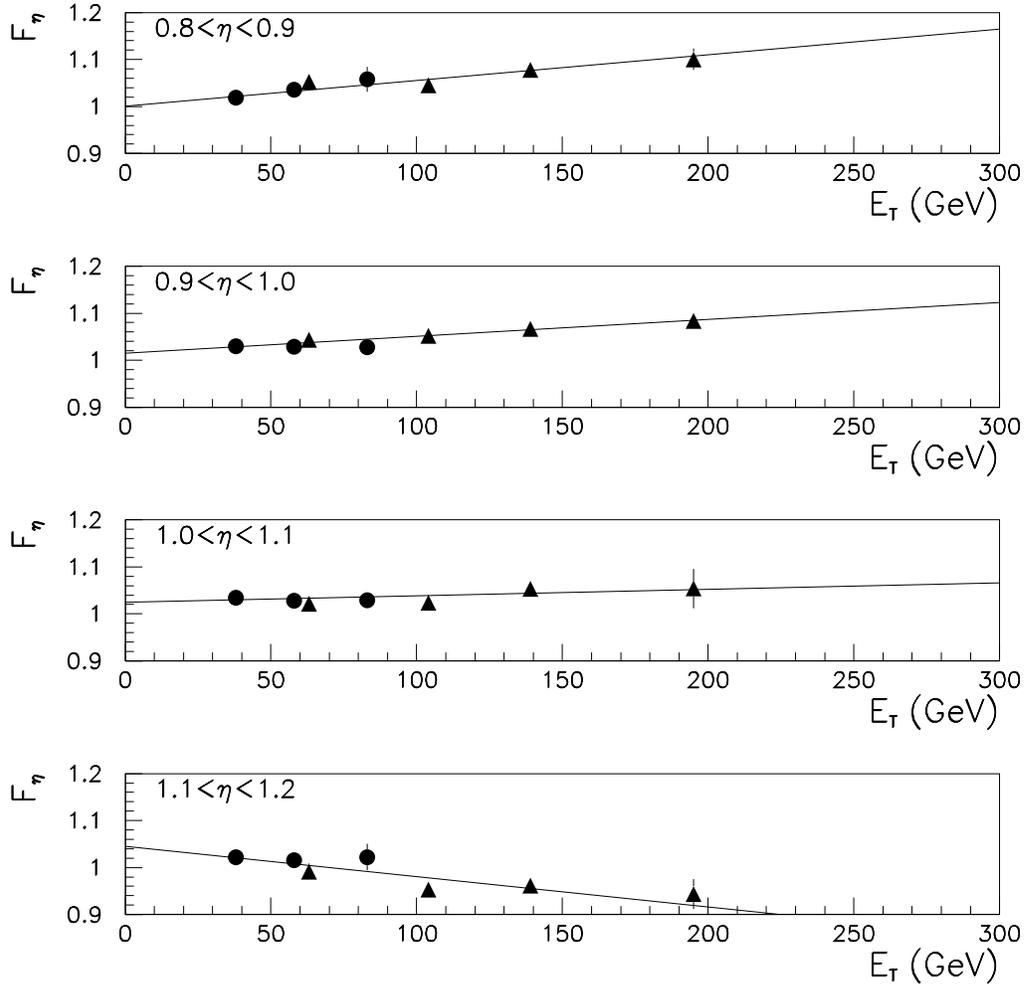,height=15cm,width=15cm}}
\caption
{Correction factor $F_{\eta}$ as a function of central jet 
$E_{T}$
for positive $\eta$ between 0.8 and 1.2.
The circles correspond to the $\gamma$-jet data and the triangles to the
jet-jet data.
The line is a linear fit through all the data points.}
\label{FIG:4}
\end{figure}

The correction is determined in bins of 0.1 in detector $\eta$.
The difference between the
fit to the ideal response and the measured value is used to obtain a
correction factor $F_{\eta}$.  Figure \ref{FIG:4} shows 
$F_{\eta}$ as
a function of the average central jet $E_{T}$ (${\mathcal R}=0.7$) for
several $\eta$ bins.
Also shown is a linear fit to the data, which is used to determine
the correction factor as a function of jet $E_{T}$.

\subsubsection{Error Analysis and Verification Studies}

\begin{figure}[p]
\centerline{\psfig{figure=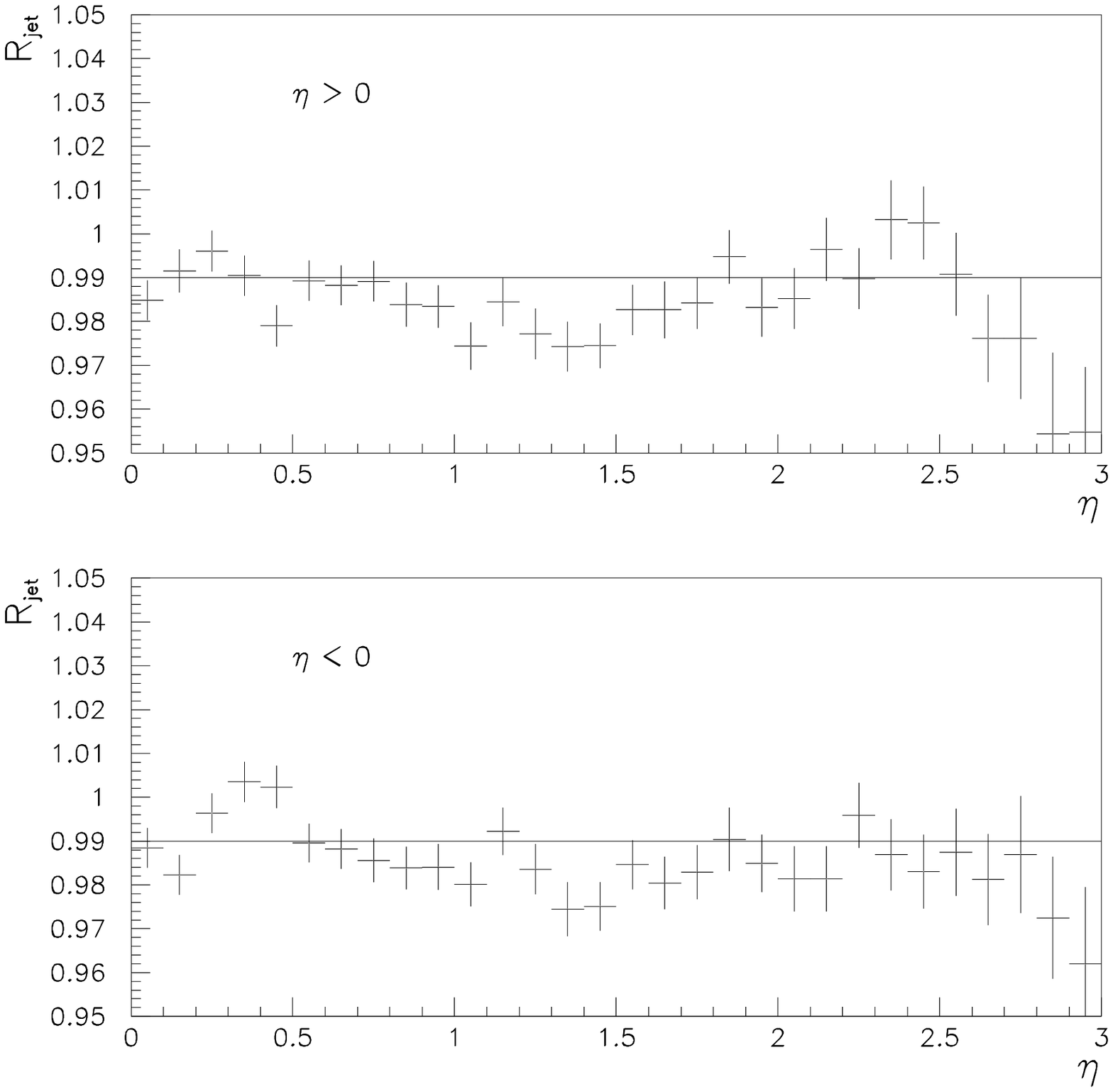,height=15cm,width=15cm}}
\caption{Response versus $\eta$ for $\gamma$-jet events after all
corrections have been applied (including the energy dependent correction).}
\label{FIG:gamma}
\end{figure}

The accuracy of the IC correction is measured
using the fact that the response versus $\eta$
should be constant after all corrections are applied. This is shown in
Fig.~\ref{FIG:gamma}, where the corrections applied to $R_{\mathrm{jet}}$ in
$\gamma$-jet data are the offset, the cryostat factor, the IC, 
and the energy dependence (see Section~\ref{sec:enedep}).
The horizontal line indicates the ideal $\eta$ dependence of the corrected
$R_{\mathrm{jet}}$. Figure~\ref{FIG:gamma} also shows that the
measured $R_{\mathrm{jet}}$ is slightly less 
than unity. The reason is that the correction to the \met\ is only applied on
found objects. As the unclustered energy is not corrected for response effects,
the average total corrected transverse energy of the event is not expected 
to be exactly zero.

\begin{figure}[p]
\centerline{\psfig{figure=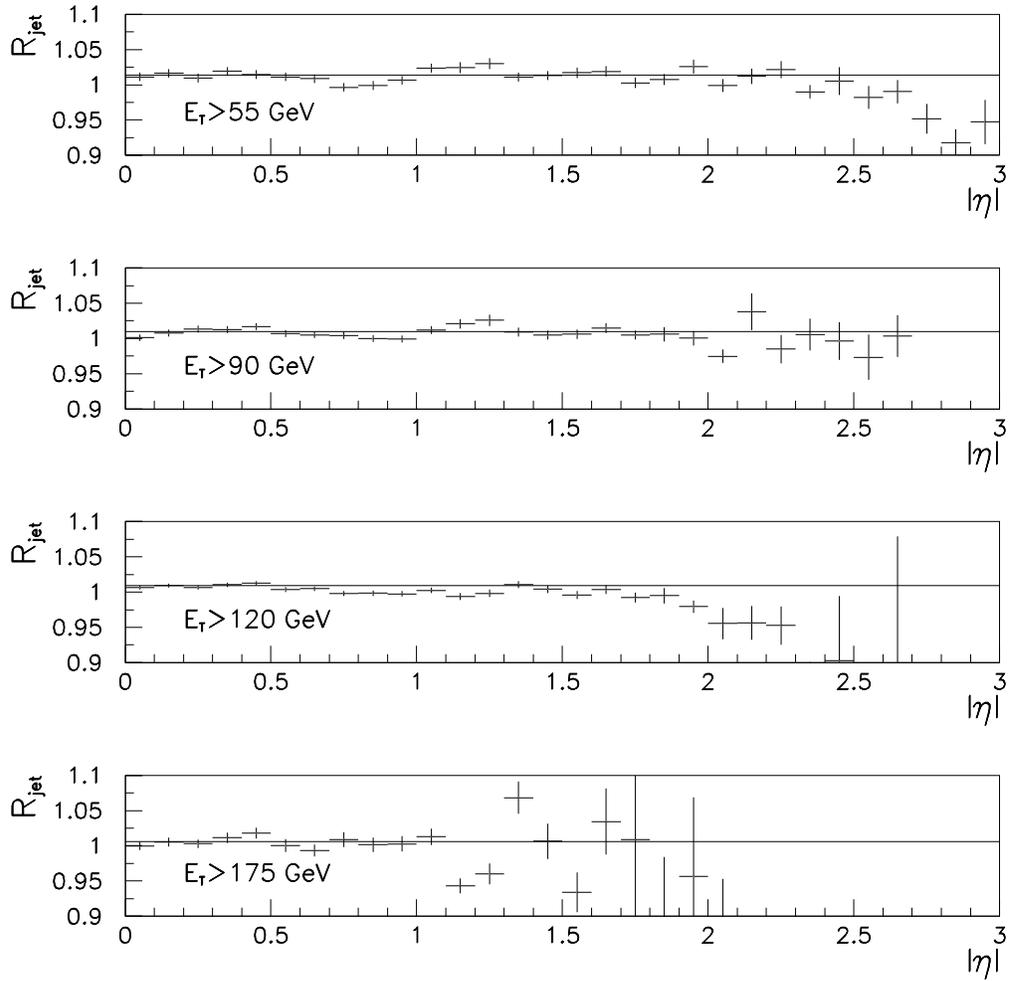,height=15cm,width=15cm}}
\caption
{Response as a function of $\eta$ from jet-jet data
after all corrections have been applied (averaged over positive and
negative $\eta$). 
The line is the ideal $\eta$
dependence. A resolution bias is evident at large $\eta$ (outside the
IC region), where no correction needs to be derived.}
\label{FIG:5}
\end{figure}

Figure~\ref{FIG:5} shows the response versus $\eta$ measured from
jet-jet data after all corrections
have been applied.
The error on $F_{\eta}$ is determined from the 
residuals of the corrected $\gamma$-jet and jet-jet data with respect to the 
ideal $\eta$ dependence. 
Table~\ref{tab:meangamma} shows the means and RMS values for the residual
distributions in $\gamma$-jet data, and Table~\ref{tab:mean}
displays the
means and RMS values for the residual distributions derived
from jet-jet data. The values measured from each sample are in good agreement.

There is some evidence of a residual $\eta$ dependence
beyond $|\eta|=2.5$ in the $\gamma$-jet measurements. 
This bias towards a lower response at high $|\eta|$
may be due to the breakdown of the constant cryostat factor approximation 
at very high pseudorapidity, or to energy lost down the beam pipe.
To account for this effect, an additional $\eta$-dependent error
is assigned. It starts from zero at $|\eta|=2.5$ and increases linearly
up to $3\%$ at $|\eta|=3$.

The IC correction
increases for decreasing cone sizes. It is determined separately  
for each cone size. Once the $\eta$-dependent corrections
are applied to both the jet energy and the event \vmet, the
response becomes constant as a function of pseudorapidity to within
$\sim$ 1\% ($|\eta|<2.5$).

\section{Response: Energy Dependence}\label{sec:enedep}

In Section~\ref{sec:calor}, the ratio $e/\pi$ was shown to 
be greater than unity and energy dependent. 
$R_{\mathrm{jet}}$ must, therefore, also be energy dependent.
The energy dependence of $R_{\mathrm{jet}}$ at low to moderate $E^{\prime}$ 
($E^{\prime} \lsim 100\;{\mathrm GeV}$) is determined from low-$E_{T}$ photons 
and CC jets ($|\eta|<0.7$). 
For high $E^{\prime}$ ($\gsim 100\;{\mathrm GeV}$), $R_{\mathrm{jet}}$ is 
measured from EC ($1.8<|\eta|<2.5$) jets, after $F_{\mathrm{cry}}$ and 
$F_{\eta}$ corrections are applied, taking advantage of the fact that the
CC and EC are similarly constructed. In other words, the EC measurement is
normalized to the CC measurement using $F_{\mathrm{cry}}$, in order to extend 
the energy reach of the global $R_{\mathrm{jet}}$ measurement. 
This Section starts with a 
discussion of the low-$E_{T}$ bias arising from reconstruction and 
resolution effects, and follows with a detailed description of the
measurement of $R_{\mathrm{jet}}$ versus energy.

\subsection{The Low-$E_{T}$ Bias}

The combination of reconstruction inefficiencies, a minimum
jet $E_{T}$ cut,
and the finite jet energy resolution produces a bias in the measured response.
At the $E_{T}$ threshold for jet reconstruction 
$E_{T}^{\mathrm{min}}$ (set at 8~GeV for all 
jet cone algorithms at \D0),
the jet fractional $E_{T}$ resolution is approximately 50\%. Thus,
the net
migration of low-$E_{T}$ jets to higher values is very large. 
At the same time, jets which fluctuate below $E_{T}^{\mathrm{min}}$ 
are not 
reconstructed. 
This low-end truncation biases the average jet $E_{T}$ to higher
values, biases the \met\ to lower values, and therefore biases 
the response to higher
values. The jet reconstruction efficiency is less than unity due to
inefficiencies in the algorithm for turning seed towers into jets. 
Consequently, the low-$E_{T}$ bias is present above the
minimum jet $E_{T}$ reconstruction threshold (up to 20~GeV), and
is both cone size and $\eta$-dependent.

Because the jet energy
response is determined from the event \vmet, $E_{T\gamma}$, and
$\hat{n}_{\gamma}$, it may be measured
without requiring any reconstructed jet in the event. 
To determine the effect of the bias, the response is measured as
a function of $E_{T\gamma}$ for those events both with and without at
least one reconstructed jet. The low-$E_{T}$ bias is then
the ratio:

\begin{equation}
  R_{\mathrm{bias}} = \frac{R_{\mathrm{jet (}\geq \mathrm{1 \, jet)}}}
  {R_{\mathrm{jet (no \, jet \, required)}}} \, . \label{eq:bias1}
\end{equation}

\begin{figure}[p]
\centerline{\psfig{figure=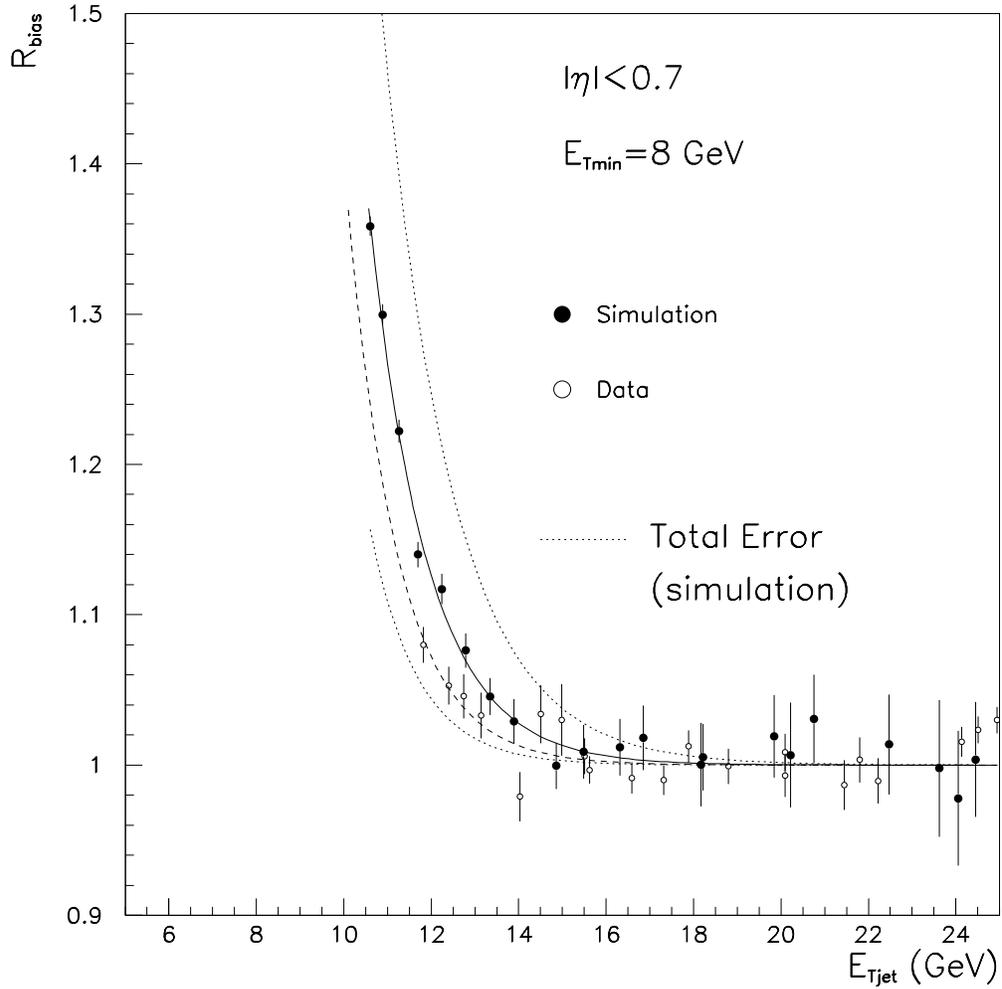,height=15cm,width=15cm}}
\caption
{Low-$E_{T}$ bias, $R_{\mathrm{bias}}$, as a function of 
$E_{T\mathrm{jet}}$ ($\mathcal R$=0.7). 
The full circles denote the simulation
and the open circles the data measurement. The solid and dashed lines are 
fits to the simulation and the data, respectively. 
The dotted band is the total systematic uncertainty in the simulation.}
\label{fig:simul_low_et}
\end{figure}

An independent estimate of the low-$E_{T}$ bias is obtained from
simulated $\gamma$-jet events, constructed as in Section~\ref{sec:mpf}.
The jet response is then ``measured'' using the missing $E_{T}$
projection fraction
method. The low-$E_{T}$ bias is determined from the ratio of
the response from a sample where the
$E_{T}^{\mathrm{min}}$ cut is modeled to the response from a sample
with no such restriction. 
This bias depends on the jet energy resolution, 
$E_{T}^{\mathrm{min}}$, and
the reconstruction efficiency, while the offset and the
$E_{T}$ dependence of the photon cross section contribute very little.
Figure~\ref{fig:simul_low_et} compares 
the simulated bias
and the data measurement for 0.7 cone jets. The results agree within 
the errors of the simulation (dotted band).
The $R_{\mathrm{bias}}$ correction is obtained from
parameterizations of the data measurements for different cone sizes, and
the uncertainty is determined from the simulation.

\subsection{Response versus $E^{\prime}$}

\begin{figure}[p]
\centerline{\psfig{figure=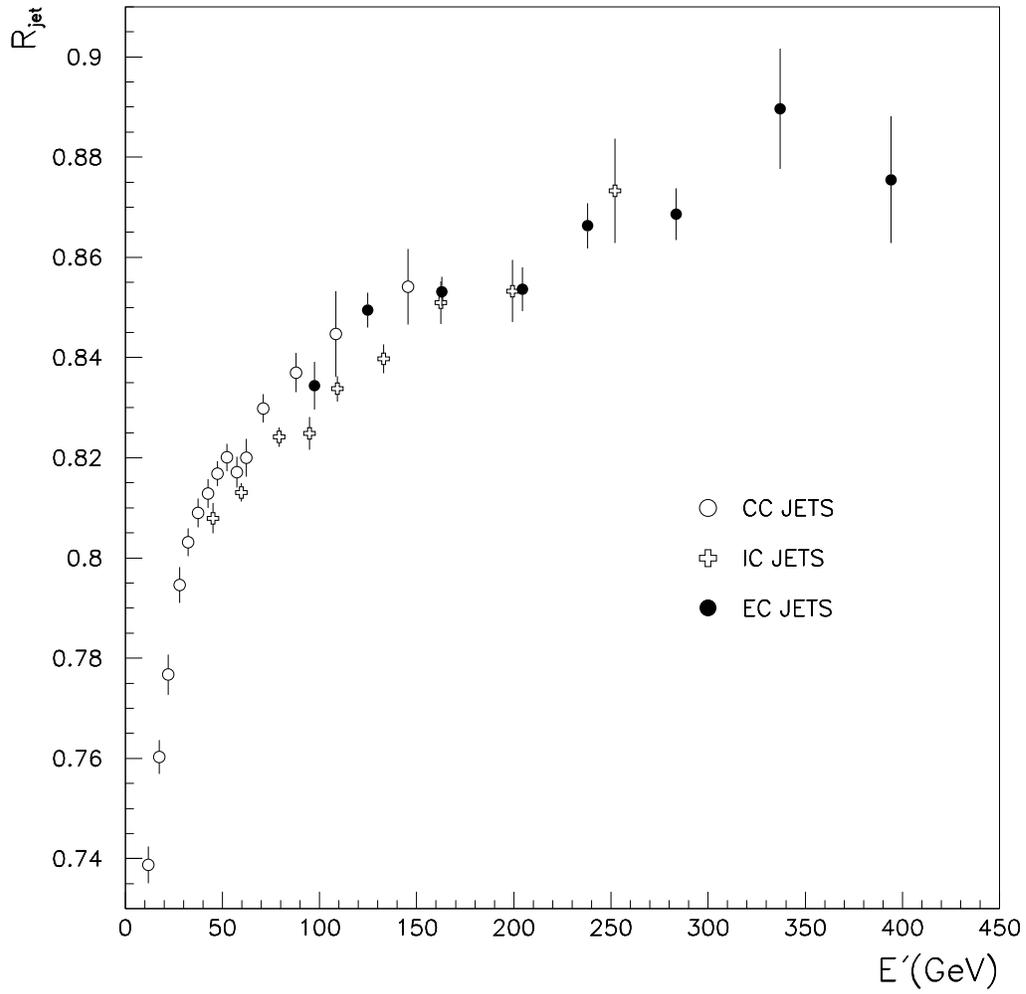,height=15cm,%
width=15cm}}
\caption
{$R_{\mathrm{jet}}$ versus $E^{\prime}$ after $\eta$-dependent corrections
are applied. CC jets correspond to $|\eta|<0.7$, IC jets
to $0.7<|\eta|<1.8$, and EC jets to 
$1.8<|\eta|<2.5$.}
\label{fig:R_v_EP_cor}
\end{figure}

After application of the low-$E_{T}$ bias, offset, cryostat factor, and
IC corrections, $R_{\mathrm{jet}}$ is recalculated as a function of 
$E^{\prime}$ (Fig.~\ref{fig:R_v_EP_cor}).
The average response in each $E^{\prime}$ bin is determined by fitting the
response distribution with either a
single or double Gaussian parameterization. A single Gaussian is sufficient
for $E^{\prime}$ bins with less than 100 entries.  
These fits provide a 
response measurement less affected by misvertexing than the arithmetic means
of the distributions because they are less sensitive to an occasional outlying
point far from the peak.
In Fig.~\ref{fig:R_v_EP_cor}, the points are plotted at the average value 
of $E^{\prime}$ within each bin.
While the $\eta$ dependence of $R_{\mathrm{jet}}$ is dramatically reduced, as
shown in Fig.~\ref{fig:R_v_EP_cor}, the
IC is slightly lower compared to the CC and
EC regions. Although the CC, IC, and EC responses are the same within the
error of the $\eta$-dependent correction, only the most accurate CC and EC 
data are used to derive the jet response. 

\begin{figure}[p]
\centerline{\psfig{figure=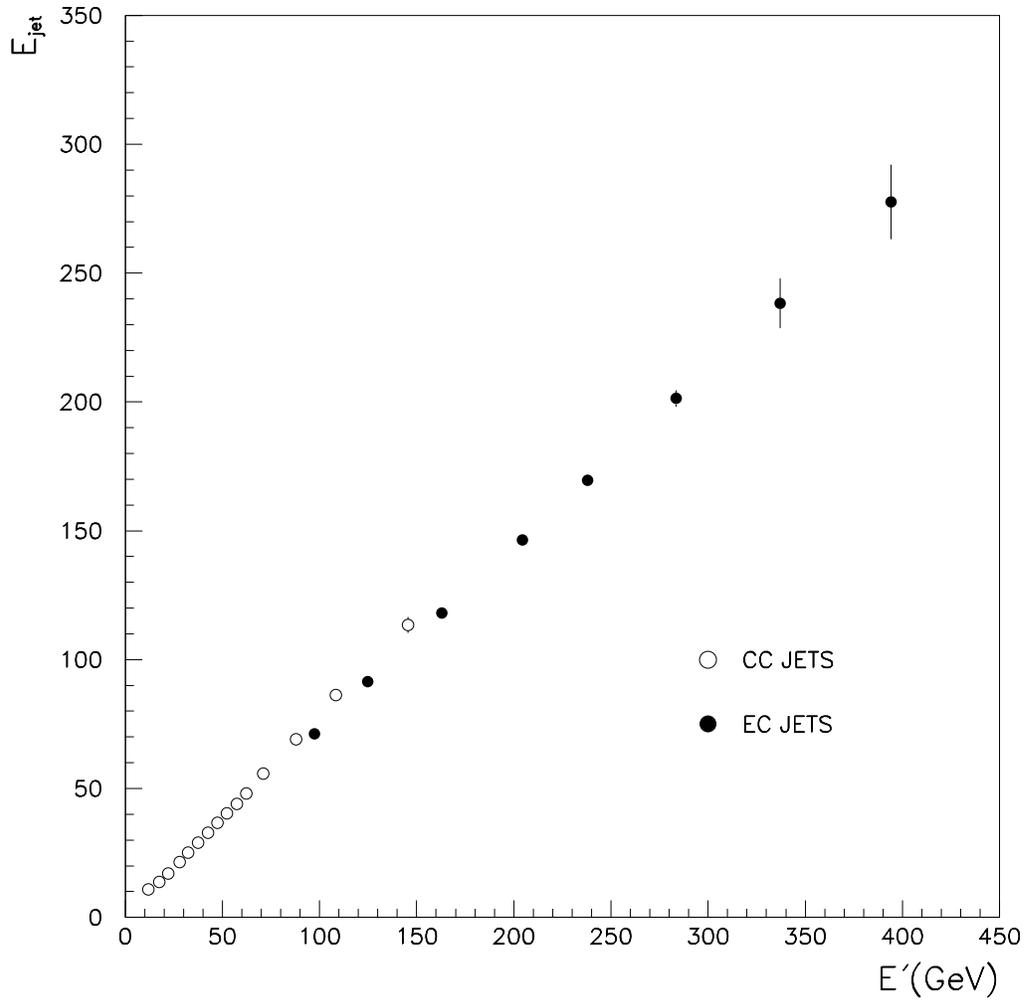,%
height=15cm,width=15cm}}
\caption{$E_{T{\mathrm jet}}^{\mathrm meas}$ (after offset and
$\eta$-dependent corrections) versus $E^{\prime}$ for the 0.7 cone jet 
algorithm. The mismatch between CC and EC jets is due to showering effects.}
\label{fig:E_v_EP_map}
\end{figure}

After determining the response versus $E^{\prime}$, a
mapping between $E^{\prime}$ and the average jet energy is obtained separately
for each jet algorithm. The mapping for 0.7 cone jets is shown in
Fig.~\ref{fig:E_v_EP_map}. The observed difference in this mapping
between EC and CC jets is consistent with showering effects, 
as will be shown in Section~\ref{sec:shower}.

\subsection{Constraining High Energy Response with Monte Carlo}

The jet energy response can be determined from the data for jets with energy up
to $\sim 300\;{\mathrm GeV}$, as shown in Fig.~\ref{fig:E_v_EP_map}. However, 
energy calibration is required for the highest energy
jets in the data, nearly 600~GeV. The inclusion of
Monte Carlo information at high energy is necessary in order to constrain the
jet response extrapolated from the available data.

\begin{figure}[p]
\centerline{\psfig{figure=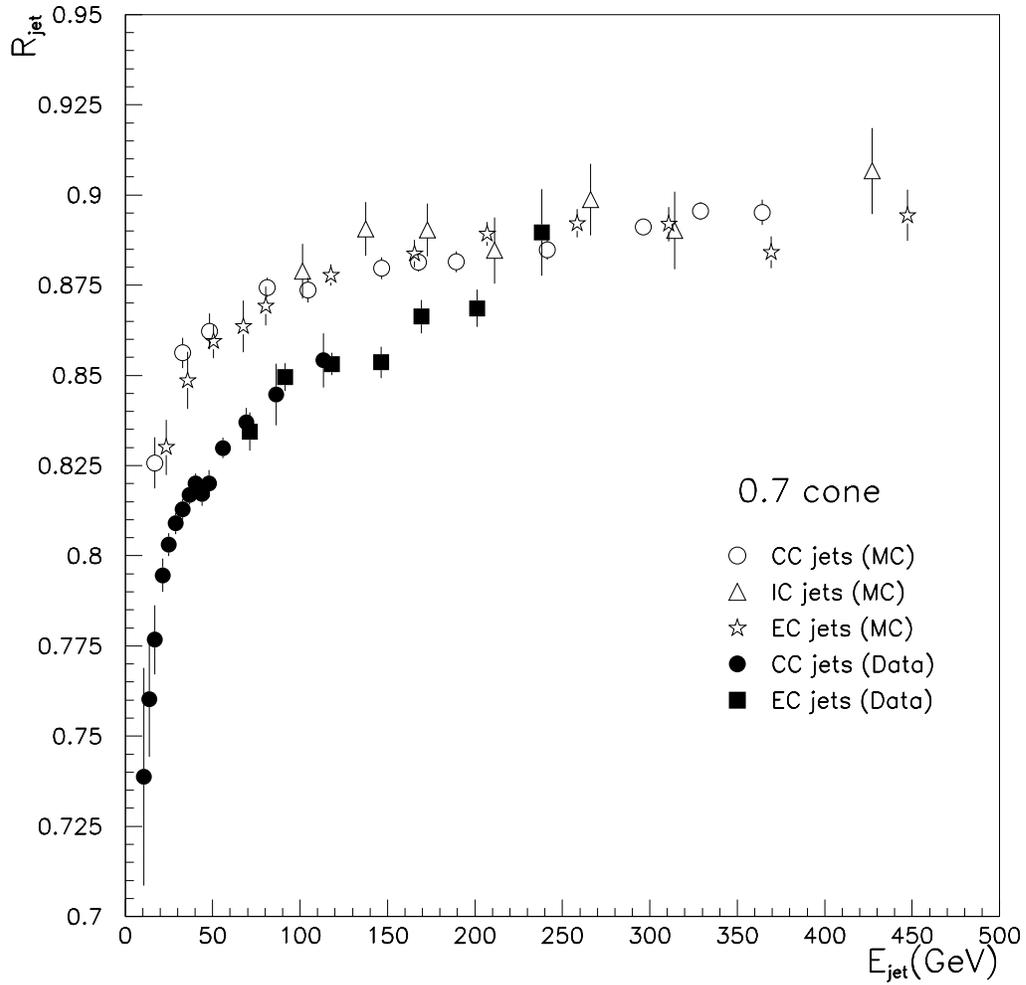,height=15cm,width=15cm}}
\caption
{Comparison between $R_{\mathrm{jet}}$ measured from data and 
$R_{\mathrm{jet}}$
determined from a {\sc showerlib} Monte Carlo sample (${\mathcal R}=0.7$).
The Monte Carlo response flattens out more rapidly
and is nearly constant above 150~GeV.}
\label{fig:compare_mc_data}
\end{figure}

A set of $\gamma$-jet events was generated using {\sc herwig}~\cite{herwig},
processed through the {\sc geant-showerlib}~\cite{showerlib} fast detector
simulation, and reconstructed with the standard photon and jet algorithms.
{\sc showerlib} is a library that contains single particle calorimeter
showers obtained using the {\sc geant} full detector simulation. Because
{\sc geant} is a time consuming computer package, the {\sc showerlib} 
approach is convenient.
The jet response as measured from this sample
is inconsistent with the response as measured from data, 
as shown in Fig.~\ref{fig:compare_mc_data}. This is because the Monte Carlo
does not correctly model the $e/\pi$ ratio (see Fig.~\ref{fig:etopi} in
Section~\ref{sec:calor}).
Therefore, a different approach is followed to generate the
Monte Carlo points at high $E_{T}$. 

\begin{figure}[p]
\centerline{\psfig{figure=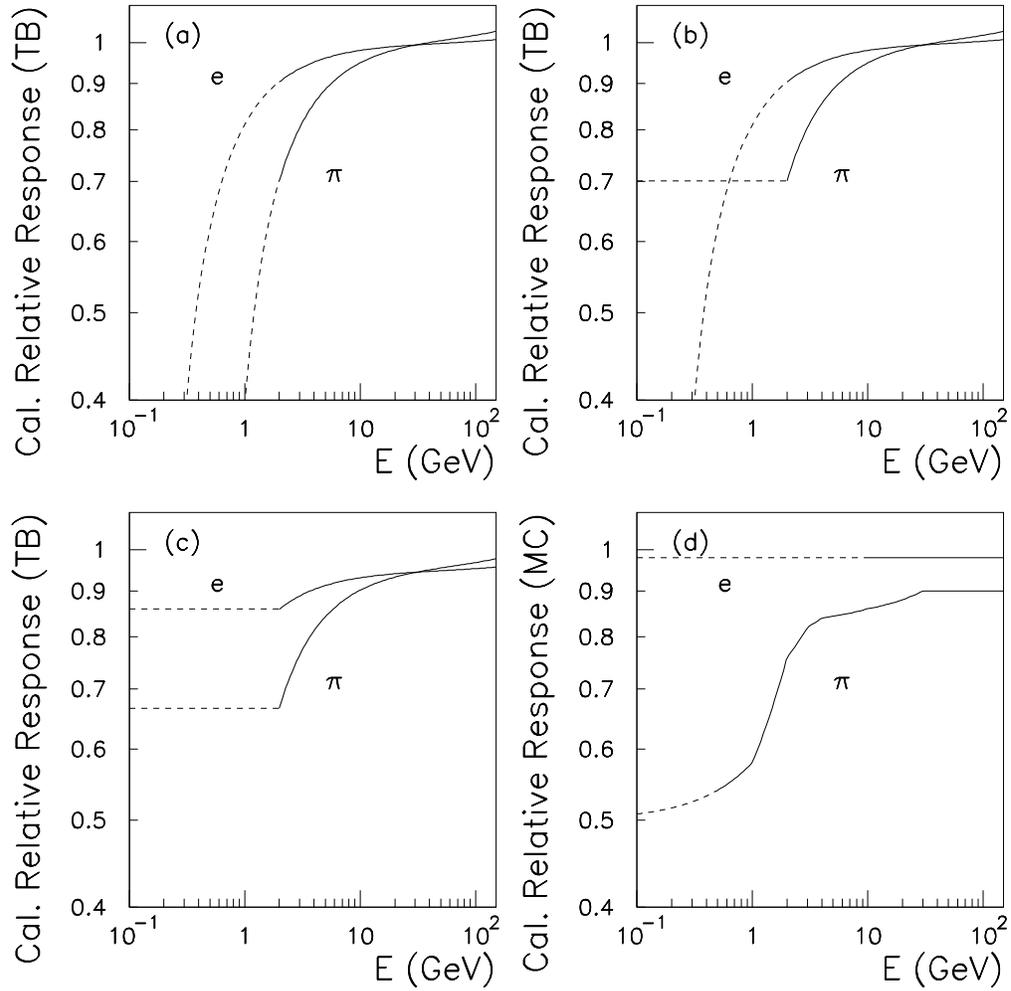,height=15cm%
,width=15cm}}
\caption
{(a-c)~Single particle responses for electrons and pions from  
test beam data.
(d) Single particle response from Monte Carlo. The test beam responses in
(a-c) are identical above 2~GeV; they differ in the assumptions made
for $E<2\;{\mathrm GeV}$ where no data are available.}
\label{fig:SINGPART}
\end{figure}

The method consists of returning to the particle level jets in the 
Monte Carlo. For a given particle jet, the sum of the energies of all 
final-state particles in the jet cone gives the total energy of the jet. 
The detector may then be simulated by convoluting the particle jet with single
particle test beam data. Each individual particle is scaled by the appropriate
response measured from the test beam, parameterized in terms of energy.
The response functions used are shown in Figs.~\ref{fig:SINGPART}a-c.
Electrons and photons are scaled by the electron response, and $\pi^{0}$
mesons
are scaled as two photons, each with half of the energy of the pion. All
other particles are scaled by the response of charged pions. Note that
there are no test beam data below 2~GeV. Figures~\ref{fig:SINGPART}a-c show 
three different assumptions for the behavior of the response in that low energy
range. For comparison, Fig.~\ref{fig:SINGPART}d shows the single particle
responses given by the {\sc showerlib} Monte Carlo. 

\begin{figure}[p]
\centerline{\psfig{figure=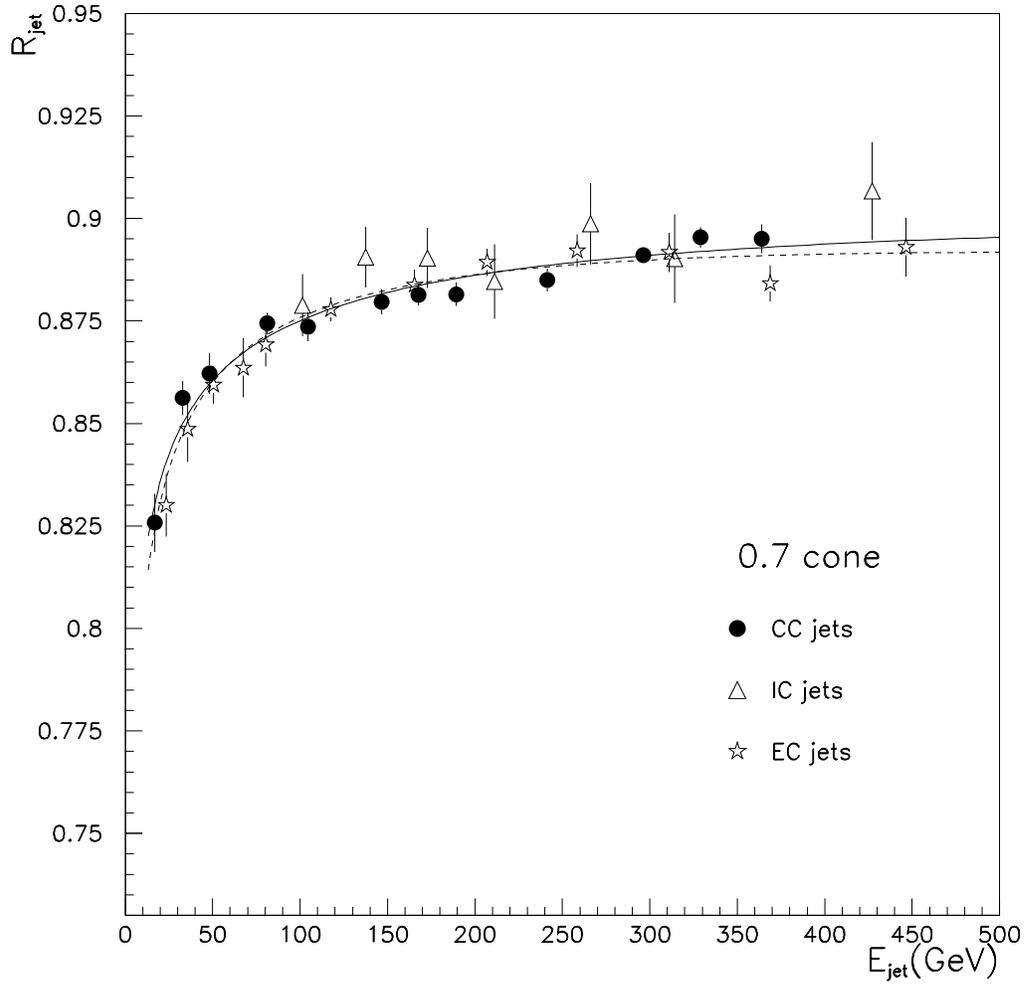,height=15cm%
,width=15cm}}
\caption{Jet energy response determined using the \met\ projection fraction
method from {\sc showerlib} Monte Carlo (circles, triangles and
stars) compared to the response obtained with
the single particle convolution approach (dashed line).  
The solid line shows a fit to the points.}  
\label{fig:MC_RESPONSE_PION}
\end{figure}

\begin{figure}[p]
\centerline{\psfig{figure=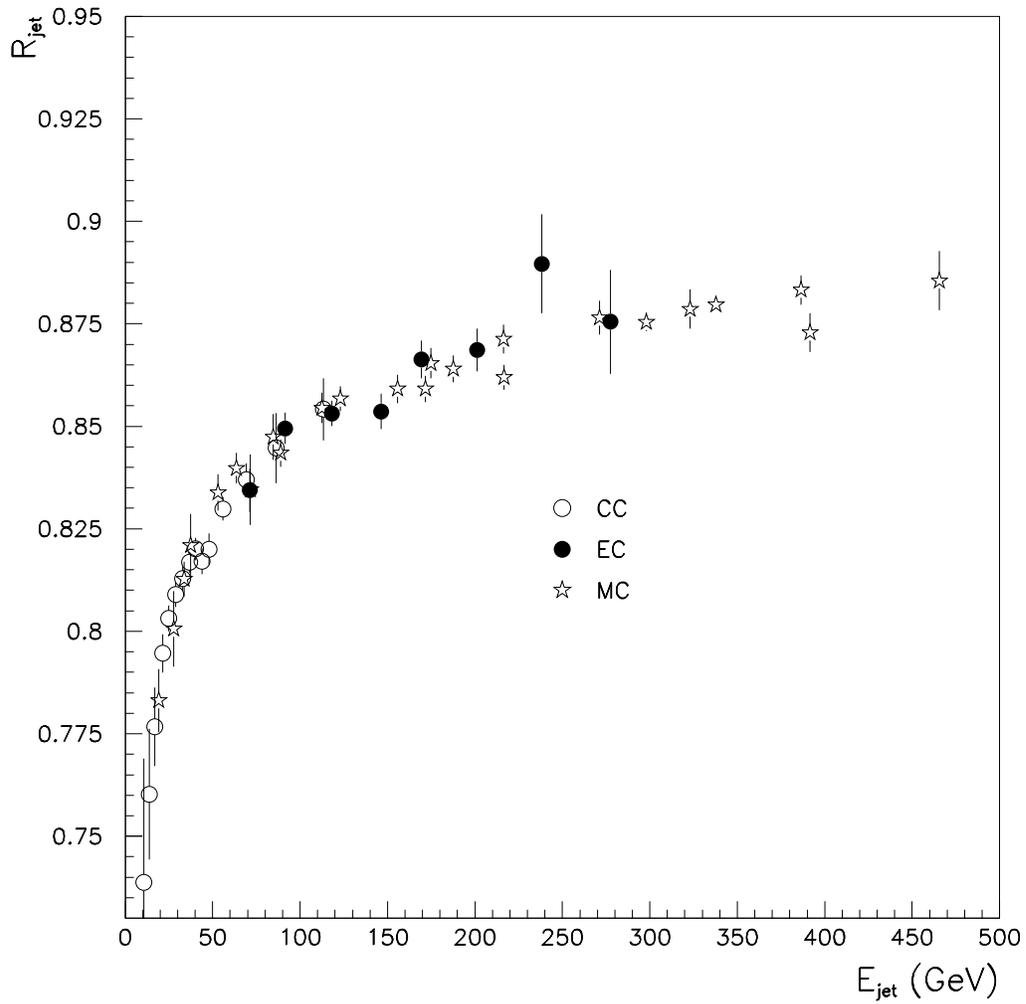,height=15cm,%
width=15cm}}
\caption
{The Monte Carlo response using the convolution method and
the best test beam model for single particle response is shown with the stars. 
It is normalized to $R_{\mathrm{jet}}$ as measured from collider data, shown
as open circles in the CC and full circles in the EC.}
\label{fig:mc_norm}
\end{figure}

As a test, illustrated in Fig.~\ref{fig:MC_RESPONSE_PION}, the {\sc showerlib} 
Monte Carlo response as determined using the \met\ projection fraction method 
is compared with the response derived from the single particle convolution
method, using the Monte Carlo response functions (Fig.~\ref{fig:SINGPART}d).
The two responses agree to within 1\%. The shapes of the response curves
derived from the three different assumptions shown in 
Figs.~\ref{fig:SINGPART}a-c are then compared with the shape of the response
derived from data. As shown in Fig.~\ref{fig:mc_norm}, the best agreement is 
found from the model in Fig.~\ref{fig:SINGPART}b (extrapolated electron 
response, fixed pion response).

\begin{figure}[p]
\centerline{
\psfig{figure=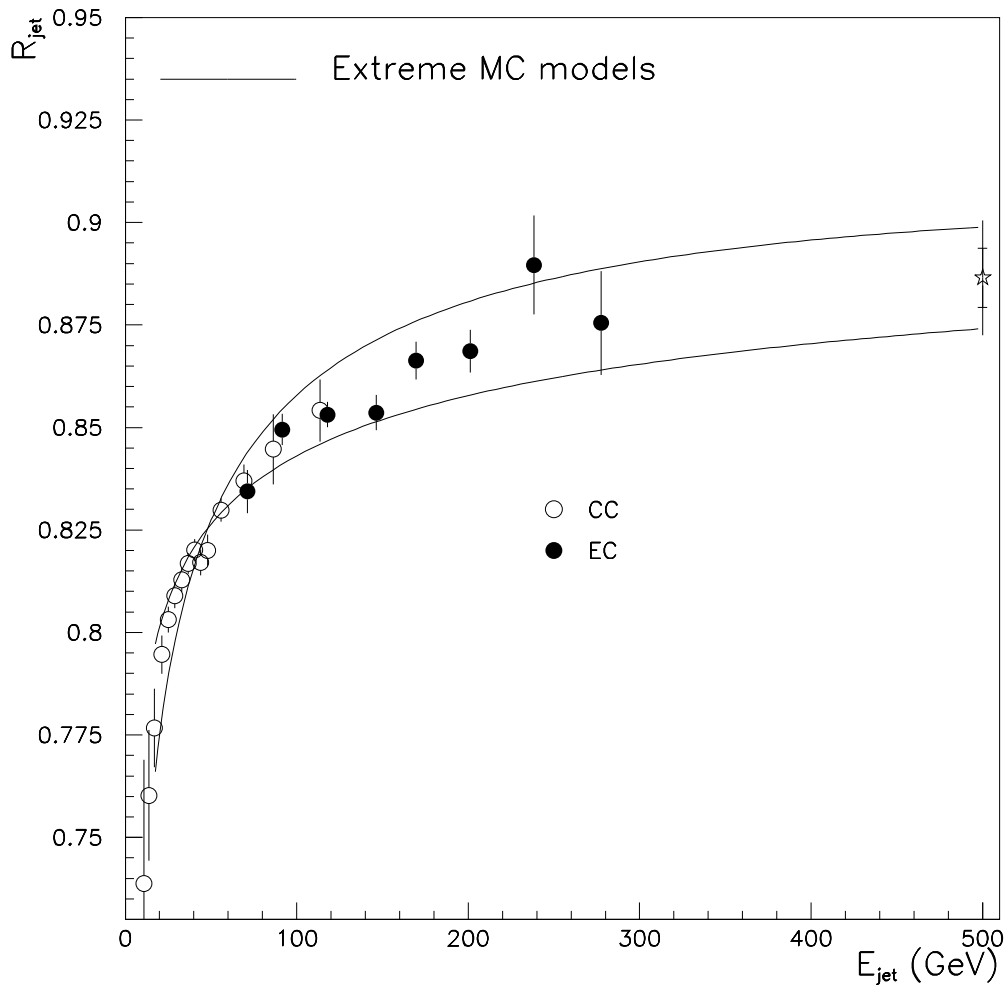,height=15cm,width=15cm}}
\caption
{The solid curves show the extremes of the
three test beam response models normalized to the collider data.
The star indicates an estimated response from the convolution method
for 500~GeV jets. The inner error bar is based on the difference between 
the two extreme models. The full error bar includes all errors, as explained 
in the text.}
\label{fig:mc_norm_err2}
\end{figure}

The model-dependent error
in the extrapolation depends upon the lower cutoff energy
for normalizing to the data. The two extreme single particle 
response models are used:
the extrapolated and the fixed 
responses (Figs~\ref{fig:SINGPART}a and \ref{fig:SINGPART}c respectively). 
The Monte Carlo curves are normalized to
the entire energy range where data exist, as shown in 
Fig.~\ref{fig:mc_norm_err2}.
The assumption is that these two extreme
models bound the true response in the region where no data are available.
The response is estimated
as the mean value of the two extreme models
at 500~GeV, and the error due to choice of model (inner error bar) 
as the difference divided by $\sqrt{12}$ ($\sim$ 0.7\%).
To this, an uncertainty of $\sim0.6$\% is added linearly, based on the closure 
test illustrated in Fig.~\ref{fig:MC_RESPONSE_PION}. 
The small normalization and fit errors are also included in 
the full error bar shown.

\subsection{Response Fits}

\begin{figure}[p]
\vskip0.5cm
\centerline{\psfig{figure=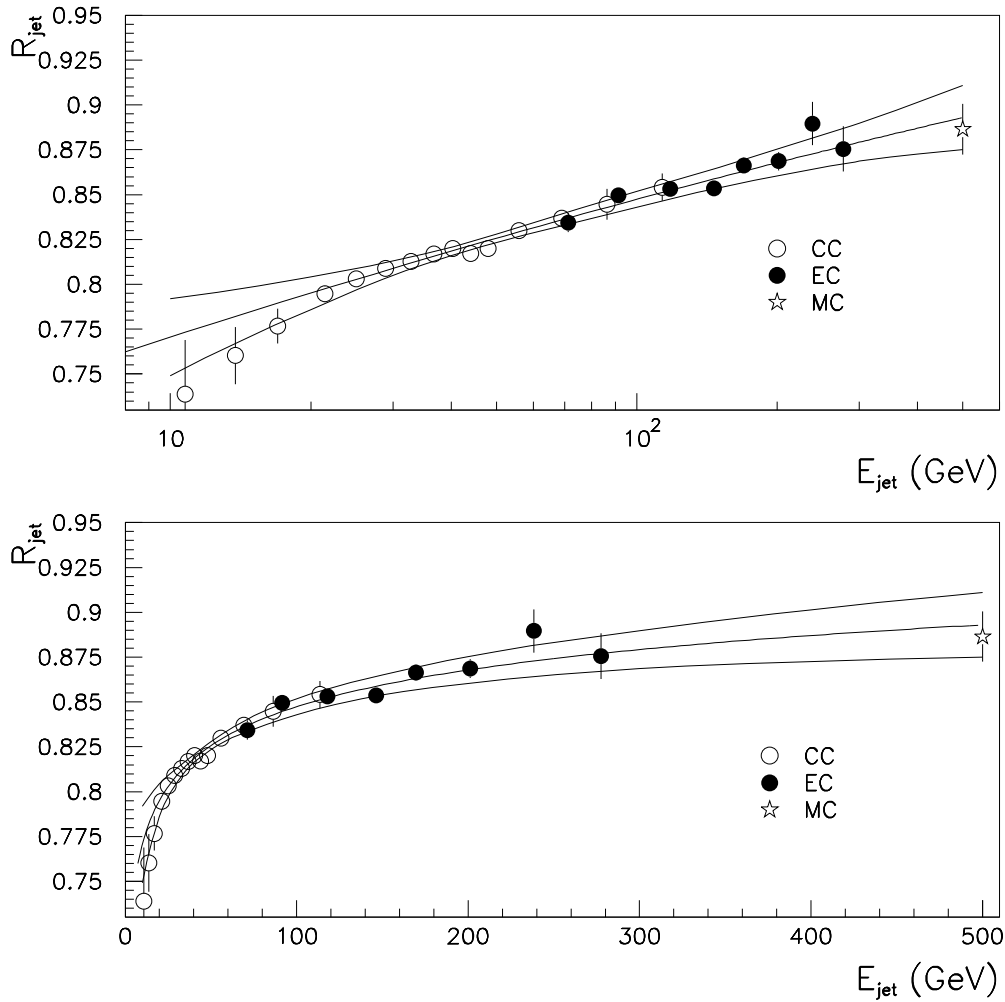,height=15cm,%
width=15cm}}
\caption
{$R_{\mathrm{jet}}$ versus energy on a logarithmic (top) and linear (bottom) 
energy scale for the 0.7 cone jet algorithm.  
The outer band shows limits on the measured response for
jets based on the region in parameter space defined by the
$\chi^2 = \chi^2_{\mathrm{min}} + 3.5$ surface.  This region corresponds to the
68\% confidence region of parameter fluctuations from the nominal values.}
\label{fig:R_v_E_err}
\end{figure}

The response versus energy for 0.7 cone jets is shown in
Fig.~\ref{fig:R_v_E_err}. The data are fit with the
functional form:

\begin{equation}
R_{\mathrm{jet}}(E) = a + b \cdot \ln E + 
             c \cdot (\ln E)^2 \, . \label{eq:enedep1}
\end{equation}

A logarithmic functional form is motivated by 
the fact that the EM content of a hadronic shower slowly increases 
with increasing shower energy~\cite{ferbel}.
In the development of such a shower some fraction 
$\langle f_{\mathrm{em}} \rangle$ of the
energy is spent in the production of $\pi^{0}$ and $\eta$ mesons.
A hadronic shower, therefore, has both an electromagnetic and a
non-electromagnetic component. As defined in Section~\ref{sec:calor}, 
$e$ and $h$
are the responses of the calorimeter to the EM and non-EM components of a 
hadronic shower. If $\pi$ is the response to charged pions,
then:

\begin{equation}
  \frac{e}{\pi} =
    \frac{1}{\frac{h}{e} \, -
    \langle f_{\mathrm{em}} \rangle (\frac{h}{e}-1)} , \, \label{eq:enedep2}
\end{equation}

where, in general, $\langle f_{\mathrm{em}} \rangle$ is reasonably well 
described by $\sim \alpha \cdot \ln E\; {\mathrm (GeV)}$. In a
completely compensating calorimeter $e/h$=1, and 
the ratio $e/\pi$ is unity for
all energies. If $e/h \neq$ 1, the ratio
$e/\pi$ tends to one at high energies but grows
exponentially as energy decreases. If $e=1$, then $\pi$ would be
$a + b \cdot \ln E$.

The logarithmic
behavior of $R_{\mathrm jet}$ versus energy is
suggested by the test beam
measured $e/\pi$ ratio as a function of beam
energy~\cite{kot}, and verified directly from the measurement of
$R_{\mathrm{jet}}$ using collider data.
The inclusion of the $(\ln E)^2$ term in Eq.~\ref{eq:enedep1}
improves the agreement of the
fit with the data in the quickly varying low energy region, while
preventing the slope of the function from rising faster than the data
at high energies.  

The three sets of points shown in Fig.~\ref{fig:R_v_E_err} correspond to
CC jet data (open circles), EC jet data (filled circles), and Monte Carlo
data (star). The three lowest energy points have large errors (nearly fully 
correlated point-to-point in energy), as a result of
the low-$E_{T}$ bias correction. They are not used in the fit which includes
points above 30~GeV.
The $\chi^2$ of the fit is 10 for 15 degrees of
freedom.
Table~\ref{tab:resp_pars} lists
the fit parameters for the various cone algorithms. The calorimeter
response is the same for all cone sizes as expected, although the 
parametrizations are slightly different because each algorithm associates
a different energy to the same cluster.  

\subsubsection{Errors from Response Fits}

The errors in the fit parameters, listed in Table~\ref{tab:resp_pars}, 
represent one standard deviation uncertainties, calculated from the 
$\chi^2 = \chi^2_{\mathrm{min}} + 1$ surface in the parameter space.
However, the probability for all fit parameters to simultaneously take on 
values within the one standard deviation region 
($\chi^2 = \chi^2_{\mathrm{min}} + 1$)
is significantly less than 68\%~\cite{MINUIT}. To account for this
reduced probability, a surface is mapped out in parameter
space containing a region with a 68\% probability for parameter
fluctuations in all the response fits.  
For three parameters this corresponds to the surface defined
by $\chi^2 = \chi^2_{\mathrm{min}} + 3.5$~\cite{MINUIT}.  The points lying on
this surface are then mapped back onto the response versus energy
plane. At each energy, the error is determined by the parameter set giving
the greatest deviation from the nominal response.
The result for the 0.7 cone algorithm is shown in Fig.~\ref{fig:R_v_E_err}.  

The band represents the error on the fit to the response, and is used in 
place of one derived from the errors in Table~\ref{tab:resp_pars}.
At high energy, it is only weakly
dependent on the region where the fit is performed.
Alternate functional forms were also investigated, giving results consistent 
with those from Eq.~\ref{eq:enedep1}. 

\subsubsection{Error Correlations in Response Fit}

The error of the fit on the jet response measurement dominates the systematic 
uncertainty over large regions of jet energy and pseudorapidity.  
In this section, the point-to-point correlations in uncertainty due to 
the fit errors described in the previous section are derived.

The error bands in Fig.~\ref{fig:R_v_E_err} show 
the 68\%-probability response at each energy based on
fluctuations of all fit parameters. However, the parameter set that produces a
given variation at one energy need not produce the same variation at any other
energy in either magnitude or sign. In general, no single
set of fit parameters may exist to produce the 68\%-probability curves. 
But it is reasonable to expect that errors for two points of similar energy 
should be largely correlated. The correlations in the fit error are 
calculated as follows:

\begin{itemize}
\item A grid of parameter sets ($N_{\mathrm{grid}}$) is generated to define the
\mbox{$\chi^2 \le \chi^2_{\mathrm{min}} + 3.5$} volume.  This volume
contains all fit parameter set fluctuations corresponding to
a 68\% probability content.
Each parameter set defines a response function contained
within the bands shown in Fig.~\ref{fig:R_v_E_err}.

\item The parameter sets are used, noting the variation
in response at 11 values of uncorrected jet
energy\footnote{In this context uncorrected energy means energy
not corrected for response. The low-$E_{T}$ bias, offset, and 
$\eta$-dependent corrections have already been applied.}
between 10 and 500~GeV.  The correlations in response
for each pair of energy values are calculated from the
parameter sets in the grid, which are used to define a correlation
matrix for the 11 energy values.  Each element of the matrix
($\rho(i,j)$) is the standard correlation coefficient between the
responses measured at each energy value, and defined as:

\begin{equation}
 \rho(i,j) = %
 \frac{ \sum_{n=1}^{N_{\mathrm{grid}}} (R_{n}(i) - 
\overline{R(i)}) %
 (R_{n}(j) - \overline{R(j)}) } %
 { [ \sum_{n=1}^{N_{\mathrm{grid}}} (R_{n}(i) - 
 \overline{R(i)})^2 %
 \sum_{n=1}^{N_{\mathrm{grid}}} (R_{n}(j) - 
 \overline{R(j)})^2 ]^\frac{1}{2} } \, , \label{eq:enedep5}
\end{equation}

\hskip-4mm where $R(i)$ is the response for the $i^{\mathrm{th}}$ energy bin.

\end{itemize}

\begin{figure}[p]
\centerline{\psfig{figure=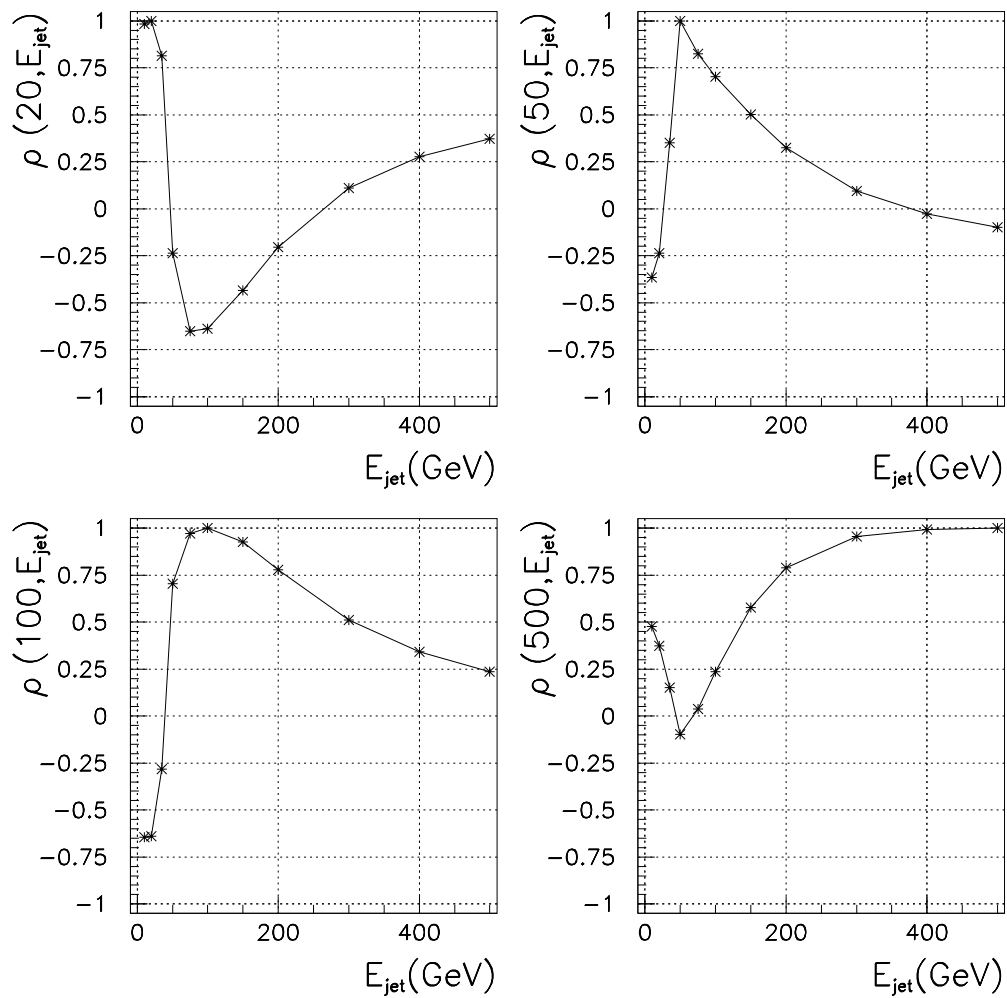,%
height=15cm,width=15cm}}
\caption
{Error correlations for 0.7 cone response fit.  Error correlation points
are shown in four slices from the full correlation matrix.  
The lines connect the points. The four curves
show the point-to-point correlation of fit errors relative to energy values of
20, 50, 100, and 500~GeV respectively.}
\label{fig:Err_corr_7b}
\end{figure}

Table~\ref{tab:cor_7b} shows
the correlation matrix for the response fit to 0.7 cone jets.
Correlations are illustrated
graphically in Fig.~\ref{fig:Err_corr_7b}, where four rows of the matrix are
plotted, showing the error correlations relative to the errors at
20, 50, 100, and 500~GeV respectively.  

\subsection{Shower Containment}

The D\O\ central and end calorimeters are more than 7.2 and 8 
interaction lengths ($\lambda_{\mathrm{INT}}$) thick, respectively.
Although the EC is thick enough to contain all expected showers, there may
be some leakage out of the CC for very high energy jets.
Because the energy not contained
within the calorimeter contributes to \met,
the measured response in principle corrects for leakage.
The response of jets with energy less than 
$\sim 100\;{\mathrm GeV}$ is measured 
using the CC and, therefore, includes the containment correction.
For jets with energy $\gsim 100\;{\mathrm GeV}$, the response is determined 
from EC data.  
Measuring the high energy response in the end cap calorimeter, 
with full containment, and applying it to the central calorimeter,
where there could be some leakage, could cause a bias at high energy.
The cryostat factor correction does not account for this bias, as it is
measured from jets with energies $\sim 100\;{\mathrm GeV}$. 
The effect of shower containment on the response measurement at high
energies is evaluated using a simulation based on Monte Carlo and experimental
data. 

\begin{figure}[p]
\centerline{\psfig{figure=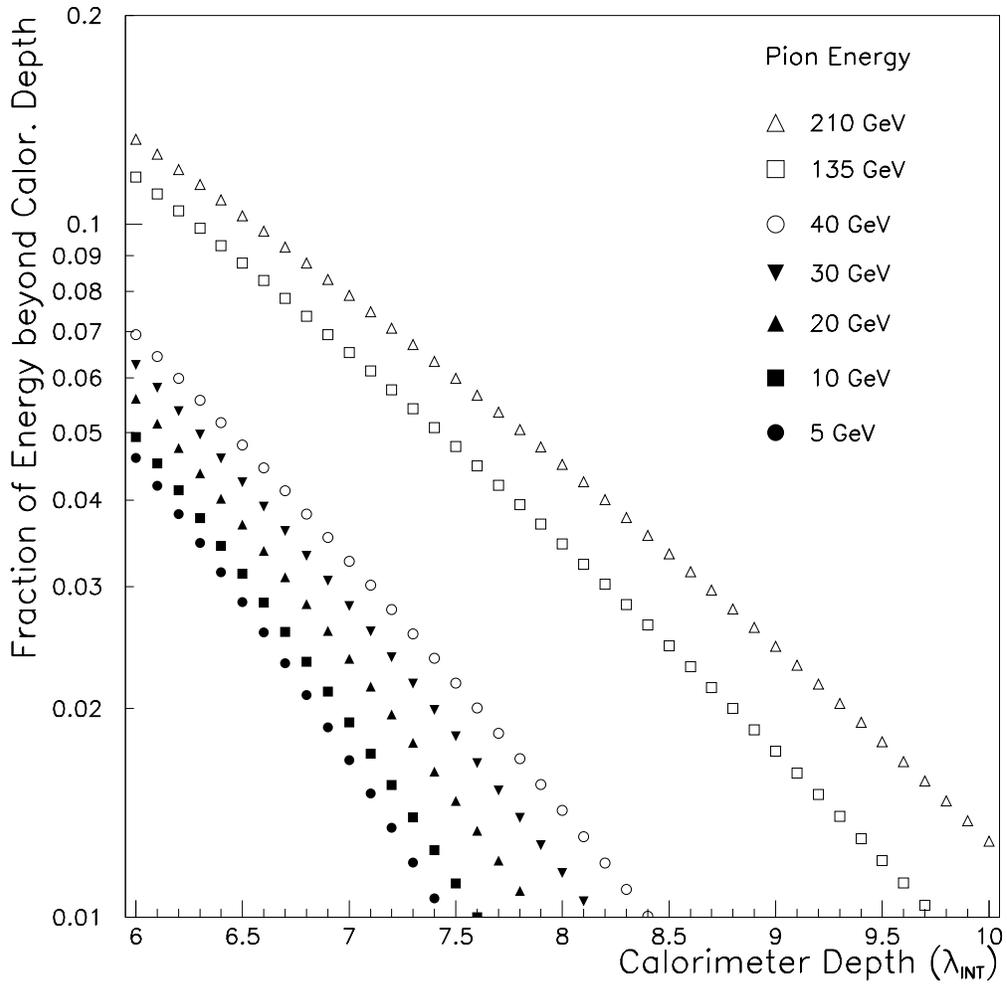,height=15cm,width=15cm}}
\caption{Fraction of pion energy escaping from the NuTeV calorimeter
as a function of depth in units of interaction lengths.}
\label{FIG:depth}
\end{figure}

The NuTeV collaboration has measured the energy 
loss of charged pions as a function of number of interaction lengths in 
their calorimeter~\cite{nutev}, which
consists of stainless steel (absorber) and scintillator (active medium). 
For pions of various incident energies,
Fig.~\ref{FIG:depth} shows the fraction of energy deposited beyond a certain
calorimeter depth as a function of that depth in units of interaction length,
$\lambda_{\mathrm{INT}}$.  
The energy loss as a function
of $\lambda_{\mathrm{INT}}$ does not depend strongly on the type of absorber.
The NuTeV data can, therefore, be used for this study despite the 
difference in the calorimeter compositions. 

\begin{figure}[p]
\centerline{\psfig{figure=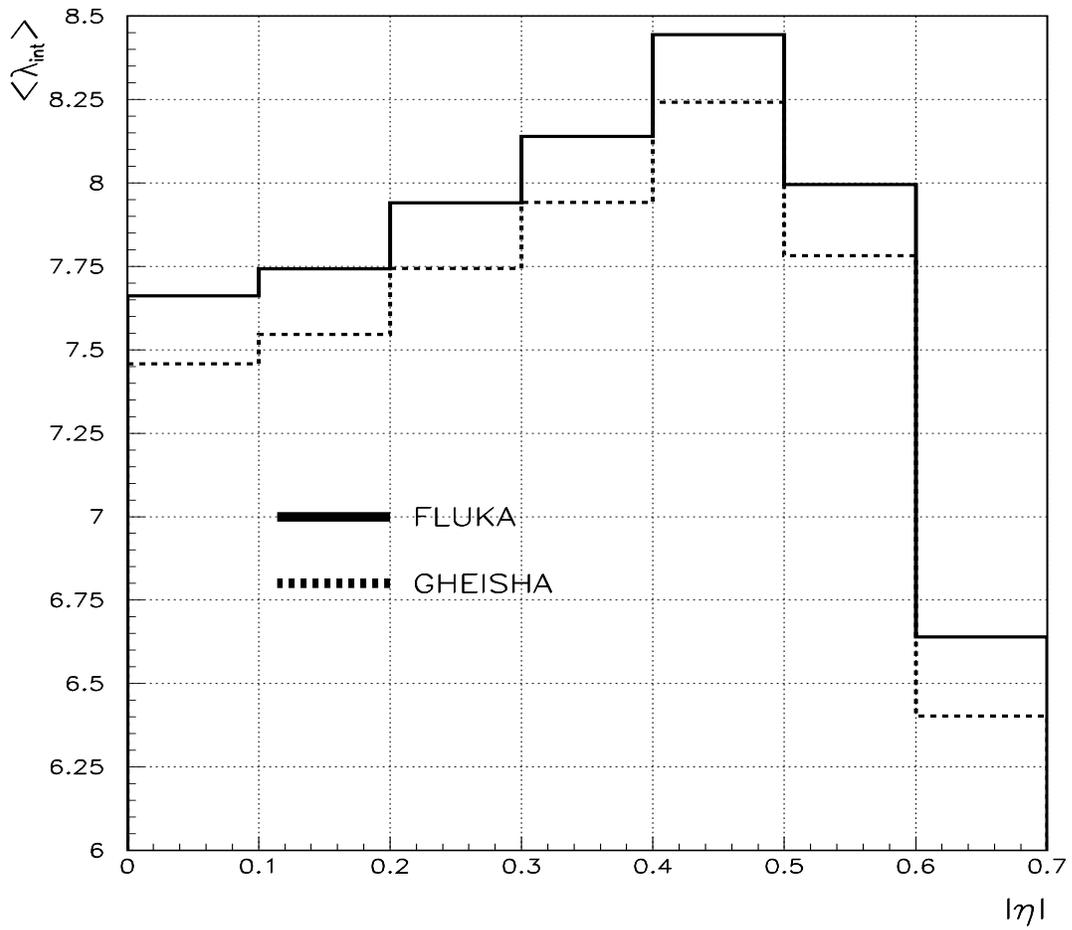,%
height=15cm,width=15cm}}
\caption{Depth of the \D0\ central calorimeter in units of
interaction lengths.}
\label{fig:lambdaint}
\end{figure}

The thickness of the D\O\ central calorimeter is determined from
a {\sc geant} simulation of the D\O\ detector.
Figure~\ref{fig:lambdaint} shows the
depth of the CC as a function of $\eta$ in units of $\lambda_{\mathrm{INT}}$.
By definition, $\lambda_{\mathrm{INT}}$ depends on the proton cross section 
for inelastic collisions with nuclei of the absorption material. 
The uncertainty in this
cross section accounts for the difference between the two sets of calorimeter
thicknesses displayed in 
Fig.~\ref{fig:lambdaint}.
{\sc geisha} and {\sc fluka} are different simulations of physics processes
implemented in {\sc geant}~\cite{geisha,fluka}.

The energy containment of jets is modeled using particle level
{\sc herwig} jets for $|\eta|<0.7$. The number of interaction lengths
each particle traverses is determined from
the $\eta$ of each particle in the jet and the information displayed
in Fig.~\ref{fig:lambdaint}. Figure~\ref{FIG:depth} is used to determine
the fraction of energy of each particle that is contained in the calorimeter.
All strongly interacting particles are treated as charged pions. 
Electromagnetically interacting particles ($\pi^{0}$,$e$,$\gamma$) are
fully contained.
The energy contained within a jet is then compared to the total particle
level jet energy to measure the fraction of energy escaping the calorimeter.

\begin{figure}
\centerline{\psfig{figure=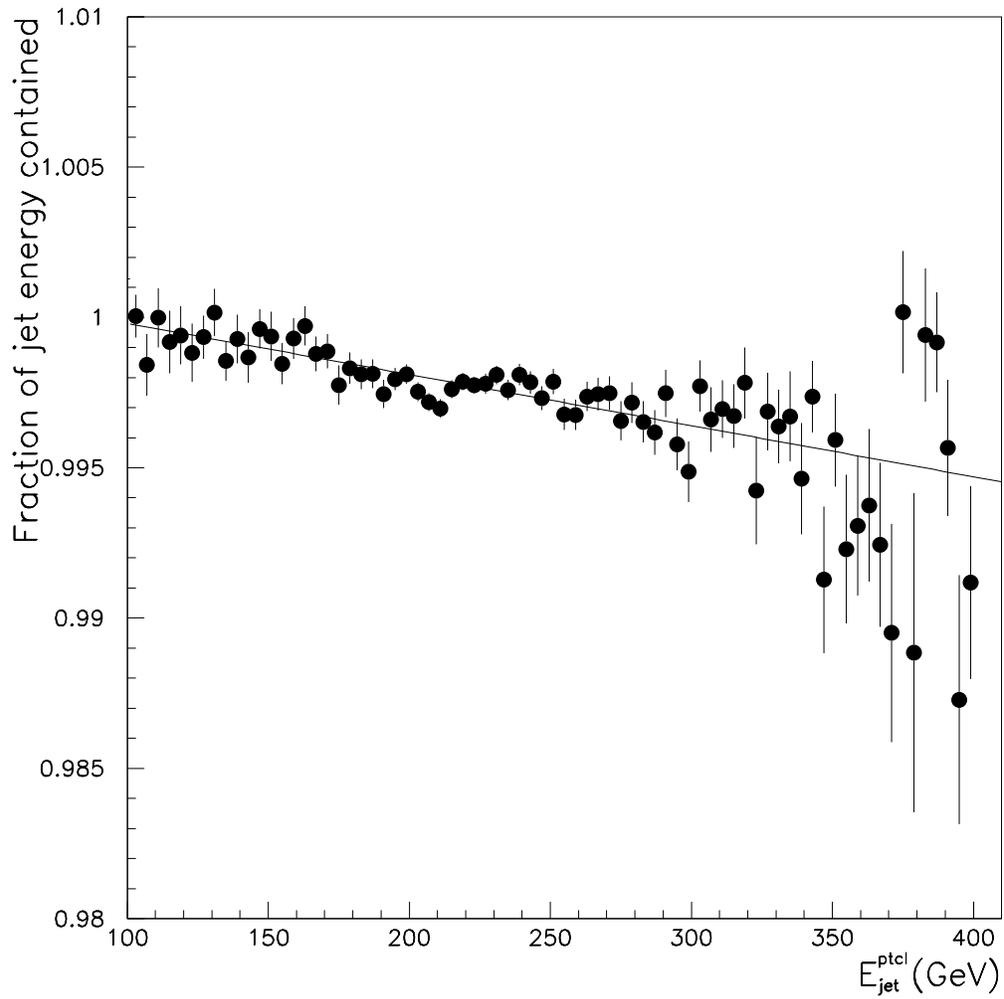,height=15cm,width=15cm}}
\caption
{Fraction of the jet energy contained within the D\O\ central
calorimeter ($|\eta|<0.7$) as a function of jet energy.
The data are normalized to unity at 100~GeV.}
\label{FIG:containment}
\end{figure}

The results of the simulation are shown in 
Fig.~\ref{FIG:containment}. The most conservative of the CC depth estimates
(7.2 $\lambda_{\mathrm{INT}}$ at $\eta$=0) is used.
The data have been normalized to 1.0 at 100~GeV in order to evaluate only the
energy loss with respect to 100~GeV jets.
The effect on the measured response due to
energy not contained by the CC is less than 0.5\% for
$E_{T}\lsim 400\;{\mathrm GeV}$.

The thickness of the calorimeter is known with a precision of 
$\sim 0.4 \lambda_{\mathrm{INT}}$. Different thicknesses between 7.2 and
$7.6 \lambda_{\mathrm{INT}}$ are, therefore, modeled. 
While the absolute energy contained varies between the different
models, the relative change as a function of jet energy is constant.
The simulation is also repeated
using a parameterization of energy
loss of charged pions as a function of $\lambda_{\mathrm{INT}}$, from
Ref.~\cite{wigmans}.
The results
are consistent with those obtained from the NuTeV parameterization.
These results determine the assignment of a 0.5\% uncertainty on the effect of 
shower containment on the CC response at high energies.

\subsection{Effects of Calorimeter Acceptance}

The event \met\ is measured from energy deposited in the individual
 calorimeter cells.
The sum is performed over all cells within the range 
$|\eta|<\eta_{\mathrm{lim}}$ with $\eta_{\mathrm{lim}}=4.5$ 
($\theta \gsim 1^{\circ}$).
To study the effects on $R_{\mathrm{jet}}$ of the large but finite calorimeter 
acceptance in the \met\ measurement, the 
\met\ is recalculated as a function of $\eta_{\mathrm{lim}}$ in a 
$\gamma$-jet Monte Carlo sample ({\sc herwig}-{\sc showerlib}).
From the measured $R_{\mathrm{jet}}$ versus $\eta_{\mathrm{lim}}$, it is 
possible to extract 
the effect of the limited acceptance
by extrapolation to $\eta_{\mathrm{lim}} \rightarrow \infty$. 

\begin{figure}[p]
\centerline{\psfig{figure=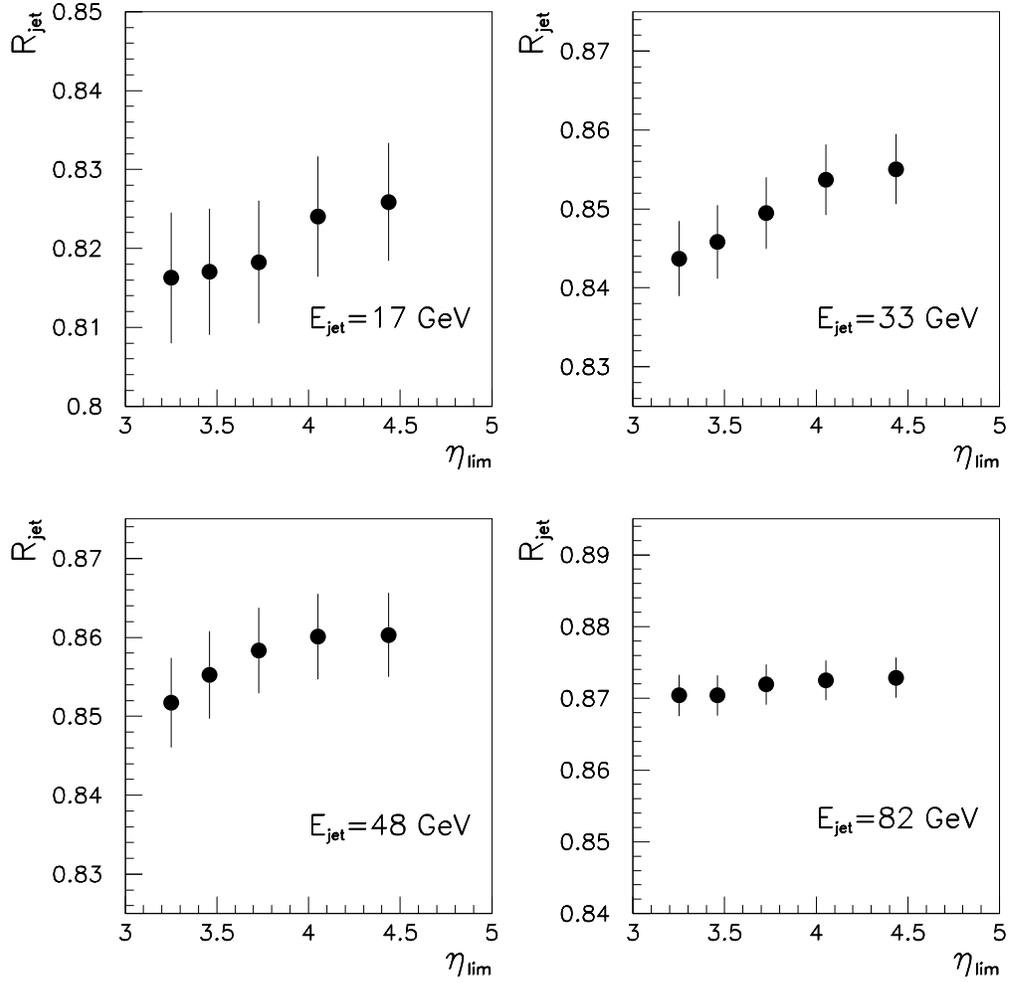,height=15cm,width=15cm}}
\caption{$R_{\mathrm{jet}}$ in the CC as a function of the 
$\eta_{\mathrm{lim}}$ used in the \met\ calculation.
The calorimeter acceptance bias is negligible
above $E_{\mathrm{jet}}\gsim 50\;{\mathrm GeV}$, given that 
$R_{\mathrm{jet}}$ is 
independent of the acceptance pseudorapidity threshold above 
$|\eta|=4$.}
\label{fig:kt1}
\end{figure}

The calorimeter acceptance bias is negligible
above $E_{\mathrm{jet}}\gsim 50\;{\mathrm GeV}$, given 
that $R_{\mathrm{jet}}$ is 
independent of 
$\eta_{\mathrm{lim}}$ above $|\eta|=4$ (Fig.~\ref{fig:kt1}).
Below this energy threshold, the bias is very small.
A 0.5\% error is assigned below $E_{T\mathrm{jet}}=15\;{\mathrm GeV}$, 
which decreases linearly to zero at $50\;{\mathrm GeV}$.

\subsection{Summary of the Systematic Errors in $R_{\mathrm{jet}}$}

In addition to the fit error, there is a 0.5\% physics background
error on $R_{\mathrm{jet}}$, as discussed in Section~\ref{sec:selec}. 
The calorimeter acceptance error
is 0.5\% below 15~GeV and decreases linearly to zero at 50~GeV.
At very low $E_{T}$'s, below $\sim 20\;{\mathrm GeV}$, the low-$E_{T}$ 
bias error
quickly becomes the dominant uncertainty. 
Above $E_{T}=100\;{\mathrm GeV}$ 
in the CC, the leakage uncertainty
increases linearly from zero to 0.5\% at 400~GeV.

There are two biases present in the response
measurement: the topology bias and the instrumental background bias.
Due to the topology bias, the measured response is 0.5\%-1\%
low; this effect is constant as a function of jet energy
(Fig. \ref{fig:dphi2}).
In addition, a constant bias of approximately 0.7\% towards a 
higher response is observed due to instrumental background.
It is difficult to accurately measure and correct for these
biases. The two effects, however, approximately cancel
each other, and therefore no correction is made. A
0.5\% error is included to account for any residual bias.
The leakage uncertainty is added linearly to this residual bias, before all
the jet energy scale error components are added in quadrature.

The $R_{\mathrm jet}$ correction is 
applied to $\sqrt{s}=630\; {\mathrm GeV}$ data
as measured in the $\sqrt{s}=1800\; {\mathrm GeV}$ $\gamma$-jet sample. 
While $R_{\mathrm jet}$ is expected to be the same at both center-of-mass 
energies, it is measured more accurately from the high statistics sample at 
$\sqrt{s}=1800\;{\mathrm GeV}$. Verification
studies~\cite{630}, based on independent measurements of 
$R_{\mathrm jet}$ in 
both data sets, show that the response is the same 
to within $0.3\%$. This number is quoted as an additional uncertainty 
for jets produced during the
low center-of-mass energy run. 

\section{Showering Correction}\label{sec:shower}

The last correction applied is the showering correction. It compensates
for the net energy flow through the cone boundary during calorimeter 
showering. Ideally, the jet energy correction should scale the energy
of the reconstructed jet to the particle level.
As the particles comprising the jet strike the
detector, they interact with the calorimeter material, 
producing a wide shower of additional particles. 
Some particles produced inside the cone
deposit a fraction of their energy outside the cone as the
shower develops, and vice versa. 

It is not possible to determine this showering effect directly
from calorimeter data. Energy outside the cone may be 
associated with gluon emission or fragmentation outside the cone at the 
particle level (``physics out-of-cone'').
Alternatively, such energy may be related to underlying
event, noise, or pile-up.
The correction is  derived using the energy per 
$\Delta \eta \times \Delta \varphi$ in
the vicinity of the jet centroid (energy density profile)
obtained from both data and particle level {\sc herwig} Monte Carlo.
The physics out-of-cone energy (from particle level Monte Carlo) 
is subtracted from the total out-of-cone energy (from data).

\subsection{Method}

The description that follows is based on ${\mathcal R}=0.7$ jets with
physics $|\eta|<0.4$.
Annuli or sub-cones of increasing sizes (steps of 0.1) are defined around 
the jet centroid.
$E_{\mathrm{jet}}(r)$ is calculated as
the sum over all cells contained within a sub-cone of radius $r$,
with $r<2$. Contributions from the
underlying event, uranium noise, and pile-up are subtracted.

The total out-of-cone ratio, $F_{\mathrm{tot}}$, is defined as 
the energy deposited inside a large cone 
taken as the limit of the cluster divided by the energy deposited inside the
algorithm cone. $F_{\mathrm{tot}}$ is measured as:

\begin{equation}
 F_{\mathrm{tot}} = \frac{E_{\mathrm{jet}}(r<1.0)}{E_{\mathrm{jet}}(r<0.7)}
  \, . \label{eq:shower1}
\end{equation}

The assumption, based on the measured energy per 
$\Delta \eta \times \Delta \varphi$ in
the vicinity of the jet centroid, is that the cluster does not extend 
beyond $r=1.0$ for central jets~\cite{note}. 
Showers produced by central and forward
jets inside the calorimeter cover approximately the same area in real space.
Pseudorapidity space, however, shrinks towards
the direction of the beam pipe. 
The cluster boundary, therefore, cannot be the same for all
pseudorapidity bins. For $|\eta|<0.4$, the limit is chosen at
$r=1.0$. This value is increased towards the beam pipe up to
$r=1.6$ for $2.5<|\eta|<3$.

The factor $F_{\mathrm{tot}}$ includes both showering loss and physics
out-of-cone. The latter, denoted as $F_{\mathrm{phy}}$, is obtained using
the same procedure as for $F_{\mathrm{tot}}$ from a {\sc herwig} particle level
jet sample reconstructed with the 0.7 cone algorithm.

The showering correction factor $F_{\mathrm{sho}}$ is
defined as:

\begin{equation}
 F_{\mathrm{sho}} = 
 \frac{E_{\mathrm{jet}}(r<0.7) + E_{\mathrm{sho}}}{E_{\mathrm{jet}}(r<0.7)} 
 \, , \label{eq:shower2}
\end{equation}

where $E_{\mathrm{sho}}$ is the amount of energy associated
with  particles emitted inside the cone at the particle level,
but deposited outside the cone in an annulus defined by
$0.7<r<1.0$ at the calorimeter level.
In the same way, $F_{\mathrm{tot}}$ and $F_{\mathrm{phy}}$ can be written as:

\begin{equation}
 F_{\mathrm{tot}} = \frac{E_{\mathrm{jet}}(r<0.7) +
             E_{\mathrm{phy}}(r>0.7)+ 
 E_{\mathrm{sho}}}{E_{\mathrm{jet}}(r<0.7)} \, , \label{eq:shower3}
\end{equation}

\begin{equation}
 F_{\mathrm{phy}} = \frac{E_{\mathrm{jet}}(r<0.7) +
             E_{\mathrm{phy}}(r>0.7)}{E_{\mathrm{jet}}(r<0.7)}  
 \, , \label{eq:shower4}
\end{equation}

where $E_{\mathrm{phy}}$ is the energy associated with the physics
out-of-cone. $F_{\mathrm{sho}}$ can be expressed in terms of 
$F_{\mathrm{tot}}$ and $F_{\mathrm{phy}}$ as:

\begin{equation}
 F_{\mathrm{sho}} = F_{\mathrm{tot}} - F_{\mathrm{phy}} +1 
 \, .\label{eq:shower5}
\end{equation}

In this paper, the fraction of the shower 
contained in the ${\mathcal R}$=0.7 cone, $S=1/F_{\mathrm{sho}}$,
is measured.

In the derivation of the showering correction described above, the offset 
is subtracted from the energy density profiles measured in data. 
After this subtraction, the energy density is still not zero outside the 
cluster limits. This contribution to the energy density profiles
is small and approximately a constant function of pseudorapidity.
It probably comes from particles produced far from the jets (in a dijet
events the two clusters are color connected) 
and constitutes one of the largest sources
of systematic uncertainty at low energies. This baseline is subtracted
from both the data and the particle level Monte Carlo, as the best
solution to avoid the bias in $S$ arising from differences 
in the baselines observed in both samples.

\subsection{Results}

\begin{figure}[p]
\centerline{\psfig{figure=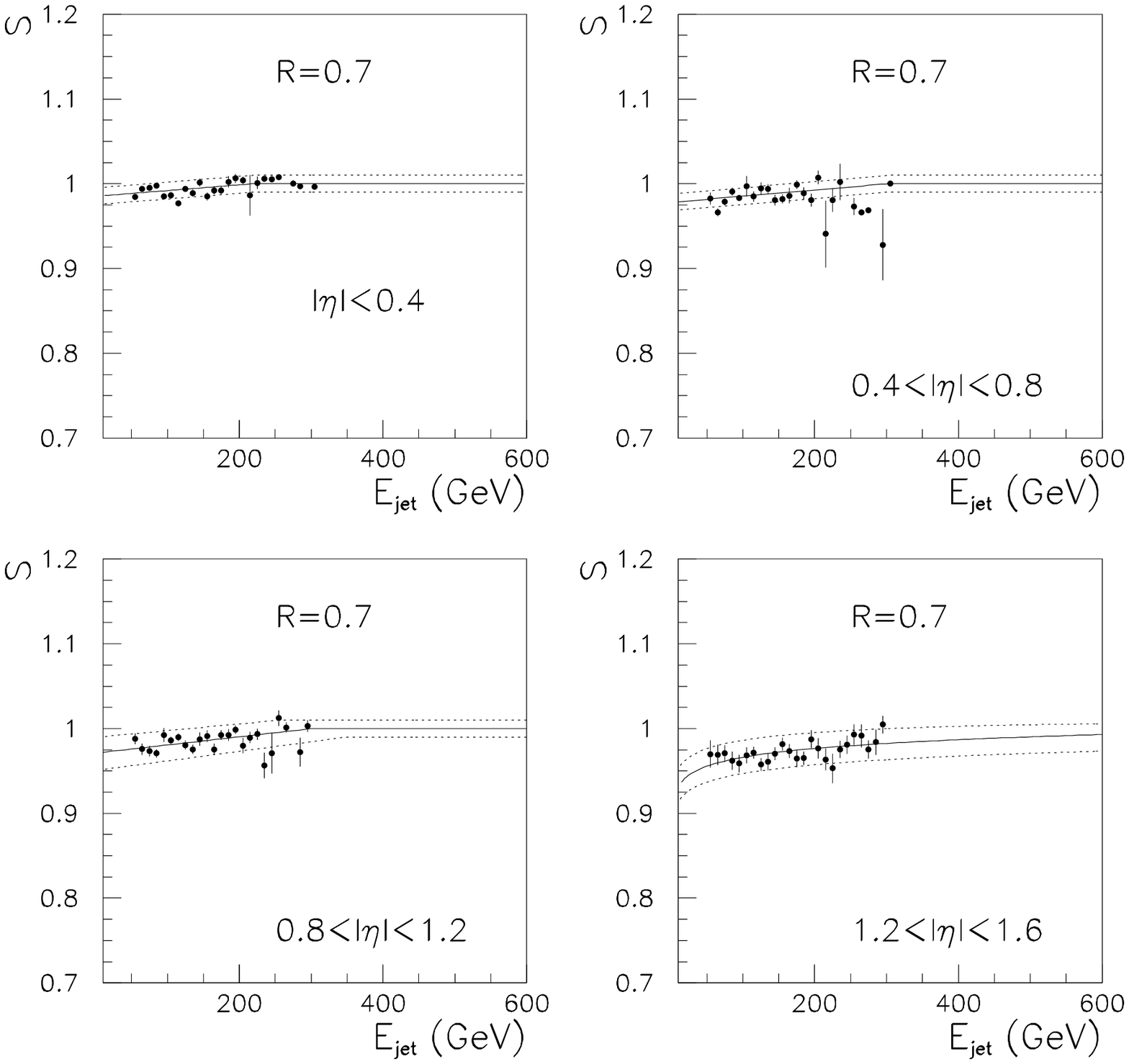,height=15cm,width=15cm}}
\caption
{$S$ versus $E_{\mathrm{jet}}$ (${\mathcal R}=0.7$). $E_{\mathrm{jet}}$ is
corrected by low-$E_{T}$ bias, offset, and response.
The solid line is a fit to the data. The dotted lines show the
systematic uncertainty.}
\label{fig:r07_shower_1}
\end{figure}

\begin{figure}[p]
\centerline{\psfig{figure=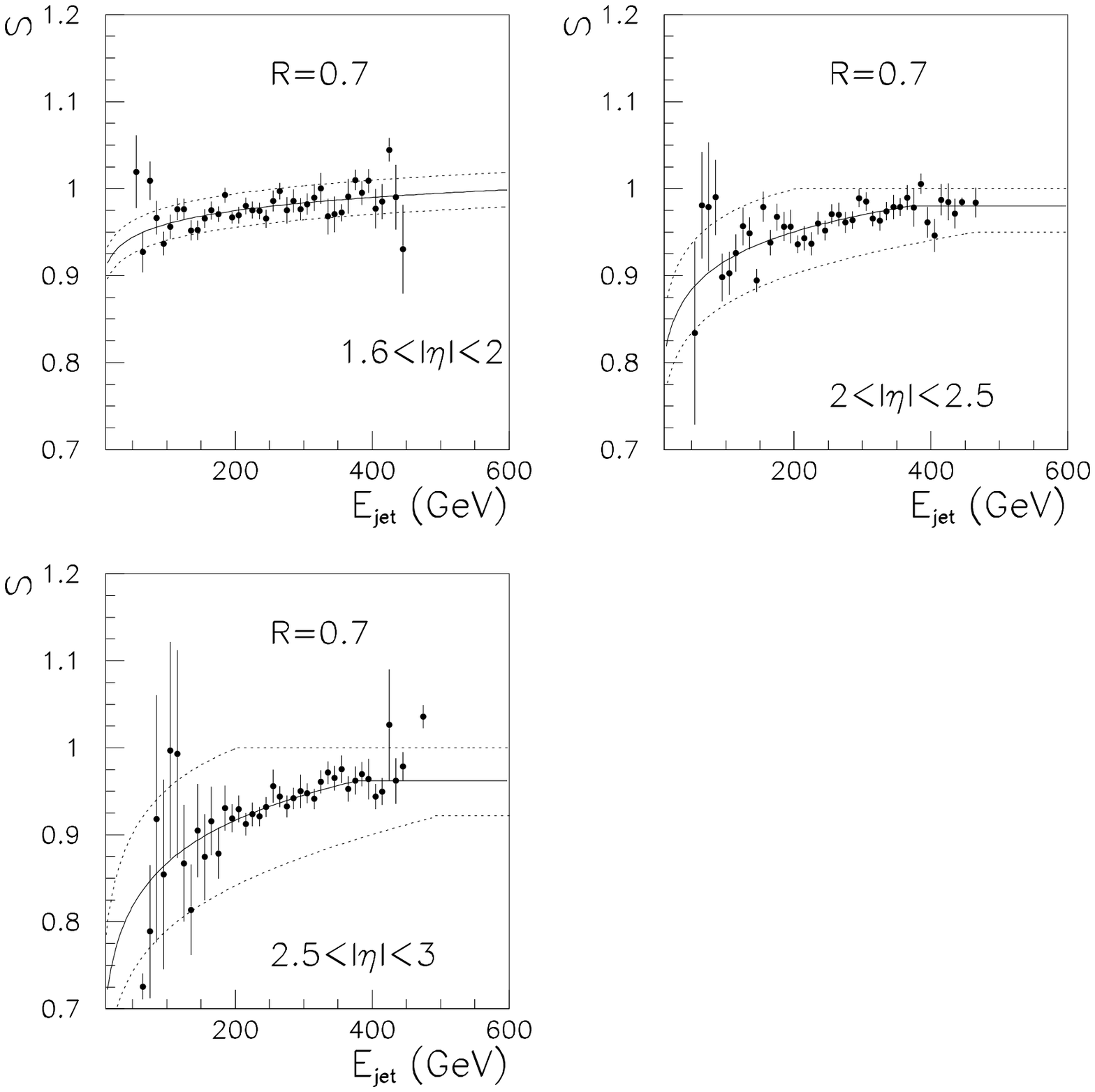,height=15cm,width=15cm}}
\caption
{$S$ versus $E_{\mathrm{jet}}$ (${\mathcal R}=0.7$). $E_{\mathrm{jet}}$ is
corrected by low-$E_{T}$ bias, offset and response.
The solid line is a fit to the data. The dotted lines show the
systematic uncertainty. }
\label{fig:r07_shower_2}
\end{figure}

\begin{figure}[p]
\centerline{\psfig{figure=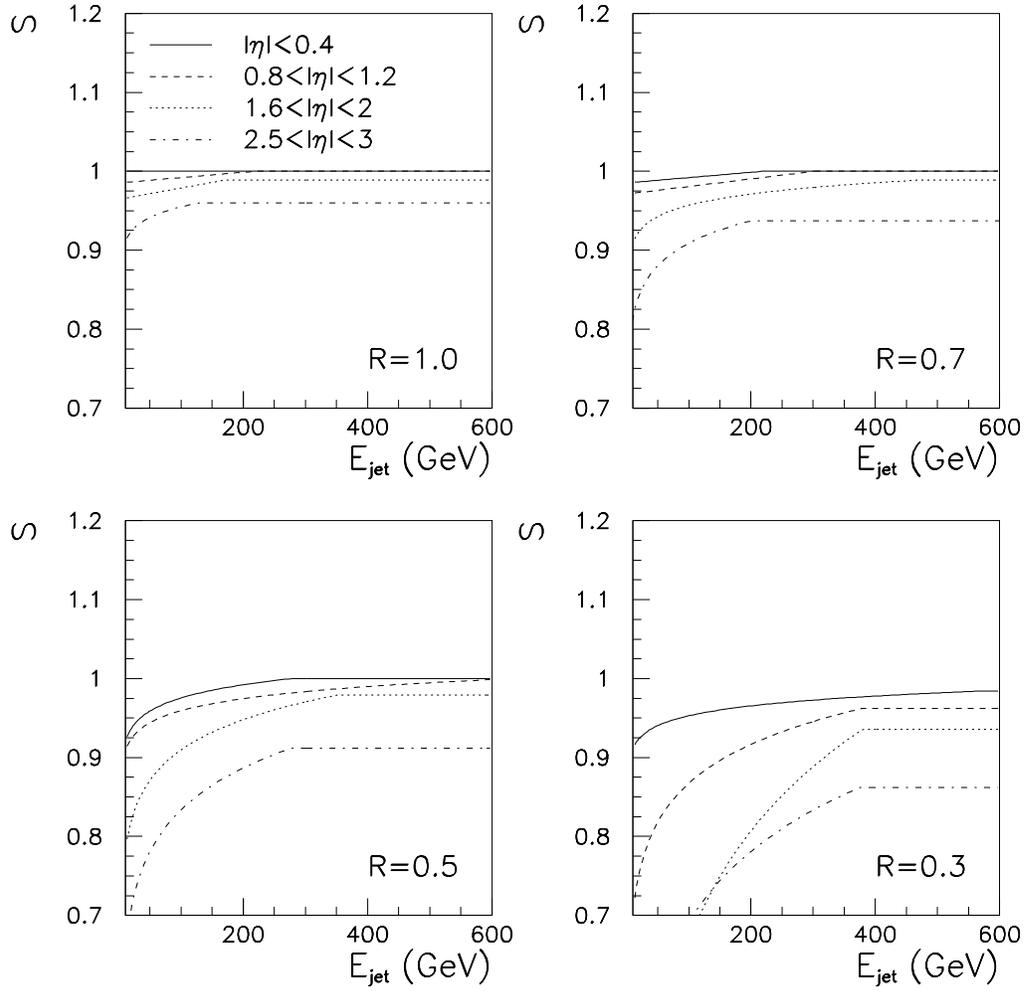,height=15cm,width=15cm}}
\caption
{$S$ versus $E_{\mathrm{jet}}$ for different cone sizes in different
$\eta$ regions. Only the parameterizations to the data points are 
shown.}
\label{fig:diff_cones_sho}
\end{figure}

Figures~\ref{fig:r07_shower_1}~and~\ref{fig:r07_shower_2} show $S$
versus $E_{\mathrm{jet}}$ for the 0.7 cone over all physics pseudorapidities. 
The solid curves are fits to $S$ versus $E_{\mathrm{jet}}$.
In the low energy range, the data 
are fit to either a logarithmic ($a + b \cdot \ln E_{\mathrm{jet}}$)
or a linear ($c + d \cdot E_{\mathrm{jet}}$) function. At high 
energies, $S$ is well described by a constant. 
The errors at low energy are dominated by the offset and
baseline subtraction, and at high energy by fit errors due to poor statistics. 
The bands between dotted lines account for the total uncertainty.
At high energies, 
the total uncertainty is 1\% (4\%) in the central (forward) 
regions for 0.7 cone jets. At very low energies, the errors
increase up to 1\% (10\%) in the central (forward) regions for
the same cone size. 
Figure~\ref{fig:diff_cones_sho} displays the
parameterizations of $S$ versus $E_{\mathrm{jet}}$ for four cone sizes 
${\mathcal R}=$1.0, 0.7, 0.5, 0.3 and different $\eta$ regions. At high
energies, the error for 0.3 cone jets is 2.5\% (5\%) in the central (forward)
region. The uncertainty increases up to 2.5\% (10\%) in the
central (forward) region at very low energies.
Note that although the errors increase rapidly for very low energy jets in the 
forward region, they are not as large in the $E_{T}$ range of
interest. This is important because the significant variable in physics 
analyses is not energy but $E_{T}$.

Showering losses depend on the jet energy profiles in $\eta$-$\varphi$
space. Different profiles would translate into different 
responses. Given that the $R_{\mathrm jet}$ measured from the
$\sqrt{s}=1800$ and $630\;{\mathrm GeV}$ samples are consistent, 
showering losses do not depend on the \PBARP\ 
center-of-mass energy for jets of the same energy and pseudorapidity.

\section{Monte Carlo Studies}\label{sec:mc}

The Monte Carlo analysis described in this section
serves two purposes. First, it provides the jet energy scale
to correct Monte Carlo jets processed through
the \D0 shower library simulation ({\sc showerlib}). 
Second, it proves that the method used
to derive the correction achieves its purpose within errors.
The analysis is based primarily on
a sample of {\sc herwig} direct photon events processed through
{\sc showerlib} for simulation of particle showers.

\subsection{Jet Energy Scale}

\begin{figure}[p]
\centerline{\psfig{figure=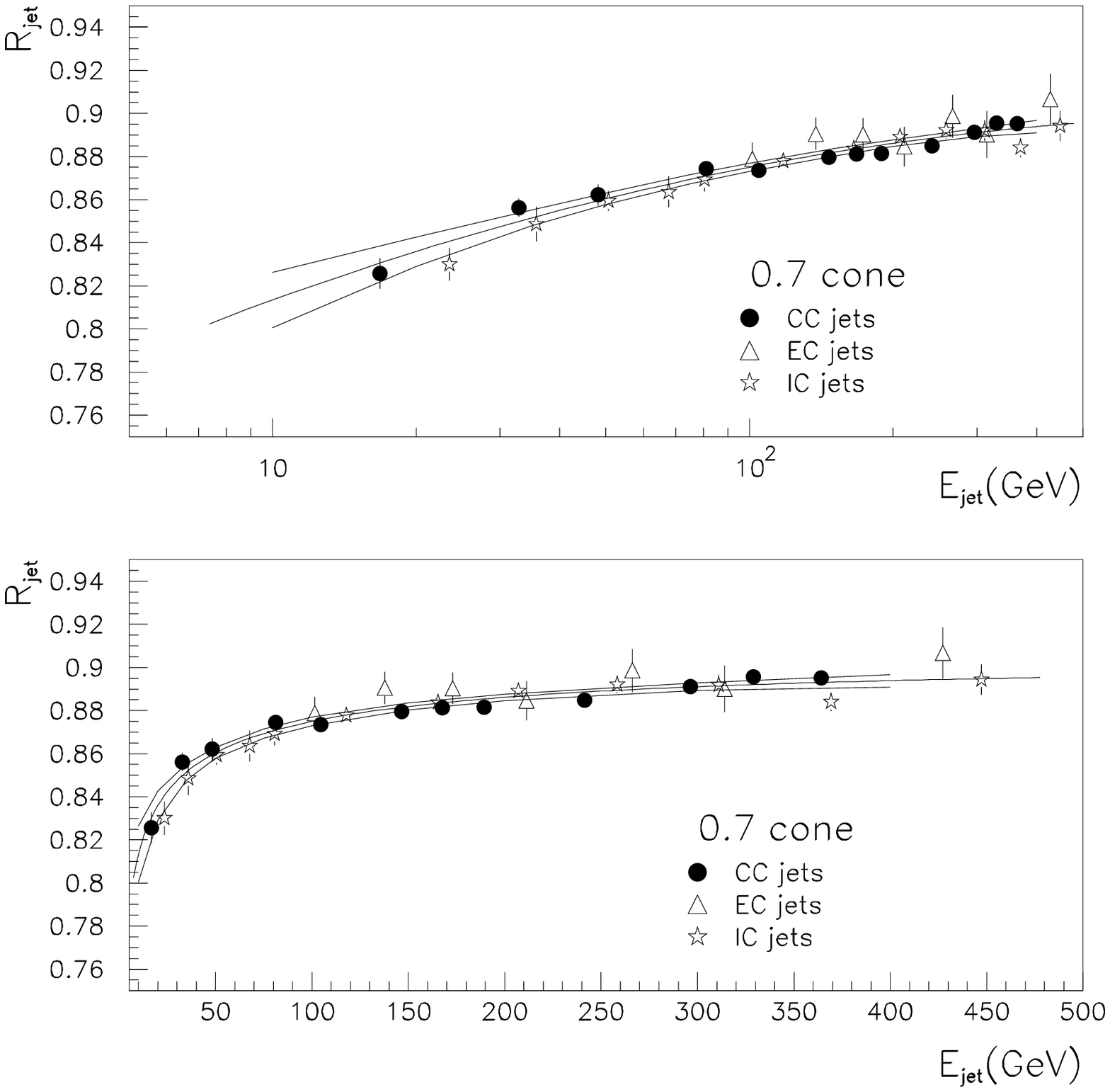,%
height=15cm,width=15cm}}
\caption
{$R_{\mathrm{jet}}$ values for ${\mathcal R}=0.7$ Monte Carlo jets 
(full dots are CC jets, triangles EC jets and
stars IC jets) along with the result of the nominal fit
and the error band ({\sc herwig}-{\sc showerlib}).}
\label{fig:response_07_banderrors}
\end{figure}

After application of
the low-$E_{T}$ bias, offset, cryostat factor, and $\eta$-dependent 
corrections,
the Monte Carlo response in the CC is determined 
using the same procedure as for the data.
Figure~\ref{fig:response_07_banderrors} shows $R_{\mathrm{jet}}$ versus 
$E_{\mathrm{jet}}$
for jets reconstructed with the 0.7 cone algorithm, along with the
associated error band. The response is now
uniform over the  whole detector.
As mentioned in Section~\ref{sec:enedep}, 
the shape of the response obtained from a {\sc herwig} sample
processed through {\sc showerlib} is different from the response
measured from data. 
In addition to a difference in overall normalization,
the Monte Carlo response increases less rapidly, remaining 
nearly constant above 150~GeV. 

Some sources of error in the data analysis do not contribute to the 
Monte Carlo jet scale uncertainty. For example,
uranium noise, pile-up, and multiple \PBARP\ interactions are not modeled in 
the Monte Carlo sample. In addition, physics backgrounds in Monte Carlo
are limited to diphoton events, which produce a negligible effect on 
the response measurement.
Luminosity and multiple interaction cuts are not needed, because
only single interactions are generated.
There is also no instrumental background in the Monte Carlo.
The topology bias is on the order of 1\% at 30~GeV and becomes negligible 
above $\sim 200\;{\mathrm GeV}$.
Limits in the calorimeter acceptance contribute a small error to 
$R_{\mathrm{jet}}$.
Finally, because the full jet shower is contained within the
calorimeter in the {\sc showerlib} approximation, the shower 
containment error is negligible.

\subsection{Closure Tests}

The Monte Carlo sample provides an opportunity to verify the method used to 
derive the jet energy scale correction. This ``closure''
test directly compares the corrected jet energy with the energy of
the associated particle jet. The ratio of these two quantities should
be unity for all values of $E_{\mathrm{jet}}^{\mathrm{ptcl}}$ and 
pseudorapidity.
For the closure test, the underlying event contribution is
subtracted from both calorimeter and particle jets. Split and 
merged
jets are removed from the sample, because splitting and merging is not 
implemented in the particle level algorithms.

\begin{figure}[p]
\centerline{\psfig{figure=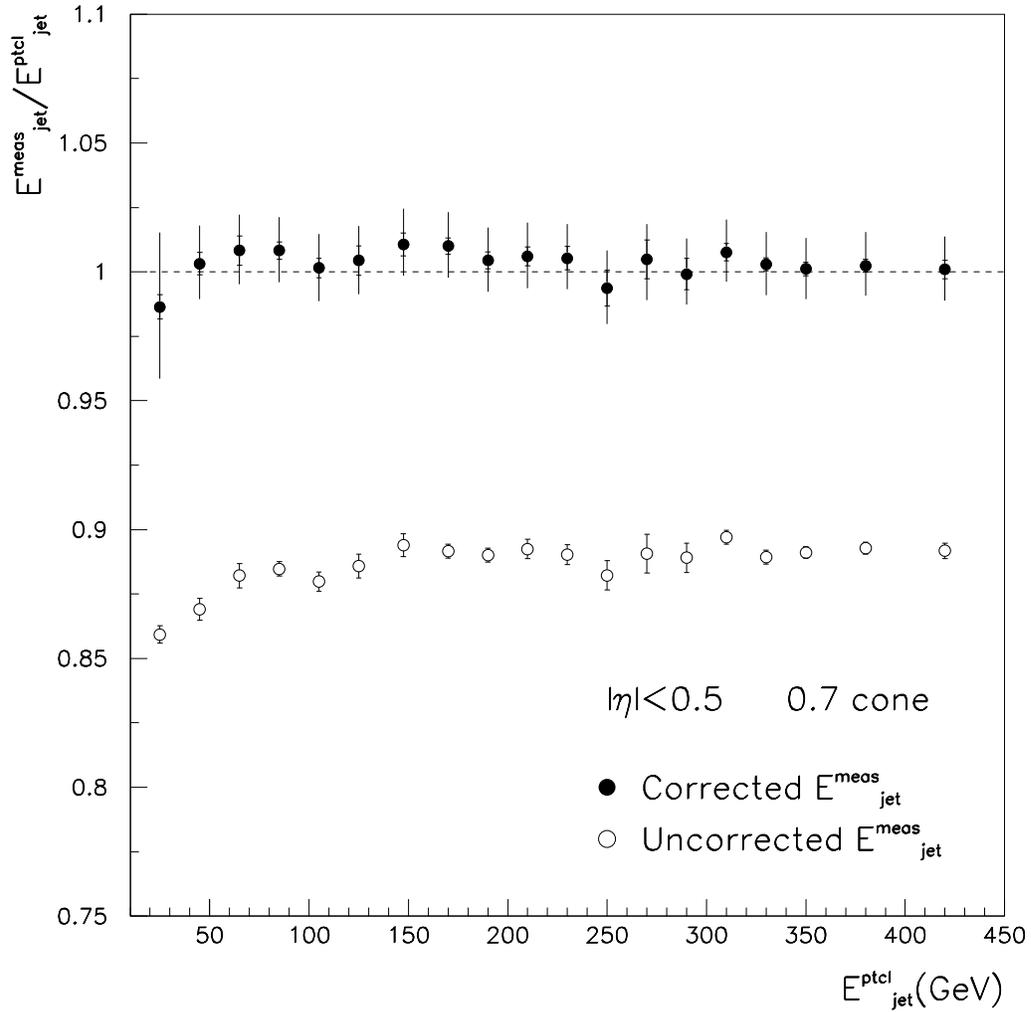,height=15cm,width=15cm}}
\caption{Ratio of calorimeter and
particle jet energy before (open circles) and after (full circles)
the jet scale correction is applied ({\sc herwig}-{\sc showerlib}). The ratio
is plotted as a function of $E_{\mathrm{jet}}^{\mathrm{ptcl}}$ for 
${\mathcal R}=0.7$ Monte Carlo jets.
Inner bars are statistical errors, and outer bars
contain statistical and systematic errors added in quadrature.}
\label{fig:clo_1}
\end{figure}

\begin{figure}[p]
\centerline{\psfig{figure=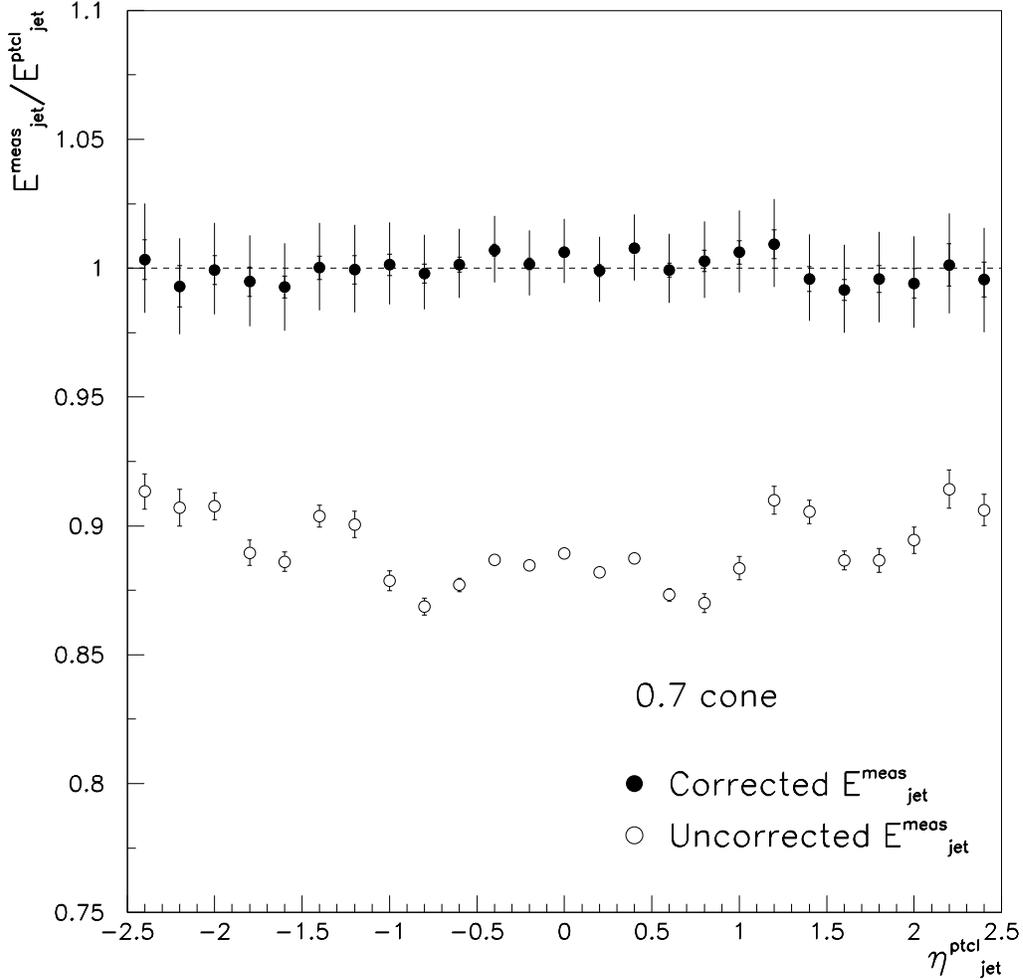,height=15cm%
,width=15cm}}
\caption
{Ratio of calorimeter and
particle jet energy before (open circles) and after (full circles)
the jet scale correction is applied ({\sc herwig}-{\sc showerlib}). The ratio
is plotted as a function of particle level $\eta$ for 
${\mathcal R}=0.7$ Monte Carlo jets.
A $E_{T}>15\;{\mathrm GeV}$ cut is applied to
remove the region where showering losses and low-$E_{T}$ bias 
effects dominate.
Inner bars are statistical errors, and outer bars
are statistical and systematic errors added in quadrature.}
\label{fig:clo_2}
\end{figure}

Figures~\ref{fig:clo_1}~and~\ref{fig:clo_2} show the ratio of 
calorimeter to
particle jet energy before (open circles) and after (full circles)
the jet scale correction, as a function
of $E_{\mathrm{jet}}^{\mathrm{ptcl}}$ and particle level 
$\eta$. Statistical and
systematic uncertainties are also shown.
Within errors, the ratio is consistent with unity after all corrections.
Thus, on average, the correction scales the energy of a 0.7 calorimeter jet
to the energy of the associated particle jet to
within $\sim$0.5\% for $|\eta|<0.5$.

\begin{figure}[p]
\centerline{\psfig{figure=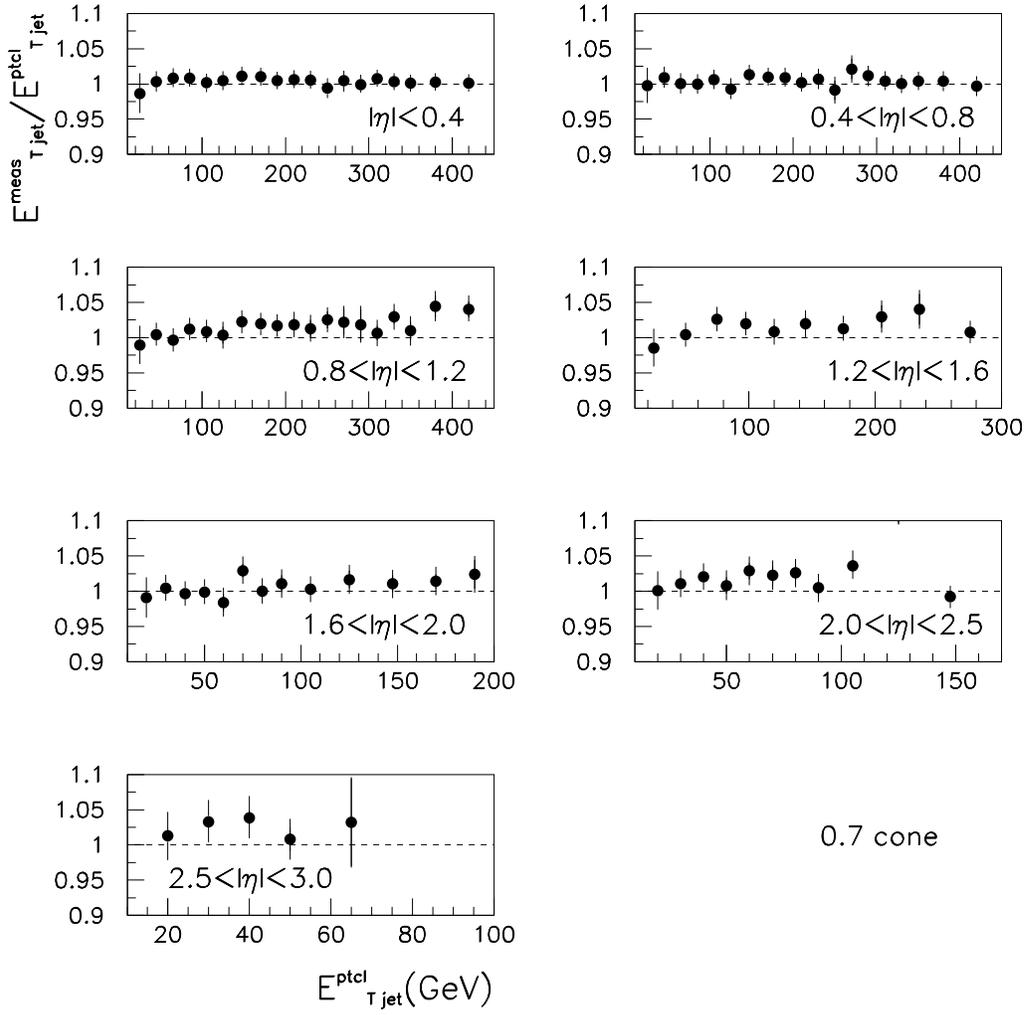,%
height=15cm,width=15cm}}
\caption
{Ratio of calorimeter and
particle jet energy after the jet scale correction is applied 
({\sc herwig}-{\sc showerlib}).
The ratio is plotted as a function of $E_{T\mathrm{jet}}^{\mathrm{ptcl}}$ for
${\mathcal R}=0.7$ Monte Carlo jets.
The bars are statistical and systematic errors added in quadrature.}
\label{fig:cloe1}
\end{figure}

The closure test is also performed for other cones and pseudorapidity
bins. Figure~\ref{fig:cloe1} shows the
$E^{\mathrm{meas}}_{\mathrm{jet}}/E^{\mathrm{ptcl}}_{\mathrm{jet}}$
ratio versus $E_{T\mathrm{jet}}$ calculated for the 0.7 cone jets in
seven $\eta$ bins within the range $|\eta|<3$. All ratios are consistent with
unity within the total systematic uncertainties.

\section{Summary and Conclusions}\label{sec:conc}

The energy calibration was performed for jets observed in \PBARP\ collisions
with the \D0 detector at Fermilab. 
The work described in this paper is based primarily on data taken by \D0 
during the 1992-1996 \PBARP\ collider run. Test beam
data and Monte Carlo samples are also used in some cases.
The corrections were derived for two center-of-mass energies of the \PBARP\
system, $\sqrt{s}=1800 \; {\mathrm GeV}$ and 
$\sqrt{s}=630 \; {\mathrm GeV}$, and
are valid in the range 
$E_{T{\mathrm jet}}>8 \; {\mathrm GeV}$ and $|\eta|<3$.
The jet energy scale compensates for spectator 
interactions, uranium noise, response, and showering loss.
The response correction is small, due to the hermeticity, linearity, and
good $e/\pi$ ratio of the \D0 calorimeters.

\begin{figure}[p]
\centerline{\psfig{figure=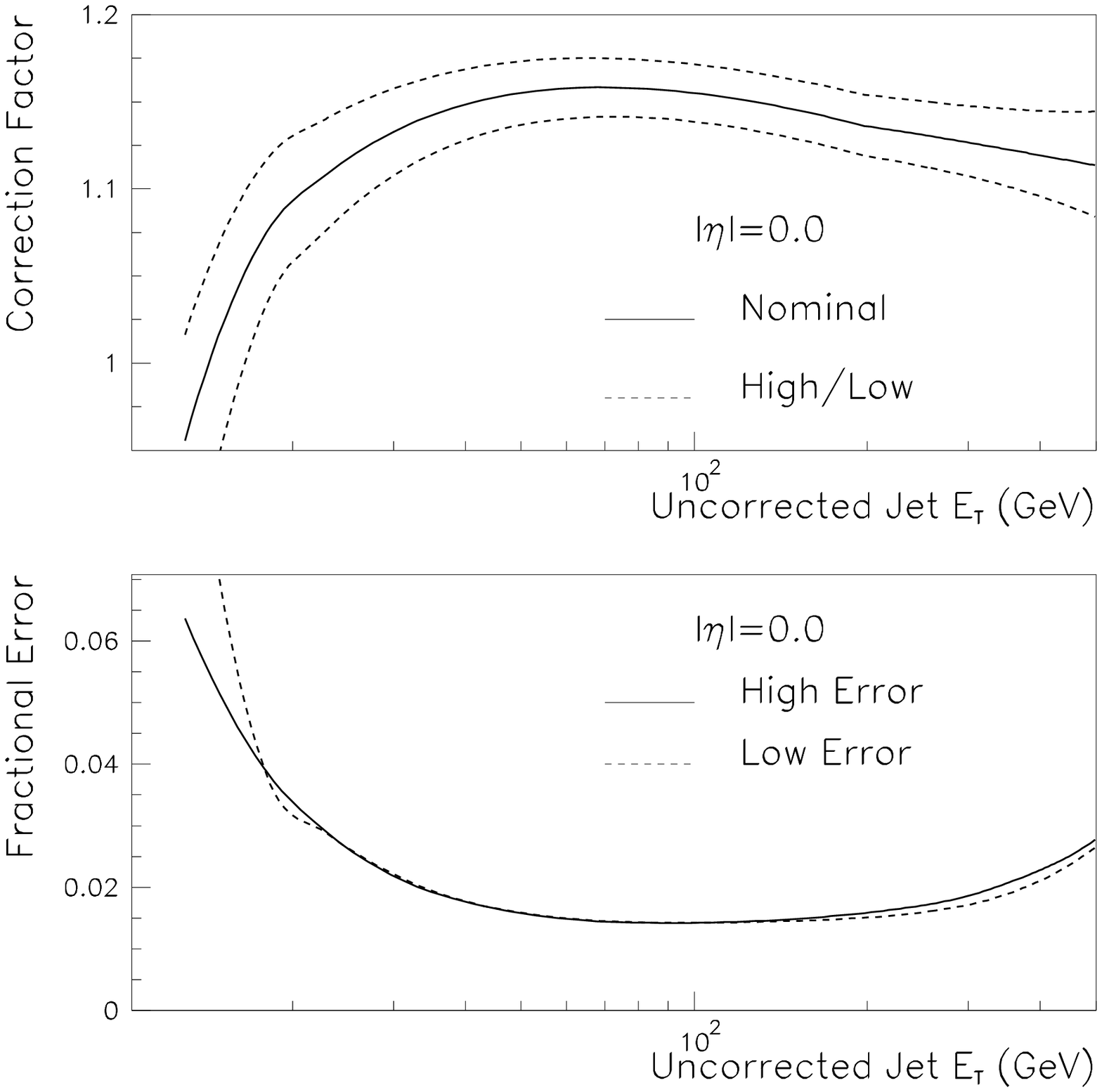,%
height=15cm,width=15cm}}
\caption
{Corrections and errors for $\eta=0$, ${\mathcal R}=0.7$.
Top: Nominal, high (nominal $+ \, \sigma$), and low (nominal $- \, \sigma$)
correction factors.  Bottom: high and low
fractional errors.}
\label{fig:mpf_note_err_tot0}
\end{figure}

\begin{figure}[p]
\centerline{\psfig{figure=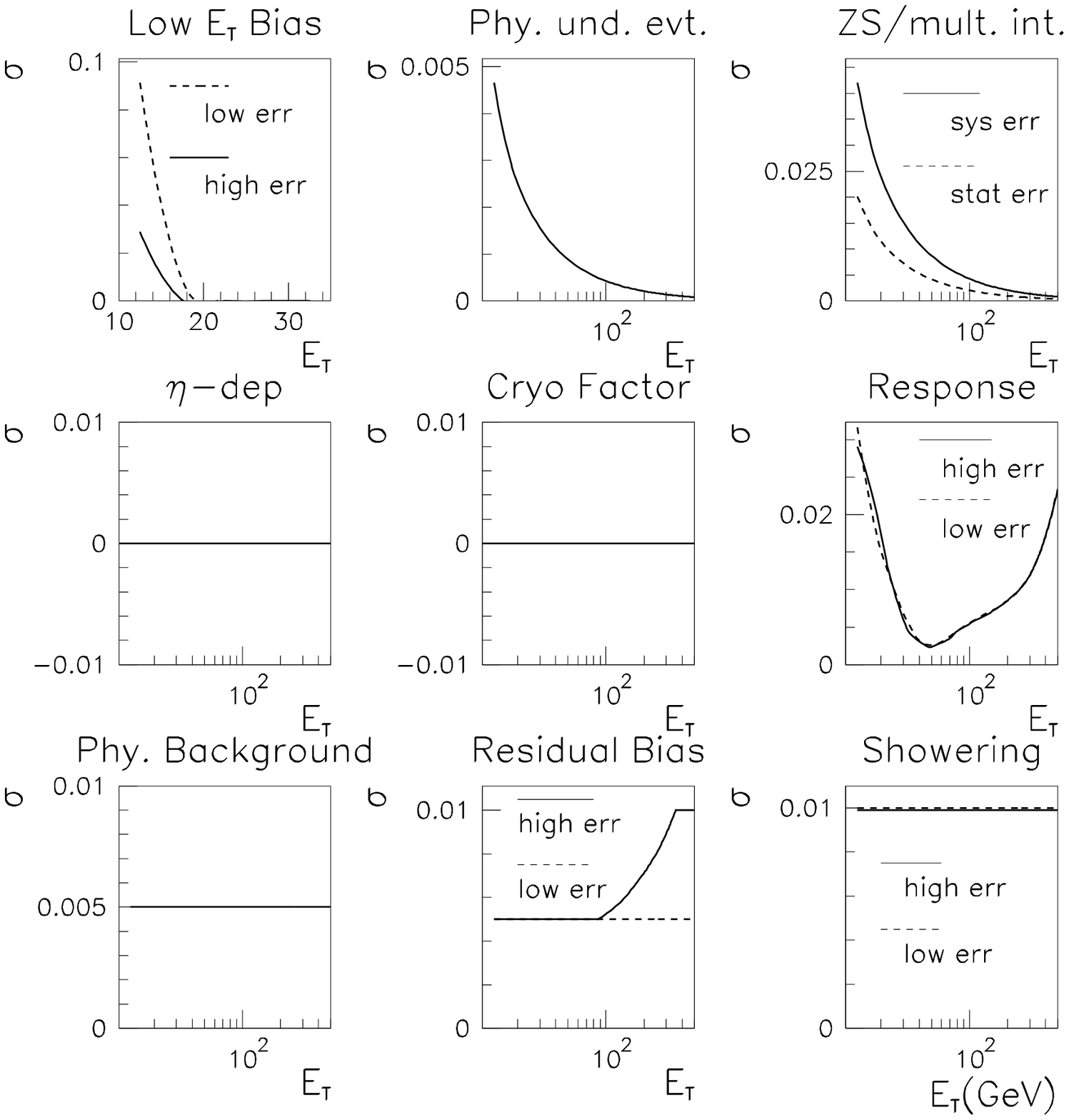,%
height=15cm,width=15cm}}
\caption{Individual components of the fractional jet scale error 
versus uncorrected jet $E_{T}$ for $\eta=0$ (${\mathcal R}=0.7$): 
physics underlying event, uranium noise (zero suppression), and multiple 
interactions (Section~\ref{sec:off});
cryostat factor and
$\eta$ dependence in the IC region (Section~\ref{sec:etadep}); 
low-$E_{T}$ bias, response fit, 
physics background, and residual bias from event topology, instrumental 
background and shower containment (Section~\ref{sec:enedep}); showering 
(Section~\ref{sec:shower}).
Some components are
zero because they do not contribute in that particular $\eta$ region.}
\label{fig:mpf_note_err_sum0}
\end{figure}

\begin{figure}[p]
\centerline{\psfig{figure=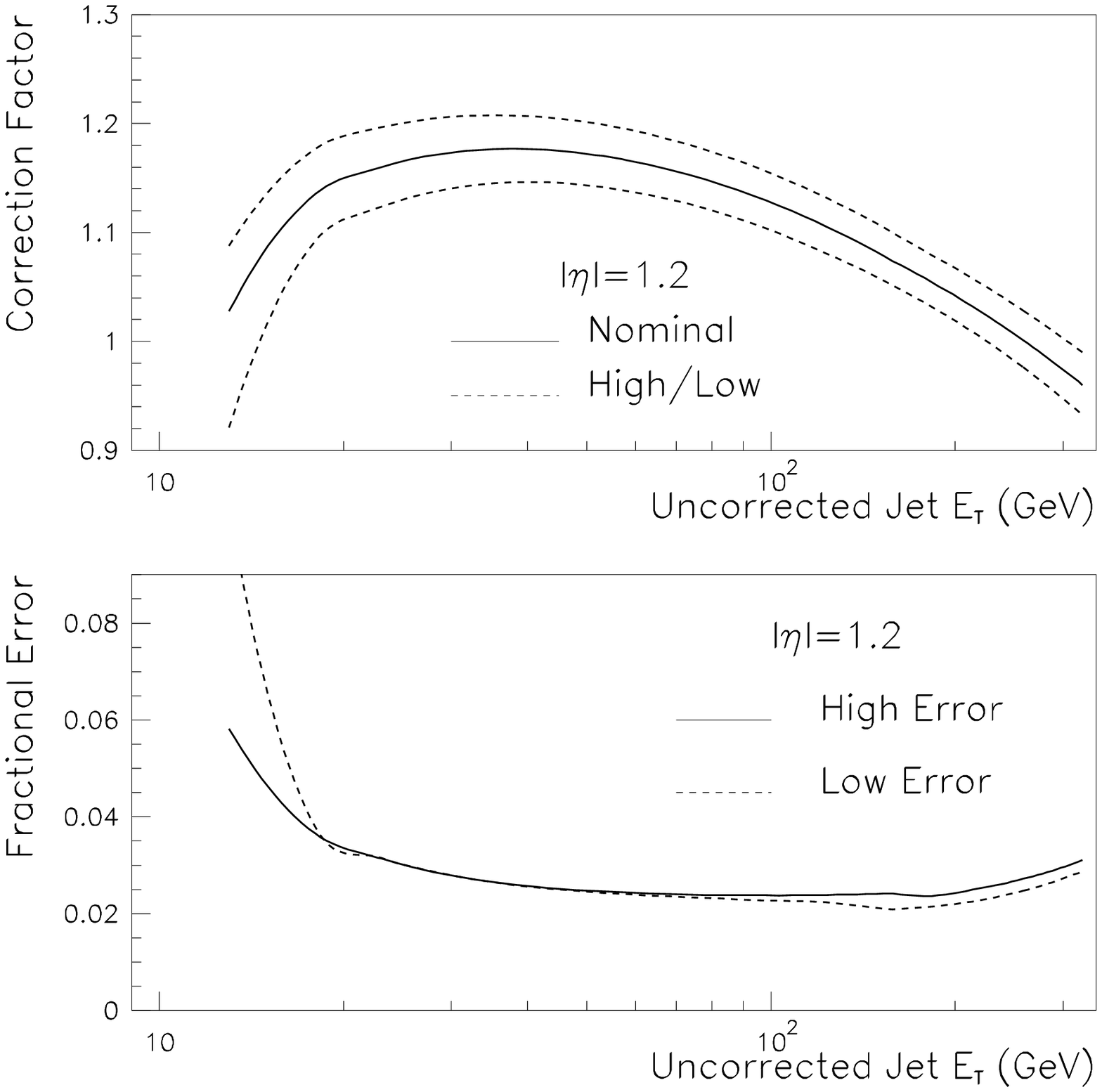,%
height=15cm,width=15cm}}
\caption
{Total Corrections and errors for $|\eta|=1.2$ (${\mathcal R}=0.7$).
Top: Nominal, high (nominal $+ \, \sigma$), and low (nominal $- \, \sigma$)
correction factors. Bottom: high and low fractional errors.}
\label{fig:mpf_note_err_tot12}
\end{figure}

\begin{figure}[p]
\centerline{\psfig{figure=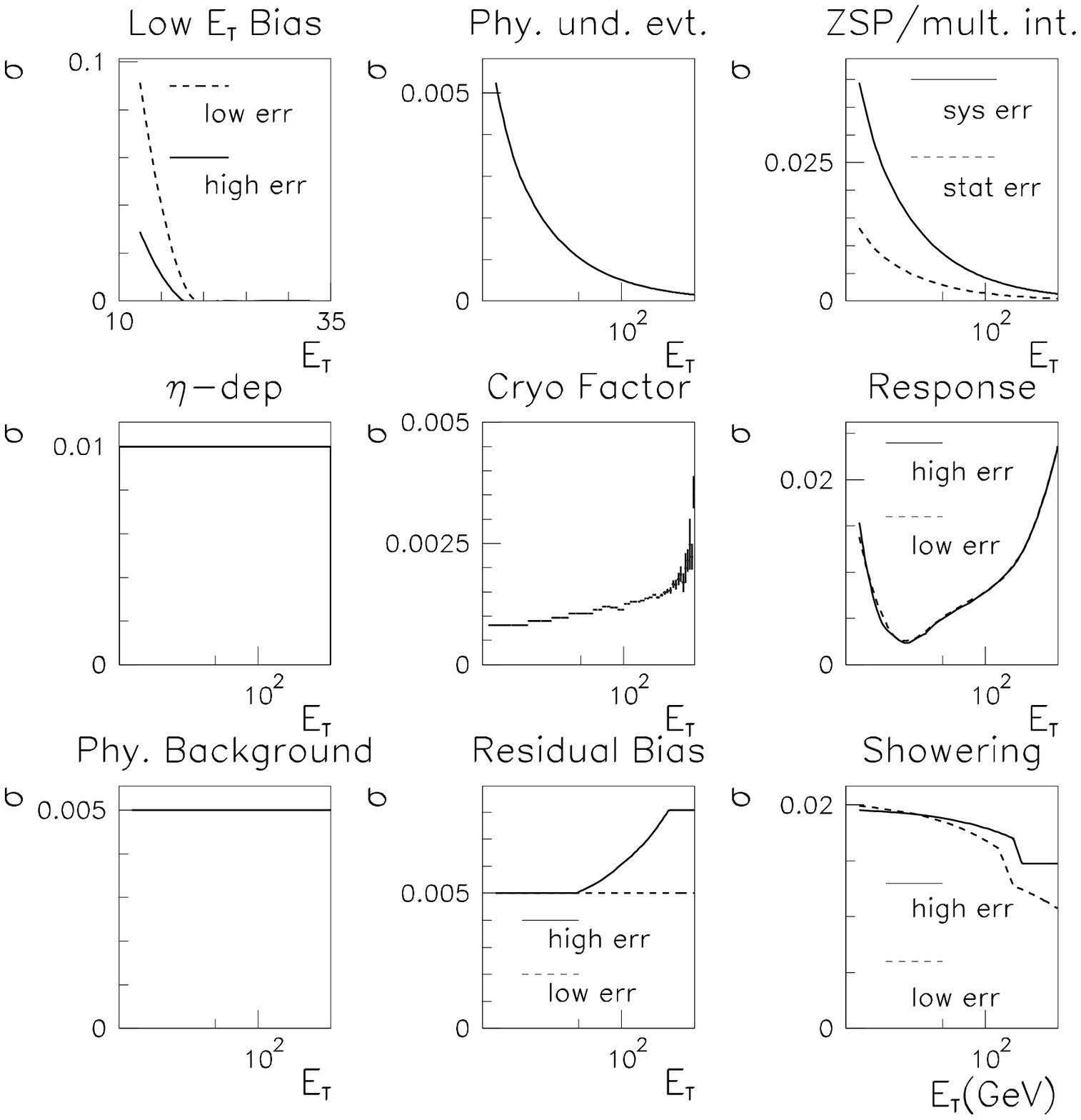,%
height=15cm,width=15cm}}
\caption
{Individual components of the fractional jet scale error 
versus uncorrected jet $E_{T}$ for $\eta=1.2$ (${\mathcal R}=0.7$): 
physics underlying event, uranium noise (zero suppression), and 
multiple interactions (Section~\ref{sec:off}); cryostat factor and
$\eta$ dependence in the IC region (Section~\ref{sec:etadep}); 
low-$E_{T}$ bias, response fit, 
physics background, and residual bias from event topology, instrumental 
background and shower containment (Section~\ref{sec:enedep}); showering 
(Section~\ref{sec:shower}).}
\label{fig:mpf_note_err_sum12}
\end{figure}

\begin{figure}[p]
\centerline{\psfig{figure=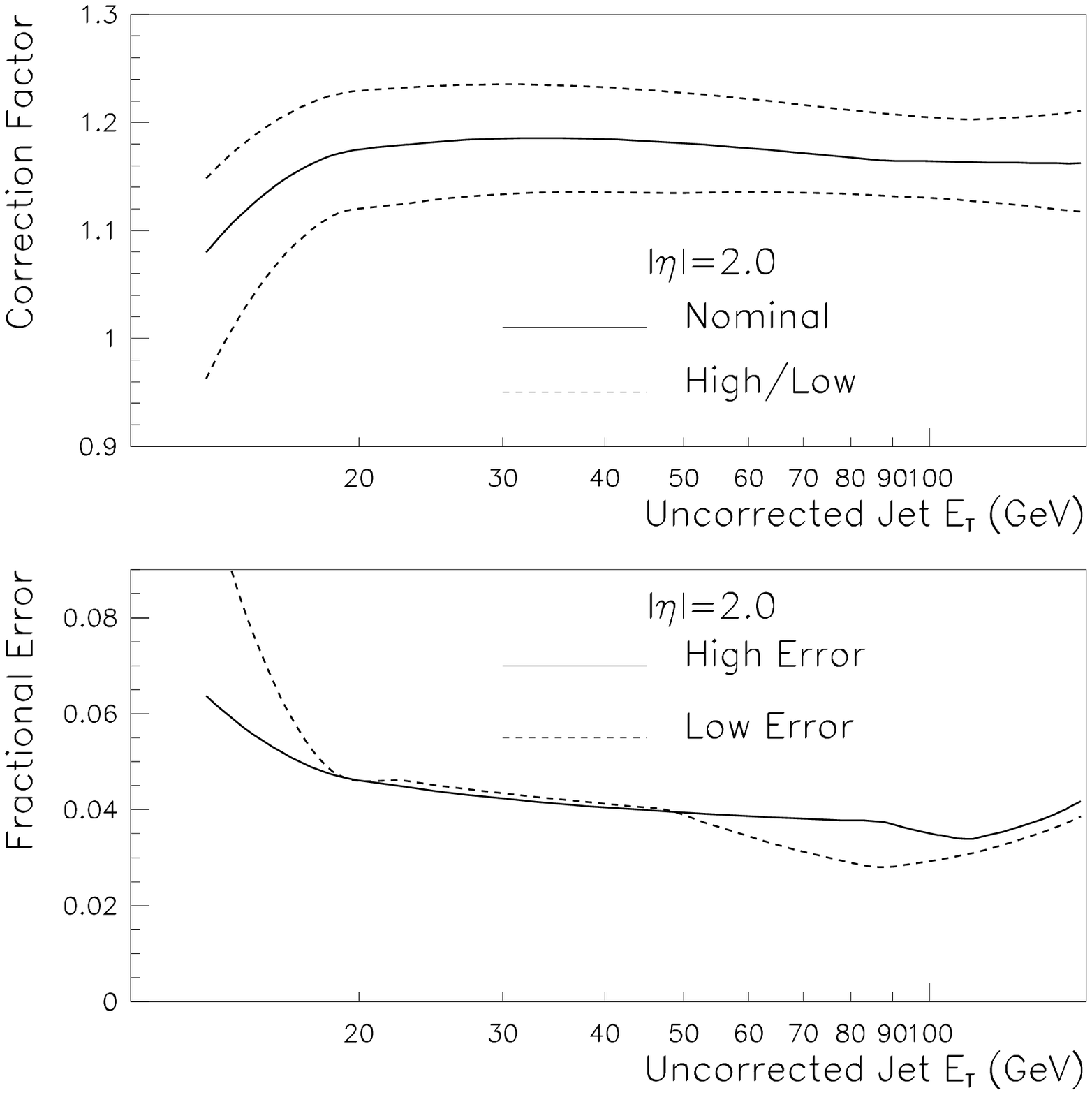,%
height=15cm,width=15cm}}
\caption
{Corrections and errors for $\eta=2$ (${\mathcal R}=0.7$).
Top: Nominal, high (nominal $+ \, \sigma$), and low (nominal $- \, \sigma$)
correction factors. Bottom: high and low fractional errors.}
\label{fig:mpf_note_err_tot2}
\end{figure}

\begin{figure}[p]
\centerline{\psfig{figure=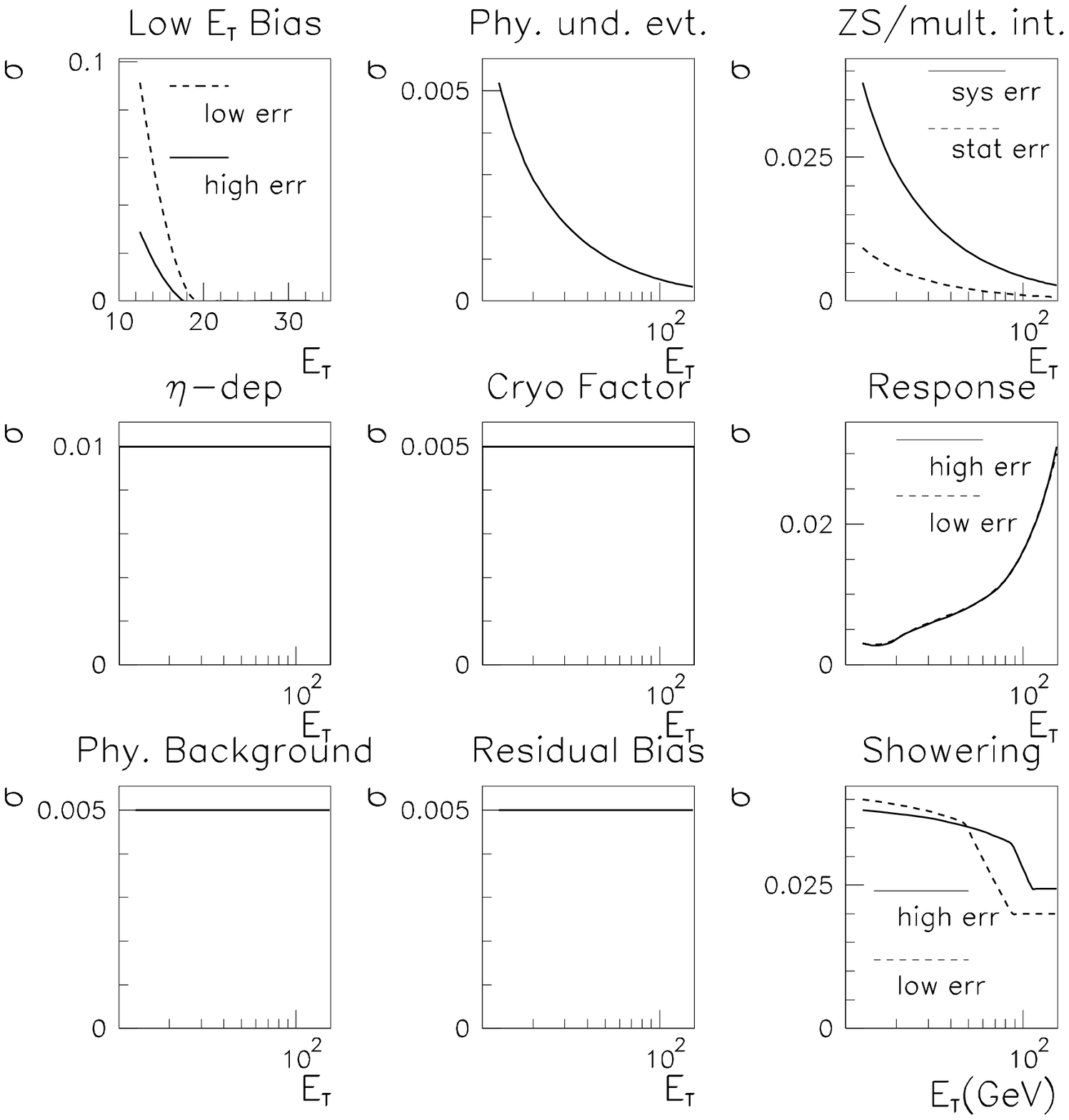,%
height=15cm,width=15cm}}
\caption
{Individual components of the fractional jet scale error 
versus uncorrected jet $E_{T}$ for $\eta=2$ (${\mathcal R}=0.7$): 
physics underlying event, uranium noise (zero suppression), and 
multiple interactions (Section~\ref{sec:off}); cryostat factor and
$\eta$ dependence in the IC region (Section~\ref{sec:etadep}); 
low-$E_{T}$ bias, response fit, 
physics background, and residual bias from event topology, instrumental 
background and shower containment (Section~\ref{sec:enedep}); showering 
(Section~\ref{sec:shower}).}
\label{fig:mpf_note_err_sum2}
\end{figure}

\begin{figure}[p]
\centerline{\psfig{figure=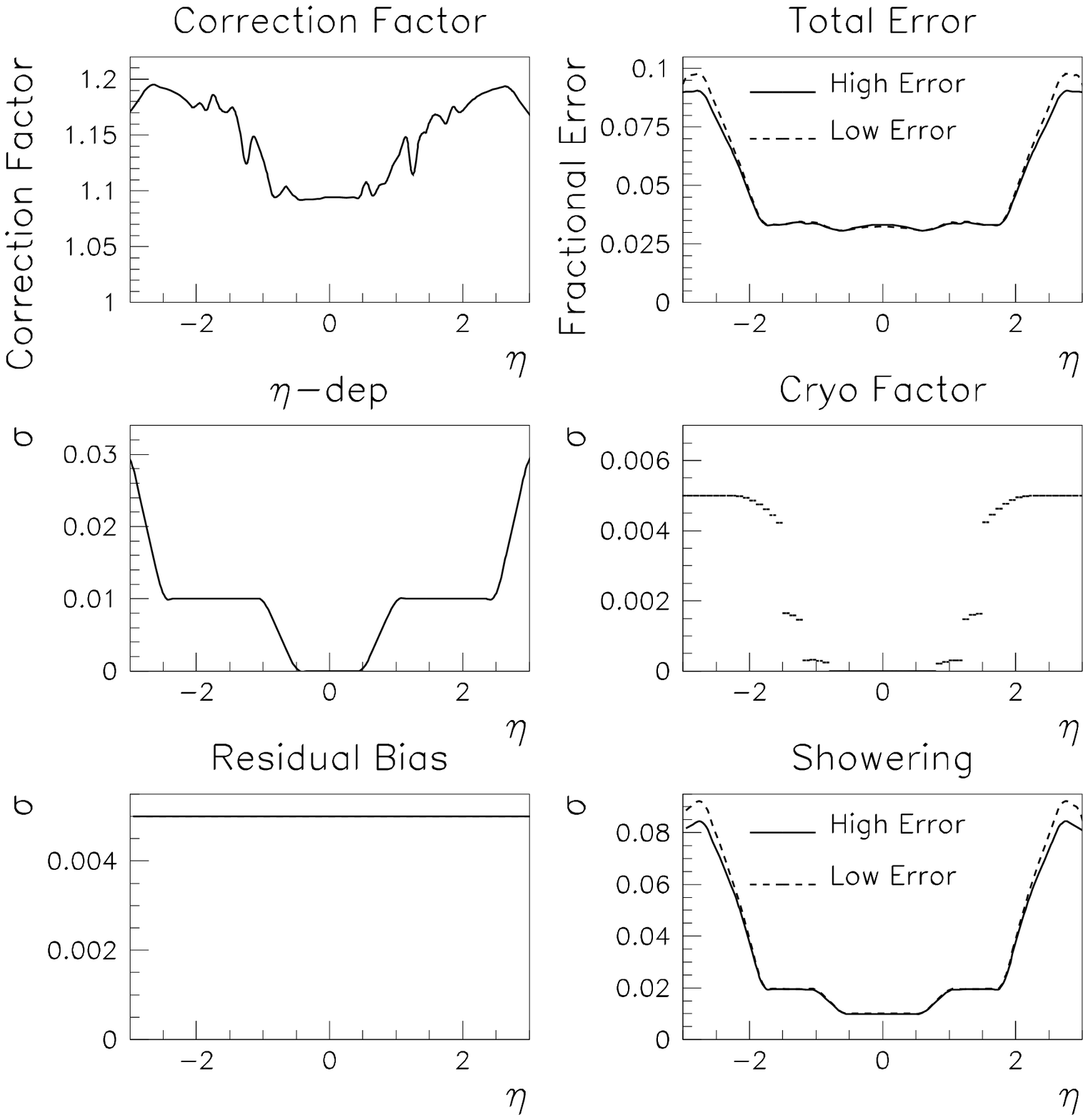,%
height=15cm,width=15cm}}
\caption
{Corrections and fractional errors versus $\eta$ (jet 
$E_{T}=20\;{\mathrm GeV}$ and ${\mathcal R}=0.7$).  
The total correction and error are both shown as well as the
$\eta$ dependence of several individual components of the jet scale error:
cryostat factor and $\eta$ dependence in the IC region 
(Section~\ref{sec:etadep}); 
residual bias from event topology, instrumental 
background and shower containment (Section~\ref{sec:enedep}); showering 
(Section~\ref{sec:shower}).}
\label{fig:mpf_note_err_eta20}
\end{figure}

\begin{figure}[p]
\centerline{\psfig{figure=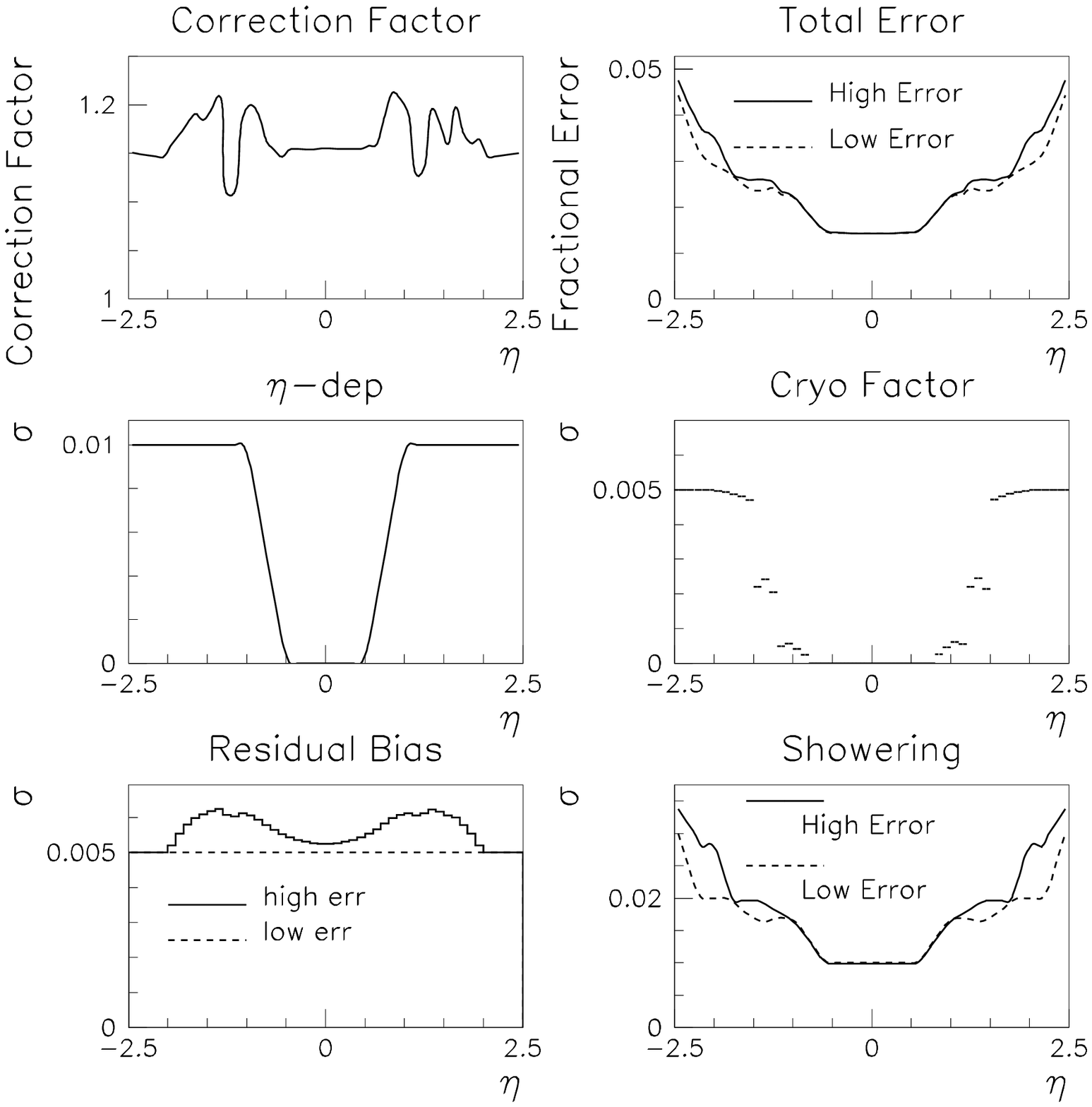,%
height=15cm,width=15cm}}
\caption{Corrections and errors versus $\eta$ (jet $E_{T}=100\;{\mathrm GeV}$
and ${\mathcal R}=0.7$).  
The total correction and fractional error are both shown as well as the
$\eta$ dependence of several individual components of the jet scale error:
cryostat factor and $\eta$ dependence in the IC region 
(Section~\ref{sec:etadep}); 
residual bias from event topology, instrumental 
background and shower containment (Section~\ref{sec:enedep}); showering 
(Section~\ref{sec:shower}).}
\label{fig:mpf_note_err_eta100}
\end{figure}

Figures~\ref{fig:mpf_note_err_tot0}-\ref{fig:mpf_note_err_eta100}
show the magnitude of the total correction and uncertainty as a function
of jet energy for pseudorapidities of 0, 1.2, and 2.
The contribution of the individual sources of systematic error are also shown.
The total uncertainty is the sum in quadrature of the individual components.
In most of the plots in this paper, 
the fixed cone algorithm with ${\mathcal R}=0.7$ is used as an
example, the data correspond to 
$\sqrt{s}=1800\; {\mathrm GeV}$ collisions,
and the luminosity is set to
$5 \times 10^{30}\;{\mathrm cm}^{-2}{\mathrm sec}^{-1}$.

The overall correction factor to jet energy in the central calorimeter is 
1.160 and 1.120 at 70 and 400~GeV, respectively. The total uncertainties at the
same energies are 0.015 and 0.023.
At lower energies, larger pseudorapidities, and smaller cone sizes, 
the corrections and errors increase.
The procedure is verified with a Monte Carlo simulation which shows that the 
jet energy is corrected by this procedure to the particle level to within 
the quoted errors.

%
We thank the NuTeV Collaboration for permission to use their unpublished data.
We also thank the staffs at Fermilab and collaborating institutions for their
contributions to this work, and acknowledge support from the 
Department of Energy and National Science Foundation (U.S.A.),  
Commissariat  \` a L'Energie Atomique (France), 
State Committee for Science and Technology and Ministry for Atomic 
   Energy (Russia),
CAPES and CNPq (Brazil),
Departments of Atomic Energy and Science and Education (India),
Colciencias (Colombia),
CONACyT (Mexico),
Ministry of Education and KOSEF (Korea),
and CONICET and UBACyT (Argentina).

\clearpage
\newpage

\begin{table}[p]
\begin{center}
\caption{Quality cuts on photon candidates to reject background.}
\label{tab:phocut}
\begin{tabular}{cccccc}
\hline                  
   Variable             & $E_{T\gamma}<15 \; {\mathrm GeV}$    
&  $E_{T\gamma}>15 \; {\mathrm GeV}$   \\  \hline
dE/dx                 & $<0.8$ or $>1.5$          
& $<0.6$ or $> 1.5$            \\  
$\varepsilon_{t}$       & $<0.25$ or $>0.75$         
& $<0.1$ or $> 0.9$            \\  
$\sigma_{\mathrm trk}$          & $>3.0$                  
& $>3.0$                    \\  
EMF                     & $>0.9$                  
& $>0.96$                   \\  
$f_{\mathrm iso}$               & $<0.5$                  
& $<0.15$                   \\  \hline
\end{tabular}
\end{center}
\end{table}

\clearpage
\newpage

\begin{table}[p]
\begin{center}
\caption{Means and RMS values for the residuals in $\eta$-dependent response
derived from $\gamma$-jet data. Residual is defined as the
difference between the response measured from data after the 
$\eta$-dependent corrections and the ideal
response.}
\begin{tabular}{ccc}
\hline
$\eta$ Region & Mean & RMS \\
\hline
$|\eta|<0.5$ & 0.0 & 0.008 \\
$0.5<|\eta|<1.0$ & 0.004 & 0.002 \\
$1.0<|\eta|<1.5$ & 0.010 & 0.005 \\
$1.5<|\eta|<2.0$ & 0.004 & 0.004 \\
$2.0<|\eta|<2.5$ & 0.0   & 0.009 \\
$2.5<|\eta|<3.0$ & 0.010 & 0.010 \\
\hline
\end{tabular}
\label{tab:meangamma}
\end{center}
\end{table}

\clearpage
\newpage

\begin{table}[p]
\begin{center}
\caption{Means and RMS values for the residuals in $\eta$-dependent response
derived from jet-jet data. Residual is defined as the difference
between the response measured from data after the $\eta$-dependent corrections
and the ideal response.}
\begin{tabular}{ccc}
\hline
$\eta$ Region & Mean & RMS \\
\hline
$|\eta|<0.5$ & 0.0 & 0.005 \\
$0.5<|\eta|<1.0$ & 0.007 & 0.005 \\
$1.0<|\eta|<1.5$ & 0.001 & 0.024 \\
$1.5<|\eta|<2.0$ & 0.010 & 0.022 \\
\hline
\end{tabular}
\label{tab:mean}
\end{center}
\end{table}

\clearpage
\newpage
 
\begin{table}[p]
\begin{center}
\caption{Parameters
of the fit to $R_{\mathrm{jet}}$ for different cone sizes.}
\begin{tabular}{ccccc} \hline
\multicolumn{5}{c}{Jet Response Parameters} \\ \hline
Cone & 1.0 & 0.7 & 0.5 & 0.3  \\ \hline
$a$  & 0.6739  & 0.6802  &  0.6807 &  0.6822      \\
     & $\pm$0.0485 & $\pm$0.0515  & $\pm$0.0506 & $\pm$0.0555      \\ \hline
$b$     & 0.0433 &  0.0422  &  0.0429 &  0.0439      \\
     & $\pm$0.0211 &  $\pm$0.0236  &  $\pm$0.0235 &  $\pm$0.0263      \\ \hline
$c$     & -0.0013 & -0.0013  & -0.0014 & -0.0015      \\
     & $\pm$0.0025  &  $\pm$0.0027  &  $\pm$0.0027 &  $\pm$0.0031    \\ \hline
\end{tabular}
\label{tab:resp_pars}
\end{center}
\end{table}

\clearpage
\newpage

\begin{table}[p]
\begin{center}
\caption{Correlation matrix for error band in jet response correction
(${\mathcal R}=0.7$).}
\begin{footnotesize}
\begin{tabular}{cccccccccccc} \hline
\multicolumn{12}{c}{Correlations for Fit Error (${\mathcal R}=0.7$)} \\ \hline
$E$ (GeV) & 10&    20&    35&    50&    75&   100&   150&  200& 300& 400&  
 500 \\ \hline
 10&  1.00&  0.98&  0.71& -0.37& -0.70& -0.65& -0.39& -0.13&  0.21&  0.38& 
 0.48 \\
 20&  0.98&  1.00&  0.81& -0.24& -0.65& -0.64& -0.43& -0.20&  0.11&  0.28& 
 0.37 \\
 35&  0.71&  0.81&  1.00&  0.35& -0.19& -0.28& -0.25& -0.15&  0.00&  0.10& 
 0.15 \\
 50& -0.37& -0.24&  0.35&  1.00&  0.83&  0.70&  0.50&  0.33&  0.09& -0.03&
 -0.10 \\
 75& -0.70& -0.65& -0.19&  0.83&  1.00&  0.97&  0.82&  0.62&  0.32&  0.14& 
 0.04 \\
100& -0.65& -0.64& -0.28&  0.70&  0.97&  1.00&  0.93&  0.78&  0.51&  0.34& 
 0.24 \\
150& -0.39& -0.43& -0.25&  0.50&  0.82&  0.93&  1.00&  0.96&  0.79&  0.67& 
 0.58 \\
200& -0.13& -0.20& -0.15&  0.33&  0.62&  0.78&  0.96&  1.00&  0.94&  0.85& 
 0.79 \\
300&  0.21&  0.11&  0.00&  0.09&  0.32&  0.51&  0.79&  0.94&  1.00&  0.98& 
 0.95 \\
400&  0.38&  0.28&  0.10& -0.03&  0.14&  0.34&  0.67&  0.85&  0.98&  1.00& 
 0.99 \\
500&  0.48&  0.37&  0.15& -0.10&  0.04&  0.24&  0.58&  0.79&  0.95&  0.99&  
1.00 \\ \hline
\end{tabular}
\end{footnotesize}
\label{tab:cor_7b}
\end{center}
\end{table}

\clearpage
\newpage

\clearpage
\newpage

\clearpage
\newpage

\clearpage
\newpage

\clearpage
\newpage

\clearpage
\newpage

\clearpage
\newpage

\clearpage
\newpage

\clearpage
\newpage

\clearpage
\newpage

\clearpage
\newpage

\clearpage
\newpage

\clearpage
\newpage

\clearpage
\newpage

\clearpage
\newpage

\clearpage
\newpage

\clearpage
\newpage

\clearpage
\newpage

\clearpage
\newpage

\clearpage
\newpage

\clearpage
\newpage

\clearpage
\newpage

\clearpage
\newpage

\clearpage
\newpage

\clearpage
\newpage

\clearpage
\newpage

\clearpage
\newpage

\clearpage
\newpage

\clearpage
\newpage

\clearpage
\newpage

\clearpage
\newpage

\clearpage
\newpage

\clearpage
\newpage

\clearpage
\newpage

\clearpage
\newpage

\clearpage
\newpage

\clearpage
\newpage

\clearpage
\newpage

\clearpage
\newpage

\clearpage
\newpage

\clearpage
\newpage

\clearpage
\newpage

\clearpage
\newpage

\clearpage
\newpage

\clearpage
\newpage

\clearpage
\newpage

\clearpage
\newpage

\clearpage
\newpage

\clearpage
\newpage

\clearpage
\newpage

\clearpage
\newpage

\clearpage
\newpage

\clearpage
\newpage

\end{document}